\pdfoutput=1
\newcommand*{\ATLASLATEXPATH}{}
\documentclass[PAPER, atlasdraft=true, texlive=2016, UKenglish, cernpreprint]{\ATLASLATEXPATH atlasdoc}
 
\usepackage{\ATLASLATEXPATH atlaspackage}
\usepackage{\ATLASLATEXPATH atlasbiblatex}
 
\usepackage{\ATLASLATEXPATH atlasphysics}
 
\graphicspath{{logos/}{figures/}}
 
\usepackage{ANA-STDM-2017-38-PAPER-defs}

\AtlasTitle{Measurements of the production cross-section for a $Z$ boson in association with $b$-jets in proton--proton collisions at $\sqrt{s} = 13$ TeV with the ATLAS detector}
 
\PreprintIdNumber{CERN-EP-2020-022}
\AtlasAbstract{
This paper presents a measurement of the production cross-section of a $Z$ boson  in association with $b$-jets, in proton--proton collisions at
$\sqrt{s} = 13$ TeV with the ATLAS experiment at the Large Hadron Collider using data corresponding to an integrated luminosity of 35.6 fb$^{-1}$.
Inclusive and differential cross-sections are measured for events containing a $Z$
boson decaying into electrons or muons and produced in association with at least one or at least two $b$-jets with transverse momentum $p_\textrm{T}>$ 20 GeV and rapidity $|y| < 2.5$.
Predictions from several Monte Carlo generators based on leading-order (LO) or next-to-leading-order (NLO) matrix elements interfaced with a parton-shower simulation and testing different flavour schemes for the choice of initial-state partons are compared with measured cross-sections.
The 5-flavour number scheme predictions at NLO accuracy agree better with data than 4-flavour number scheme ones. The 4-flavour number scheme predictions underestimate data in events with at least one b-jet.}
 
\author{The ATLAS Collaboration}
 
\AtlasRefCode{STDM-2017-38}

\AtlasCoverSupportingNote{Measurement of a $Z$ boson produced in association with heavy flavour jets in 36 fb$^{-1}$ at $\sqrt{s}=13$ TeV}{https://cds.cern.ch/record/2293067}
 
\AtlasCoverCommentsDeadline{1 March 2020}
 
\AtlasCoverAnalysisTeam{
Valentina Cairo, Valerio Dao, Chiara Debenedetti, Laura Fabbri, Benedetto Giacobbe, Kevin Hildebrand, Vivek Jain, Marco Aurelio Lisboa Leite,
Gennady Lykasov, Josh Mcfayden, Evelin Meoni, Mark Oreglia, Jacob Martin Pasner, Jose Luis La Rosa Navarro, Douglas Schaefer, Federico Sforza, Stewart Swift, Semen Turchikhin, Camilla Vittori, Francesco Giuli}
 
\AtlasCoverEdBoardMember{Ulrike Blumenschein (chair)}
\AtlasCoverEdBoardMember{Yuji Enari}
\AtlasCoverEdBoardMember{Kristof Schmieden}
 
\AtlasCoverEgroupEditors{atlas-STDM-2017-38-editors@cern.ch}
 
\AtlasCoverEgroupEdBoard{atlas-STDM-2017-38-editorial-board@cern.ch}
 
\AtlasJournal{JHEP}
\hypersetup{pdftitle={ATLAS document},pdfauthor={The ATLAS Collaboration}}
 
\begin{document}
\AtlasJournalRef{JHEP 07 (2020) 044}
\AtlasDOI{10.1007/JHEP07(2020)044}
 
\maketitle
 
\tableofcontents
 
\section{Introduction}
 
The measurement of the production rate of a $Z$ boson in  association with jets originating from
\bquarks \footnote{Unless otherwise mentioned, it is implicitly assumed that \bquark\ refers to both $b$-quark and $\bar{b}$-antiquark.} (\Zbjets)
in proton--proton ($pp$) collisions provides an important test of perturbative quantum chromodynamics (pQCD).
Current predictions for \Zbjets\ production are known at  next-to-leading-order (NLO) accuracy in pQCD, and they can be derived
in either a 4-flavour number scheme (4FNS) or a 5-flavour number scheme (5FNS)~\cite{Vbteo1,Vbteo2,Vbteo3,Vbteo5}.
In the 4FNS,  \bquarks\  do not contribute to the parton distribution functions (PDFs) of the proton and, in QCD, they only appear in a massive final state due to  gluon splitting ($g \to bb$).
In the 5FNS, \bquark\ density is allowed in the initial state via a \bquark\ PDF, with the \bquark\ typically being massless.
Therefore, in the 5FNS the \Zbjets\  cross-section is sensitive to the \bquark\ PDF and can be used to constrain it.
The ambiguity among the schemes is an intrinsic property of the calculation and is expected to reduce with the inclusion of higher order perturbative corrections~\cite{Vbteo3}.

Furthermore, the measurement of \Zbjets\ production provides a benchmark to test predictions from Monte Carlo (MC) simulations. These are commonly used to estimate the background contribution of $Z+b$-jet events to other topologies, such as the production of a Higgs boson decaying into a \bquark\ pair in association with a $Z$ boson, or in searches for  physics beyond the SM with signatures containing leptons and \bjets\ in the final state.
 
The  \Zbjets\ processes occur more rarely than the production of \Zbos\ events with inclusive jets (\Zjet s) and they are more challenging to measure.
The \bjets\ are identified by exploiting the long lifetime of $b$-hadrons produced in the quark hadronisation, and a higher level of background affects the measurement.
The background is mainly composed of events with a $Z$ boson associated with light-flavour jets or $c$-jets\footnote{A
$c$-jet is a jet originating from a \cquark.}, misidentified as \bjets, and events from the dileptonic decay of a $t \bar{t}$ pair.

Inclusive and differential cross-sections of \Zbjets\ production have been measured in proton--antiproton collisions at the centre-of-mass energy of $\sqrt{s} = 1.96$~\TeV\ by the CDF and D0 experiments~\cite{Aaltonen:2008mt, Abazov:2013uza, Aaltonen:2009qi, D0:2012qt} and at the
Large Hadron Collider (LHC)~\cite{Evans:2008zzb} in $\sqrt{s} = 7$~\TeV\ $pp$
collisions by the ATLAS and CMS experiments~\cite{STDM-2012-11, STDM-2012-15, Chatrchyan:2014dha, Chatrchyan:2013zja, Chatrchyan:2013uza, Chatrchyan:2012vr}, as well as in $\sqrt{s} = 8$~\TeV\ $pp$ collisions by the CMS experiment~\cite{Khachatryan:2016ipq, Khachatryan:2016iob}.
The CMS experiment also recently released a measurement of the ratio of \Zbjet s to  \Zjet s cross-sections and the ratio of \Zcjet s
to \Zbjets\ cross-sections for events with at least one \bjet\ or one \cjet\ in $\sqrt{s} = 13$~\TeV\ $pp$ collisions~\cite{CMSNEW}.
 
This paper presents a measurement of the inclusive and differential production cross-sections of a $Z$ boson, decaying into electrons or muons, in association with at least one or at least two \bjets\ using  35.6~fb$^{-1}$ of  $pp$ collision data collected by the ATLAS experiment at $\sqrt{s} = 13$~\TeV\  in 2015 and 2016.
For events with at least one \bjet, the differential cross-sections are presented as a function of the transverse momentum\footnote{ATLAS uses a right-handed coordinate system with its origin at the nominal interaction point (IP) in the centre of the detector and the $z$-axis along the beam pipe. The $x$-axis points from the IP to the centre of the LHC ring, and the $y$-axis points upwards. Cylindrical coordinates $(r,\phi)$ are used in the transverse plane, $\phi$ being the azimuthal angle around the $z$-axis. The pseudorapidity is defined in terms of the polar angle $\theta$ as $\eta = -\ln \tan(\theta/2)$. Angular separation is measured in units of $\Delta R \equiv \sqrt{ (\Delta \eta)^2 + (\Delta \phi)^2}$. When dealing with massive jets and particles, the rapidity $y$ = $\frac{1}{2}\ln\frac{E+p_z}{E-p_z}$ is used, in which $E$  is the jet or particle energy and $p_z$ is the $z$-component of the jet or particle momentum.} (\pt) and the absolute value of the rapidity ($|y|$) of the leading \bjet,  the \pt\ and the $|y|$ of the $Z$ boson ($Z$ $\pt$ and $Z$ $|y|$), and as a function of
observables correlating the $Z$ boson with  the leading \bjet, namely the azimuthal angle between them ($\Delta \phi_{Zb}$), the absolute value of their rapidity difference  ($\Delta y_{Zb}$), and their angular separation ($\Delta R_{Zb}$).
For events with at least two \bjets, the differential cross-sections are presented as a function of  the \pt\ of the $Z$ boson
and as a function of observables  built using the two leading \bjets, namely
their \pt\ (\ptbb),  their invariant mass (\mbb), \ptbb\ divided by their invariant mass (\ptmbb),  the azimuthal angle between them ($\Delta \phi_{bb}$),
the absolute value of their rapidity difference  ($\Delta y_{bb}$), and their angular separation ($\Delta R_{bb}$).
The higher $\sqrt{s}$ leads to a large increase in the measured cross-section in comparison with previous ATLAS publications.
This allows more extreme regions of phase space to be explored and new measurements to be performed in the rare two-\bjets\ configuration (i.e.\ \ptbb\ and \ptmbb). Previous ATLAS measurements were compared with MC predictions based on leading-order matrix elements interfaced with a parton-shower simulation, which showed substantial mismodelling. Recent advances in this field  permit this paper to compare the data with the latest MC predictions using next-to-leading-order matrix elements, which are expected to provide a better description of the data.

The experimental apparatus is described in Section~\ref{sec:detector}, and details of the data sample and  the MC simulations are provided in Section~\ref{sec:DataMC}.
The object definitions and the event selection at detector level are presented in Section~\ref{sec:EvtSel}.
Backgrounds that do not contain  a real $Z$ boson are estimated via MC simulations and validated in control regions  in data or via  data-driven techniques, while backgrounds containing a real $Z$ boson
and jets not originating from  \bquarks\ are estimated with a fit to data  distributions sensitive to the flavour of the jet (flavour fit); both are described in Section~\ref{sec:bkgEstimate}. Distributions of the kinematic variables are presented in Section~\ref{sec:KinPlots}.
After background subtraction, the data are unfolded to particle level  in a fiducial phase space, which is  detailed in Section~\ref{sec:Unfolding}. Systematic uncertainties in the unfolded data are discussed in
Section~\ref{sec:UnfoldingSys}. The results are presented in Section~\ref{sec:result}, and conclusions are drawn in Section~\ref{sec:conclusion}.

 
\section{The ATLAS detector}
\label{sec:detector}
The ATLAS detector~\cite{PERF-2007-01} at the LHC covers nearly the entire solid angle around the collision point.
It consists of an inner tracking detector surrounded by a thin superconducting solenoid, electromagnetic and hadronic calorimeters, and a muon spectrometer incorporating three large superconducting toroidal magnets.
 
The inner-detector system (ID) is immersed in a \SI{2}{\tesla} axial magnetic field  and provides charged-particle tracking in the range \(|\eta| < 2.5\).
The high-granularity silicon pixel detector covers the vertex region and provides four measurements for most tracks, the first hit normally being in the insertable B-layer~\cite{ATLAS-TDR-2010-19,PIX-2018-001}.
It is followed by the silicon microstrip tracker, which provides eight measurements per track. These silicon detectors are complemented by the transition radiation tracker (TRT), which enables radially extended track reconstruction up to \(|\eta| = 2.0\).  The TRT also provides electron identification information  based on the fraction of hits (typically 30 in total) with an energy deposit above the transition-radiation threshold.
 
The calorimeter system covers the pseudorapidity range \(|\eta| < 4.9\). Within the region \(|\eta|< 3.2\), electromagnetic calorimetry is provided by barrel and
endcap high-granularity lead/liquid-argon (LAr) calorimeters, with an additional thin LAr presampler covering \(|\eta| < 1.8\) to correct for energy loss in material upstream of the calorimeters.
Hadronic calorimetry is provided by the steel/scintillator-tile calorimeter, segmented into three barrel structures within \(|\eta| < 1.7\), and two copper/LAr hadronic endcap calorimeters.
The solid angle coverage is completed with forward copper/LAr and tungsten/LAr calorimeter modules optimised for electromagnetic and hadronic measurements, respectively.
 
The muon spectrometer (MS) comprises separate trigger and high-precision tracking chambers measuring the deflection of muons in a magnetic field generated by the superconducting air-core toroid magnets.
The field integral of the toroid magnets ranges between \num{2.0} and \SI{6.0}{\tesla\metre} across most of the detector.
The precision chambers cover the region \(|\eta| < 2.7\) with three layers of monitored drift tubes, complemented by cathode-strip chambers in the forward region, where the background is highest.
The muon trigger system covers the range \(|\eta| < 2.4\) with resistive-plate chambers in the barrel, and thin-gap chambers in the endcap regions.
 
Interesting events are accepted by the first-level trigger system implemented in custom hardware, followed by selections made by algorithms implemented in software in the high-level trigger~\cite{TRIG-2016-01}.
The first-level trigger accepts events from the \SI{40}{\MHz} bunch crossings at a rate below \SI{100}{\kHz}, which the high-level trigger further reduces in order to record events to disk at about \SI{1}{\kHz} rate.


\section{Data set and simulated event samples}
\label{sec:DataMC}
\subsection{Data set description}

The data used in this measurement were recorded in 2015 and 2016 with the ATLAS detector
at the LHC in $pp$ collisions at $\sqrt s = 13$~\TeV.
The candidate events were selected by either a single-electron or single-muon trigger
that imposed a minimum transverse energy (transverse momentum) threshold for the electron (muon) channel and quality and isolation requirements, which depended on the LHC running conditions.
The threshold in 2015 was 24 (20)~\GeV\ for the electrons (muons), satisfying loose isolation requirements. Due to the higher instantaneous luminosity in 2016, the threshold was increased to 26~\GeV\ for both the electrons and the muons, and a more restrictive isolation requirement was imposed on both leptons along with more restrictive identification requirements for electrons.
Triggers with  higher thresholds but with no isolation requirement or with loosened identification criteria were also used to increase the efficiency.
Crossings of proton bunches occurred every 25~ns,  the collisions achieved a peak instantaneous luminosity of \mbox{$1.37\times 10^{34}$~cm$^{-2}$s$^{-1}$}, and the mean number of $pp$ interactions per bunch crossing (pile-up) was $\langle \mu \rangle = 24$.
After applying criteria to ensure good ATLAS detector operation, the total integrated luminosity amounts to $35.6~\mathrm{fb}^{-1}$.
The uncertainty in the combined 2015--2016 integrated luminosity is $2.1\%$~\cite{ATLAS-CONF-2019-021}, obtained using the LUCID-2 detector~\cite{LUCID2} for the primary luminosity measurements.
 
\subsection{Simulated event samples for signal and background processes}\label{subsec:MC}
 
MC simulations  are used to describe signal events, to estimate the contribution of background processes, to unfold the data yield  to the particle level, to estimate systematic uncertainties, and to compare predictions with the unfolded data distributions.
 
An overview of all signal and background processes and the generators used for the production of nominal results
is given in Table~\ref{tab:generators} together with the theory uncertainty in the normalisation cross-sections corresponding to PDFs and scale variations.
 
Inclusive $Z(\to \ell \ell, \ell = e, \mu)$
production in association with both light- and heavy-flavour jets was simulated using the \sherpa~v2.2.1~\cite{Bothmann:2019yzt}  generator.
In this set-up, matrix elements at NLO for up to two partons, and matrix elements at LO for up to four partons, were calculated with the Comix~\cite{Gleisberg:2008fv} and
OpenLoops~\cite{Cascioli:2011va,Denner:2016kdg} libraries.
They were matched with the \sherpa parton shower~\cite{Schumann:2007mg} using the MEPS@NLO prescription~\cite{Hoeche:2011fd,Hoeche:2012yf,Catani:2001cc,Hoeche:2009rj}.
\sherpa  uses the 5FNS with massless $b$- and $c$-quarks in the matrix element, but massive quarks in the parton shower.
Samples were generated using  the  NNPDF3.0nnlo PDF set~\cite{Ball:2014uwa}, along with the dedicated set of tuned parton-shower parameters developed by the \sherpa authors.
In Section~\ref{sec:result}, where several predictions are compared with the unfolded data, these samples are shown with their uncertainties and are referred to as \sherpa \textsc{5FNS (NLO)}.
The uncertainties account for missing higher orders and are evaluated~\cite{Bothmann:2016nao} using seven variations of the QCD factorisation and renormalisation scales in the matrix elements by factors of $0.5$ and $2$ and avoiding variations in opposite directions.

Additional $Z(\to \ell \ell)$ samples  were produced with the LO matrix-element generator \alpgen~v2.14~\cite{Mangano:2002ea}, interfaced with \pythia~v6.426~\cite{Sjostrand:2006za} to model parton showers, using the parameter values of the Perugia2011C tune~\cite{Skands:2010ak}  for simulating the underlying event, and the CTEQ6L1 PDF set~\cite{Pumplin:2002vw}.
Matrix elements were calculated for up to five partons, and merged using the MLM prescription~\cite{Alwall:2007fs} with a matching scale of 15~\GeV.
\alpgen uses the 4FNS with massive $b$- and $c$-quarks in the matrix element and in the parton shower of \pythia. The matrix elements for the production of $Z+b\bar{b}$ and $Z+c\bar{c}$ events are explicitly included and a heavy-flavour overlap procedure is used to remove the double counting, between the matrix element and the parton shower, of heavy quarks from gluon splitting.
The  properties of  $b$- and $c$-hadron decays were simulated with   EvtGen~v1.2.0~\cite{Lange:2001uf}, as was done in all generated samples where the parton shower was simulated with \pythia.
Photos++~v3.52~\cite{Golonka:2005pn,Davidson:2010ew} was used to simulate QED final-state radiation (FSR).
The \alpgen samples are used in the analysis to estimate systematic uncertainties in the unfolding procedure and in backgrounds containing a genuine $Z$ boson.
In Section~\ref{sec:result} these samples are  referred to as \alpgen + \pythiasixshort  \textsc{4FNS (LO)}.
Samples of $Z(\to \tau \tau)$, $W (\to \ell \nu)$, and $W (\to \tau \nu)$  events were simulated with \sherpa, using the same set-up adopted for the signal samples.
 
The $Z$-boson and $W$-boson samples are normalised to the inclusive next-to-next-to-leading-order (NNLO) cross-section predictions provided by the FEWZ~3.1 program~\cite{Anastasiou:2003ds,Fewz2,Fewz21,Fewz21ewc} with the CT14 PDF set.
The $K$-factor applied to the $Z$ samples  to match the NNLO prediction is 0.975 for \sherpa and  1.196 for \alpgen.
 
The production of $t\bar{t}$  events with at least one $W$ boson decaying leptonically was modelled using the \POWHEGBOX
~\cite{Frixione:2007nw,Nason:2004rx,Frixione:2007vw,Alioli:2010xd}~v2
generator at NLO with the NNPDF3.0NLO~\cite{Ball:2014uwa} PDF set. The \hdamp\ parameter,
which regulates the high-\pt\ emissions against which the $t\bar{t}$ system recoils, is set to 1.5~\mtop~\cite{ATL-PHYS-PUB-2016-020}.
The events were interfaced with \pythia~v8.230~\cite{Sjostrand:2014zea} using the A14 tune~\cite{ATL-PHYS-PUB-2014-021}.
The $t\bar{t}$ sample is normalised to  the theory prediction at NNLO
in QCD including the resummation of next-to-next-to-leading logarithmic (NNLL) soft-gluon terms~\cite{Beneke:2011mq,Cacciari:2011hy,Baernreuther:2012ws,Czakon:2012zr,Czakon:2012pz,Czakon:2013goa,Czakon:2011xx}.
Four additional $t\bar{t}$ samples were simulated to evaluate the uncertainty in this process, as described in~\cite{ATL-PHYS-PUB-2016-020}.
One sample was produced with \mgamc~\cite{MG52014} and the same parton-shower model of the nominal $t\bar{t}$ sample in order to  estimate the uncertainty due to the modelling of the hard scattering process.
A second \POWHEGBOX sample  showered with \HERWIGV{7.13}~\cite{Bahr:2008pv,Bellm:2015jjp} was generated to evaluate the uncertainty due to the modelling of the parton shower and hadronization processes.
A third sample was produced to simulate higher energy radiation with the factorisation and renormalisation scales changed by a factor of 0.5 while simultaneously increasing
the \hdamp\ value to 3.0~\mtop\ and using the upper variation of the initial state radiation (ISR) from the A14 tune. The last sample simulates the lower energy radiation.
It was generated with the renormalisation and factorisation scales varied by a factor of 2.0 while keeping the \hdamp\ value at 1.5~\mtop\ and using the ISR downward variation in the parton shower.
The last two samples are also used to estimate the impact of FSR with  parton-shower weights that vary the renormalisation scale for QCD emission in the FSR by factors of 0.5 and 2.0.

Single-top-quark events in the $Wt$-, $s$- and $t$-channels were generated using the \POWHEGBOXV{v1} generator interfaced with \pythia~v6.4~\cite{Sjostrand:2006za}; the latter simulates  parton showers, fragmentation,
and the underlying event using the Perugia 2012 tune~\cite{Skands:2010ak}. The CT10 PDF set was used~\cite{CT10}.
The single-top  samples for the $t$- and $s$-channels are normalised to cross-sections from NLO predictions~\cite{StXS1,StXS2}, while the $Wt$-channel  sample is normalised to cross-sections from approximate NNLO predictions~\cite{tWXS}.
 
Diboson processes ($WW$, $WZ$, and $ZZ$) with  one of the bosons decaying hadronically and the other leptonically were generated using \sherpa~v2.2.1 with the CT10nlo PDF set.
The matrix element includes up to one parton at NLO and up to three additional partons at LO.
The samples are normalised to the NLO predictions~\cite{Gleisberg:2008ta}.
 
Simulated events for $qq\to VH (\to b \bar {b})$ with $V= W$ or $Z$  plus zero or one jet production at NLO were generated with the \POWHEGBOXV{v2} + GoSam + MiNLO generator~\cite{Alioli:2010xd,H1,Hamilton:2012np,H3}
with the NNPDF3.0NLO PDF set. The contribution from $gg \to ZH  (\to b \bar {b})$  production was simulated using the LO \POWHEGBOXV{v2} matrix-element generator.
The samples of simulated events include all final states where the Higgs boson decays into $b\bar{b}$ and the vector boson into a leptonic final state.
The mass of the Higgs boson is set to 125~\GeV\ and the $H\to b\bar{b}$ branching fraction is set to 58\%.
The $qq\to VH (\to b \bar {b})$ cross-section is calculated at  NNLO (QCD) and NLO (EW), while the $gg \to ZH $ cross-section is calculated at NLO+NLL (QCD).
 
\begin{table}[t]
\begin{center}
\begin{small}
\begin{tabular}{ c | l l c c }
\toprule
\raisebox{-0.4ex}{Process} & \raisebox{-0.4ex}{Generator}  &\raisebox{-0.4ex}{Order of} & \raisebox{-0.4ex}{Reference} & \raisebox{-0.4ex}{Normalisation}  \\
&         &    \raisebox{-0.4ex}{cross-section}  & \raisebox{-0.4ex}{normalisation}  & \raisebox{-0.4ex}{cross-section}\\
&         &    \raisebox{-0.4ex}{calculation}  &                     & \raisebox{-0.4ex}{uncertainty}\\
\hline
\hline
$Z\rightarrow \ell \ell$ ($\ell = e,\mu,\tau$ )& \sherpa &  NNLO & \cite{Anastasiou:2003ds,Fewz2,Fewz21,Fewz21ewc} & $5\%$\\
with $66<m_{\ell\ell}<116$~\GeV                  &         &       &                     &        \\\hline
$W\rightarrow \ell \nu$ \;\;($\ell = e,\mu,\tau$) & \sherpa  & NNLO & \cite{Anastasiou:2003ds,Fewz2,Fewz21,Fewz21ewc} & $5\%$\\\hline
$t\bar{t}$                                        &   \POWHEGBOX       & NNLO + NNLL  & ~\cite{Beneke:2011mq,Cacciari:2011hy,Baernreuther:2012ws,Czakon:2012zr,Czakon:2012pz,Czakon:2013goa,Czakon:2011xx}& $6 \%$ \\
&          &  ($\mtop =172.5$~\GeV) & & \\ \hline
Single top               & \POWHEGBOX  &  NLO & & $6\%$ \\
($t$-, $Wt$-, $s$-channel)              &  ($\mtop =172.5$~\GeV)  & &  &  \\  \hline
Dibosons  & & &  & \\
$Z(\to\ell\ell) + Z(\to qq)$, & \sherpa & NLO &\cite{Gleisberg:2008ta} & $5\%$\\
$W(\to\ell\nu) + W(\to qq)$ )& &  & & \\\hline
Higgs     & &     &  &    \\
$qq \to Z(\to\ell\ell) + H(\to b\bar{b})$ &\POWHEGBOX  & NNLO QCD + NLO EW & ~\cite{Brein:2004Higgs1,Denner2012,Altenkamp2013} & $3\%$ \\
$gg \to Z(\to\ell\ell) + H(\to b\bar{b})$ &            & NLO + NLL          &  &  \\
$qq \to W(\to\ell\nu) + H(\to b\bar{b})$ &            & NNLO QCD + NLO EW  &    & \\
\bottomrule
\end{tabular}
\end{small}
\caption{Signal and background MC samples:
the generator programs used in the simulation are listed in the second column, the order of the QCD calculation and the reference used for the calculations of the normalisation cross section are reported in the third and fourth columns.
The normalisation cross-section uncertainty in the final column corresponds to PDFs and scale variations.
\label{tab:generators}}
\end{center}
\end{table}
 
Generated events were processed with the ATLAS detector simulation~\cite{SOFT-2010-01}, based on $\GEANT4$~\cite{Agostinelli:2002hh}, to simulate the detector response to final-state particles. To account for the effects of pile-up, multiple overlaid $pp$ collisions were simulated with the soft QCD processes of \pythia~v8.186 using the A2 tune~\cite{ATLAS-PHYS-PUB-2012-003} and the MSTW2008LO PDF set~\cite{Martin:2009iq}.
The distribution of the average number of interactions per bunch crossing in the simulation is weighted to reflect that in the data.
Simulated events are processed with the same reconstruction algorithms as for the data.
 
\subsection{Theoretical predictions}
 
In addition to particle-level predictions from the fully simulated \sherpa and  \alpgen samples described above, unfolded results from data are compared  with six other  predictions listed in Table~\ref{tab:theopred}.
 
Two particle-level predictions (using specific parton-shower and matching predictions) were produced with the \sherpa~v2.2.7 generator using NLO matrix elements~\cite{Sherpa226}.
The first sample,  referred to as \sherpaZbbNLO, includes $Z+b\bar{b}$ events generated in the 4FNS at NLO with massive $b$-quarks.
It is interesting to compare this sample, which contains two $b$-quarks in the matrix elements,  with the unfolded data even in the case of distributions with at least one $b$-jet, to understand if there are regions of the phase space that can be described with such a configuration.
The second sample, referred to as \sherpaF \textsc{4FNS+5FNS (NLO)}, contains the matrix elements at NLO for up to two partons, and matrix elements at LO for up to three partons.
It includes both $Z+b\bar{b}$ events generated in the 4FNS at NLO with massive $b$-quarks, and  $Z+$jets events generated in the 5FNS at NLO. They are combined according to the procedure described in Ref.~\cite{SherpaFusing}.
The combination is achieved by means of a dedicated heavy-flavour overlap removal procedure, the fusing technique, that acts as an additional step after the multijet merging algorithms.
This procedure combines the advantages of inclusive 5FNS calculations with the higher precision of 4FNS calculations in regions of phase space where the $b$-quark mass sets a relevant scale.
The two \sherpa samples use the NNPDF3.0nnlo PDF set with $\alphas(m_Z)=0.118$ and the corresponding number of active quark flavours. Masses of $c$- and $b$-quarks are taken into account in the parton shower in all \sherpa samples.
 
Results are also compared with predictions from the LO matrix-element generator  \mgamc v2.2.2~\cite{MG52014} interfaced with \pythia~v8.186~\cite{Sjostrand:2014zea} with the A14 tune~\cite{ATL-PHYS-PUB-2014-021} to model the parton shower and underlying event.
The matrix element includes up to four partons. Additional jets are produced by the parton shower, which uses the CKKW-L merging procedure~\cite{Hesketh:2041288}, with a matching scale of 30~\GeV.  \mgamc uses the 5FNS with massless $b$- and $c$-quarks in the matrix element, and massive quarks in the parton shower.
The NNPDF3.0nlo PDF set is used with $\alphas(m_Z)= 0.118$. This prediction is referred to as \textsc{MGaMC} + \pythiaeightshort \textsc{5FNS (LO)}.
 
Two additional predictions were produced with \mgamc v2.6.2, using matrix-element calculations with NLO accuracy.
The first sample includes \Zjet s events generated in the 5FNS with up to one parton at NLO, and massless $b$- and $c$-quarks;
the second sample includes $Z+b\bar{b}$ events generated in the 4FNS at NLO, and massive $b$-quarks. Both samples were generated using the NNPDF3.0nnlo PDF set with $\alphas=0.118$. They were interfaced to the \pythia~v8.186 parton shower using the FxFx merging scheme~\cite{Frederix:2012ps}, with a matching scale of 25~\GeV. As in the previous case, massive $c$- and $b$-quarks are produced in the parton shower. The first sample is referred to as \textsc{MGaMC} + \pythiaeightshort \textsc{5FNS (NLO)}; the second is referred to as \textsc{MGaMC} + \pythiaeightshort \textsc{Zbb 4FNS (NLO)}.
 
An additional \alpgen prediction is used to test the sensitivity of the measurements to the parton structure of the proton. The \alpgen samples presented in Section~\ref{subsec:MC}  are reweighted to the NNPDF3.0lo PDF set, using the prescriptions reported in Ref.~\cite{Butterworth:2015oua}.
These predictions are referred to as
\alpgen + \pythiasixshort (rew. NNPDF3.0lo). The predictions of LO MC generators, such as \alpgen + \pythiasixshort  \textsc{4FNS (LO)} and  \textsc{MGaMC} + \pythiaeightshort \textsc{5FNS (LO)},
with up to four or five partons in the matrix element, are still an interesting case to study as they allow comparison with the predictions of MC generators at NLO accuracy and with a smaller number of partons in the matrix element.
Furthermore, they provide a benchmark in common with past analyses, such as in Ref.~\cite{STDM-2012-15}.

\begin{table}[t]
\begin{center}
\begin{small}
\begin{tabular}{l|c|c|c|c|l}
\toprule
\raisebox{-0.4ex}{Generator} & \multicolumn{2}{c|}{$N_\mathrm{max}^\mathrm{partons} $ }  & FNS     &   PDF & Parton \\
&  NLO        &  LO                                 &           & set &          Shower \\
\hline
\hline
\multicolumn{5}{c}{ $Z$+jets (including $Z$+$b$ and $Z$+$bb$)} \\
\hline
\sherpa \textsc{5FNS (NLO)}       & 2 & 4  & 5 &  NNPDF3.0nnlo &\sherpa\\
\sherpaF \textsc{4FNS+5FNS (NLO)} & 2   & 3  & ~~~~~~5 (*)  &   NNPDF3.0nnlo& \sherpa            \\
\alpgen  + \pythiasixshort  \textsc{4FNS (LO)}&- &5 &4 & CTEQ6L1 &\pythia~v6.426\\
\alpgen  + \pythiasixshort  (rew. NNPDF3.0lo)&- &5 &4 & NNPDF3.0lo&\pythia~v6.426\\
\textsc{MGaMC} + \pythiaeightshort \textsc{5FNS (LO)} &- &4 &5   & NNPDF3.0nlo&  \pythia~v8.186 \\
\textsc{MGaMC} + \pythiaeightshort \textsc{5FNS (NLO)} &1 &- &5  &NNPDF3.0nnlo&\pythia~v8.186\\
\hline
\hline
\multicolumn{5}{c}{ $Z$+$bb$} \\
\hline
\sherpaZbbNLO & 2 &- &4 &  NNPDF3.0nnlo&\sherpa\\
\textsc{MGaMC} + \pythiaeightshort \textsc{Zbb 4FNS (NLO)} &2 &- & 4&NNPDF3.0nnlo &\pythia~v8.186 \\
\bottomrule
\end{tabular}
\end{small}
\caption{
Summary of theoretical predictions for the signal, including the maximum number of partons at each order in $\alphas$, the flavour number scheme (FNS),
the PDFs set and the parton shower.
(*)   Details of the merging between 4FNS and 5FNS in \sherpaF \textsc{4FNS+5FNS (NLO)} are available in Ref.~\cite{SherpaFusing}.
\label{tab:theopred}
}
\end{center}
\end{table}
 
 
\section{Event selection}
\label{sec:EvtSel}
Events selected in this analysis are required to have  a signature consistent with a $Z$ boson,
decaying into two electrons or two muons, in association with at least one or at least two \bjets.
Candidate events are required to have a primary vertex (PV), defined as the vertex with the highest sum of track $\pT^2$ with at least two associated  tracks measured in the ID (ID tracks), each with $\pt > 400$~\MeV.
 
Electron candidates are  reconstructed by matching a cluster of energy deposited in the EM calorimeter  to a well-reconstructed ID track. Electrons are identified using a likelihood function based on variables describing the shape of the electromagnetic showers in the calorimeter, track properties, and track-to-cluster matching quantities~\cite{PERF-2017-01}. Electrons must satisfy the `tight' likelihood requirement.
Electron candidates are required to have $\pt > 27$~\GeV\ and  $|\eta| < 2.47$. Candidates in the transition region between the
barrel and endcap electromagnetic calorimeters, $1.37 < |\eta|  < 1.52$, are excluded.
 
Muon candidates are reconstructed by fitting a unique trajectory through the hits associated with a pair of
matching tracks which are reconstructed separately in the ID and the MS; the energy loss in the calorimeter is taken into account in the combination procedure.
Muons must satisfy the `medium' identification criterion based on requirements on the number of hits and on the quality of the combined fit~\cite{PERF-2015-10}. Muon candidates are required to have $\pt > 27$~\GeV\
and  $|\eta| < 2.5$.
 
To select leptons originating from the primary $pp$ interaction, the lepton tracks are required to have a longitudinal impact parameter ($z_0$) satisfying $|z_0 \sin(\theta)|<0.5$~mm relative to the  PV.
The transverse impact parameter significance ($d_0/\sigma_{d_0}$) of the electron (muon) candidates must satisfy  $d_0/\sigma_{d_0}< 5~(3)$.
In order to further suppress leptons from non-prompt processes or leptons from hadrons in jets,
both the electron and muon candidates are required to satisfy \pt-dependent cone-based isolation requirements~\cite{PERF-2015-10}, which use information from ID tracks.
The isolation requirements are set so that the scalar sum of the transverse momenta of
the tracks  in the isolation cone\footnote{The $\Delta R$ parameter of the isolation cone is defined by $\Delta R= \min(10~\GeV/\pt,0.3 )$ where \pt\ is the transverse momentum of the lepton candidate.}
around the lepton is less than 6$\%$ of the lepton \pt.

Jets are reconstructed, using the anti-$k_t$ algorithm~\cite{Cacciari:2008gp,Fastjet} with radius parameter $R = 0.4$, from topological clusters of energy deposits in the calorimeter~\cite{PERF-2014-07}.
Jets are calibrated using a simulation-based calibration scheme, followed by in situ corrections to account for differences between simulation and data~\cite{PERF-2016-04}.
Events with jets arising from detector noise or other non-collision sources are discarded~\cite{ATLAS-CONF-2015-029}.
Furthermore, to eliminate jets containing a large energy contribution from pile-up, jets with $\pT< 60$~\GeV\ and $|\eta|<2.4$ are
required to have a significant fraction of their tracks with origin compatible with the primary vertex, as defined by a jet vertex tagger discriminant (JVT)~\cite{PERF-2014-03}.
Selected jets  must have $\pt > 20$~\GeV\ and rapidity  $|y|<2.5$.
 
An overlap removal procedure is applied to electron, muon and jet candidates to prevent double counting.
Any jet whose axis lies within $\Delta R = 0.2$ of an electron is removed. If a jet is reconstructed within $\Delta R = 0.2$  of a muon and the jet has fewer than three associated tracks or the muon energy constitutes most of the jet energy, then the jet is removed.
Any electron or muon of a given \pt\ reconstructed within  $\Delta R = \min (0.4, 0.04 + 10~\GeV / \pt )$ of the axis of any surviving jet is removed.
Jets that survive the overlap removal procedure are  removed if they are within $\Delta R = 0.4$ of the selected leptons.
 
The $b$-jets, defined as the jets containing at least one $b$-hadron,
are identified using a multivariate algorithm, MV2c10~\cite{PERF-2012-04, ATL-PHYS-PUB-2016-012}.
This algorithm uses the impact parameter and reconstructed secondary vertex information of the tracks associated with the jets.
Its output  lies in the range $[-1,+1]$. A value close to $+1$ denotes a higher probability for the jet to be a $b$-jet.
The $b$-jet candidates are selected if their MV2c10 output is greater than 0.8244.
This selection corresponds to an efficiency of 70$\%$ for selecting jets containing $b$-hadrons, and misidentification rates of  0.26\% and 8.3\%, respectively, for light-flavour ($u$-, $d$-, $s$-quark and gluon) jets and $c$-jets,  as estimated from a sample of simulated $t\bar{t}$ events.
Other working points are defined by different $b$-tagging discriminant output thresholds; they are used to define control regions and to define the bins used in the flavour fit, as detailed in Section~\ref{sec:FlavFit}.

In simulation, reconstructed jets are labelled as $b$-jets if they lie within $\DR=0.3$ of one or more weakly decaying $b$-hadrons with $\pt > 5$~\GeV. Reconstructed jets not identified as $b$-jets are considered to be $c$-jets if they lie within $\DR=0.3$ of any $c$-hadron with $\pt > 5$~\GeV. All other jets are classified as light-jets.
Simulated \Zjet s events are sequentially categorised  depending on the labels of the jets, starting from \bjets,  as follows:
$Z+b$ when they have  exactly one \bjet, $Z+bb$ when
they have at least two \bjets, $Z+c$ when they have at least one \cjet, $Z+l$ when they have only light-jets.
A similar classification is adopted for simulated \Wjet s events.
In the distributions with at least one $b$-jet, the sum of $Z+b$ and $Z+bb$ samples is used to define the signal, and the \Zjet s background is constituted by the sum of the $Z+c$ and $Z+l$ samples.
In the distributions with at least two $b$-jets, the $Z+bb$ samples alone constitute the signal, while the sum of the $Z+b$, $Z+c$, and $Z+l$ samples form the \Zjet s background.
 
The missing transverse momentum (\MET), which may correspond to  a neutrino escaping interaction with the detector,
is defined as the negative vector sum of the transverse momentum of all identified hard physics objects (electrons, muons, jets),
as well as an additional track-based soft term defined in Ref.~\cite{PERF-2016-07}.
 
Events are required to have exactly two leptons\footnote{At least one of the lepton candidates is required to match the lepton that triggered the event.} of the same flavour ($ee$ or $\mu \mu$) but of opposite charge with their dilepton invariant mass in the range \SI{76}{\GeV}$<m_{\ell\ell}<$\SI{106}{\GeV}. Events with \mbox{$\ensuremath{p_{\mathrm{T}}}^{\ell\ell} < 150$~\GeV} must also have $\MET<$~\SI{60}{\GeV}.
The requirement on the $\MET$ value reduces by about 55\% the background from $t\bar{t}$ events with dileptonic decay, while the signal is reduced by about 5\%.
Events passing the above selection and having at least one or at least two jets belong to the region referred to as the pre-tag region.
The signal region is a subset of the pre-tag region.
Events belonging to the signal region are assigned to two regions: those with at least one \bjet, referred to as the 1-tag region; and those with at least two \bjets, referred to as the 2-tag region, which is a subset of the 1-tag region.
 
A summary of the object selection and the event selection used in the analysis
to define the signal regions  and the validation regions for the main backgrounds, which are presented in Section~\ref{sec:bkgEstimate}, is given in Table~\ref{tab:ZBJetSelection}.

\begin{table}[!h]
\begin{small}
\centering
\begin{tabular}{l|c|c}
\toprule
& Electron channel & Muon channel \\
\hline
Trigger &  Single electron &  Single  muon\\
\hline
&  Tight    &  Medium \\
& Isolated & Isolated \\
Leptons        & {\small PV association:\,$|d_0/\sigma_{d_0} |<5$, $|z_0 \sin \theta|<0.5$\,mm} &{\small PV association:\,$|d_0/\sigma_{d_0} |< 3$, $|z_0 \sin \theta|< 0.5$\,mm} \\
&  \pt\ $>$ \SI{27}{GeV} &  \pt\ $>$ \SI{27}{GeV} \\
&   $|\eta| < 1.37$ or $1.52<|\eta| < 2.47$ &   $|\eta| < 2.5$ \\
\hline
Jets    &  \multicolumn{2}{c}{ $\pT$  $>$ \SI{20}{\GeV} and $|y|$ < 2.5}\\
&  \multicolumn{2}{c}{  $\Delta R$(jet, $\ell$)     > 0.4}\\
\hline
$b$-jet	& \multicolumn{2}{c}{ $\pT$  $>$ \SI{20}{\GeV} and $|y|$ < 2.5}  \\
\bottomrule
\end {tabular}
\begin{tabular}{l|c|c|c|c}
\toprule
\multicolumn{5}{c}{Regions}\\
\hline
& Pre-tag  & Signal  & \Zjet s & $t \bar t$  \\
&  region &  regions &  Validation Region &   Validation Region\\
\hline
Leptons    & \multicolumn{3}{c|}{ 2 same-flavour, opposite-charge} &  1 $e$, 1 $\mu$, opposite-charge\\
\hline
$m_{\ell\ell}$           & \multicolumn{4}{c}{\SI{76}{GeV} $< m_{\ell\ell} <$ \SI{106}{\GeV} }\\
\hline
\MET           & \multicolumn{4}{c}{ \MET$<$ \SI{60}{GeV} if $p_{\textrm{T}}^{\ell\ell}< 150~\GeV$ }\\
\hline
Jets          & \multicolumn{4}{c}{$\geq 1$ or $\geq 2$  jets}\\
\hline
$b$-tagging efficiency    & -      & 70\%  &  $\geq$ 1 \bjet\ at   77\%--70\%   & 70\%  \\
working point selection   &        &         &                               & \\
\hline
Number of & - & $\geq 1$ $b$-jets (1-tag region)  &  $\geq 1$ $b$-jets  &  \\
\bjets   &   & $\geq 2$ $b$-jets (2-tag region)  &                   &  $\geq 2$ $b$-jets  \\
\bottomrule
\end{tabular}
\caption{Summary of object and event selections defining the signal regions and the validation regions for the main backgrounds  of the analysis at
detector level.}
\label{tab:ZBJetSelection}
\end{small}
\end{table}

\subsection{Correction factors applied to simulation and corresponding uncertainties}
\label{sec:mcCorr}
 
Corrections are applied to simulated samples in order to ensure
that the object selection efficiencies and the energy and momentum calibrations agree with data within the uncertainties associated with the corrections.
 
The electron and muon trigger efficiencies are estimated in data and simulation in order to determine simulation-to-data correction factors and their corresponding uncertainties.
The average per-event correction factor is about 0.98 (0.93) for electron (muon) triggers; they are known with an uncertainty below 1\%~\cite{PERF-2017-01,PERF-2015-10}.
Corrections to efficiencies for lepton reconstruction, identification, isolation and association with the PV in simulated samples are derived from data.
Each per-lepton correction factor
is close to unity and known with a precision that is better than 1\% in the kinematic range considered~\cite{PERF-2017-01,PERF-2015-10}.
 
The energy scale of the electrons and the momentum scale of the muons in simulation are adjusted with
correction factors that deviate from unity at the per-mil level and the resolutions are adjusted with correction factors
that deviate from unity at the per-cent level in order to match lepton \pt\ and $m_{\ell\ell}$ distributions in data; the
corresponding uncertainties are negligible.
 
The jet energy scale (JES)  is calibrated  on the basis of the simulation including in situ corrections obtained from data~\cite{PERF-2016-04}.
The JES uncertainties are estimated using a decorrelation scheme comprising a set of 21 independent parameters, the largest of which may reach several per cent in specific corners of the phase space.
The jet energy resolution (JER) uncertainty is derived by over-smearing the jet energy in the simulation by
about 4\% at $\pt = 20$~\GeV\ to about 0.5\% at a \pt\ of several hundred \GeV~\cite{PERF-2011-04}.
Simulation-to-data corrections and relative uncertainties are also applied to adjust the efficiency of the JVT requirement following the prescriptions of Ref.~\cite{PERF-2016-06}.
The uncertainty in the scale and resolution of \MET\  is estimated by propagating the uncertainties in the transverse momenta of reconstructed objects
and an uncertainty to account for soft hadronic activity in the event, as described in Ref.~\cite{PERF-2016-07}.
 
Flavour-tagging efficiencies in simulation are scaled to match those measured in data for jets of all flavours as a function of
the different $b$-tagging discriminant output thresholds, and of the jet \pt\ (and $\eta$ for light-jets), using weights derived from control samples enriched in jets of each flavour~\cite{PERF-2016-05}.
In the case of \bjets, correction factors  and their uncertainties are estimated from data using dileptonic  $t\bar{t}$ events~\cite{PERF-2016-05}.
The correction factors for \bjets\ are close to unity.
The uncertainties, described by a set of 28 independent parameters, are as low as 3\% for jet \pt\ of about 60~\GeV, but reach 10\% for jet \pt\ of about 20~\GeV\ and up to 20\% beyond 300~\GeV.
In the case of $c$-jets, correction factors are derived using jets from $W$-boson decays in $t\bar{t}$ events~\cite{ATLAS-CONF-2018-001}.
The correction factors for $c$-jets range from about 1.2 to about 1.6. Their uncertainties, described by a set of 28 independent parameters, are about 20\%--30\% in the bulk of the phase space, but up to 100\% for large jet \pt\ and for the $b$-tagging discriminant output threshold closest to +1.
In the case of light-flavour jets, correction factors are derived using dijet events~\cite{ATLAS-CONF-2018-006}.
The correction factors for light-jets  range from about 2 to about 3, with uncertainties described by a set of 36 independent parameters and ranging from 50\% to 100\%.
An additional uncertainty of 30\% is applied to the efficiency of $b$-tagging for simulated jets originating from pile-up interactions, which are less than 1\% of the selected jets.
 
A variation in the pile-up reweighting of simulated events (referred to as pile-up uncertainty) is included to  account for the uncertainty in the ratio of the predicted and measured inelastic cross-sections in the fiducial volume~\cite{STDM-2015-05}.
 
\section{Background estimation}
\label{sec:bkgEstimate}
The main background in the 1-tag region is constituted by
events with  a $Z$ boson produced in association with jets, where either a light-jet or a $c$-jet is misidentified as a \bjet; it is determined  using a fit to data as detailed in Section~\ref{sec:FlavFit}.
Dileptonic $t\bar{t}$ events dominate in the 2-tag region.
Smaller background contributions from the production of dibosons, a Higgs boson, a single top quark, a $Z \to \tau \tau$, or a $W \to \ell \nu$ are  estimated using simulation, as described in Section~\ref{subsec:MC}.
Uncertainties in the normalisation cross-section of these predictions
range from  4\% to 6\% depending on the process, as detailed in Table~\ref{tab:generators}.
Background contributions from multijet events are estimated with a data-driven technique and found to be negligible, as described below.
 
The $t\bar{t}$ contribution is estimated using  simulated events generated with \POWHEGBOX + \pythia
normalised to the theoretically predicted cross-section, as discussed in Section~\ref{subsec:MC}.
An uncertainty of about  6\% is assigned to the inclusive $t\bar{t}$  cross-section (see Table~\ref{tab:generators}),
following the variation of the renormalisation and factorisation scales by a factor of 2.0, and the variation of the PDFs within their uncertainties.
In addition, uncorrelated systematic uncertainties in the modelling of the distributions are derived by comparing the predictions from the nominal $t\bar{t}$ sample with the ones from the alternative samples described in Section~\ref{subsec:MC}.

\begin{figure}[h!]
\begin{center}
\includegraphics[width=.49\linewidth]{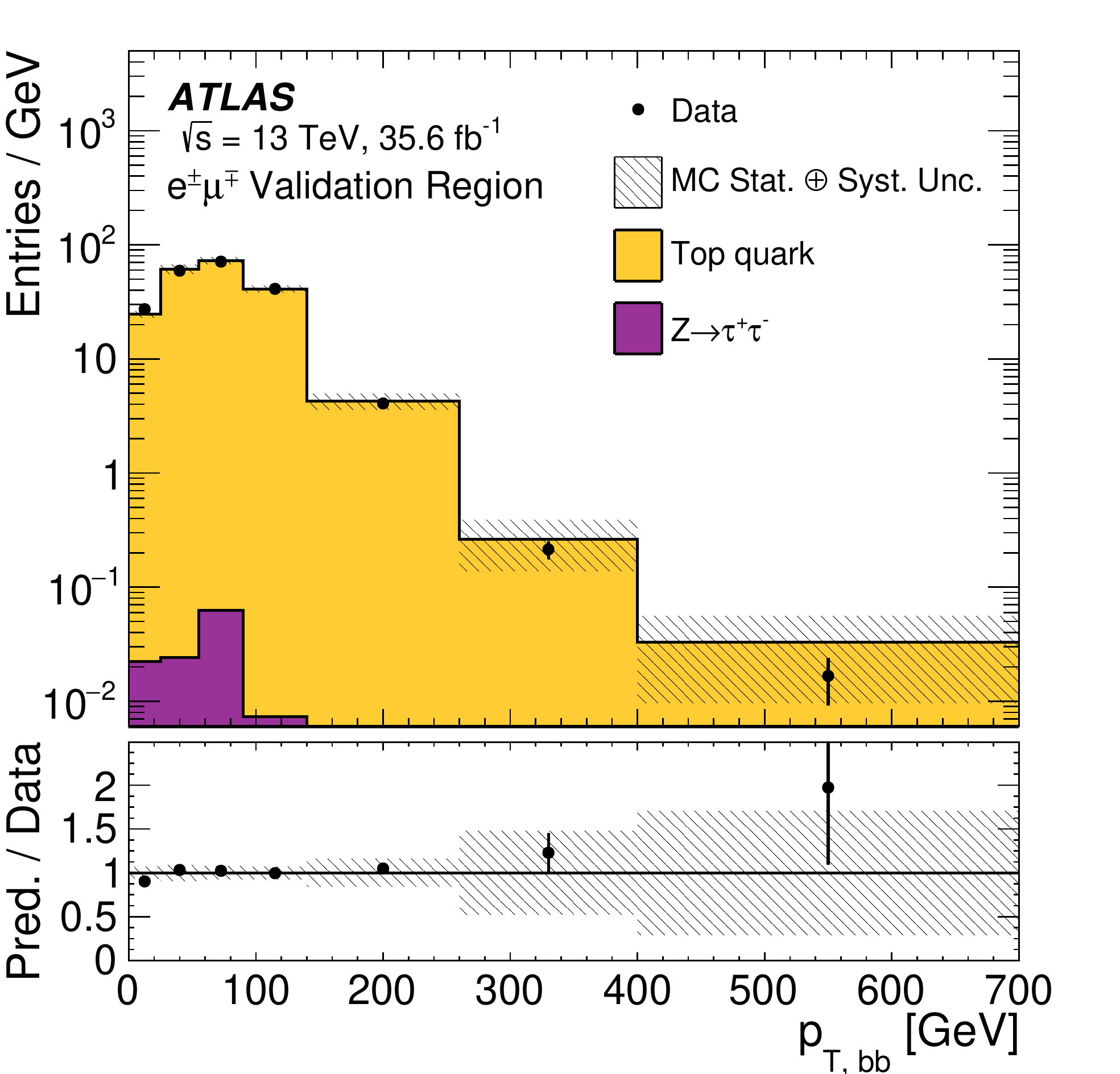}
\includegraphics[width=.49\linewidth]{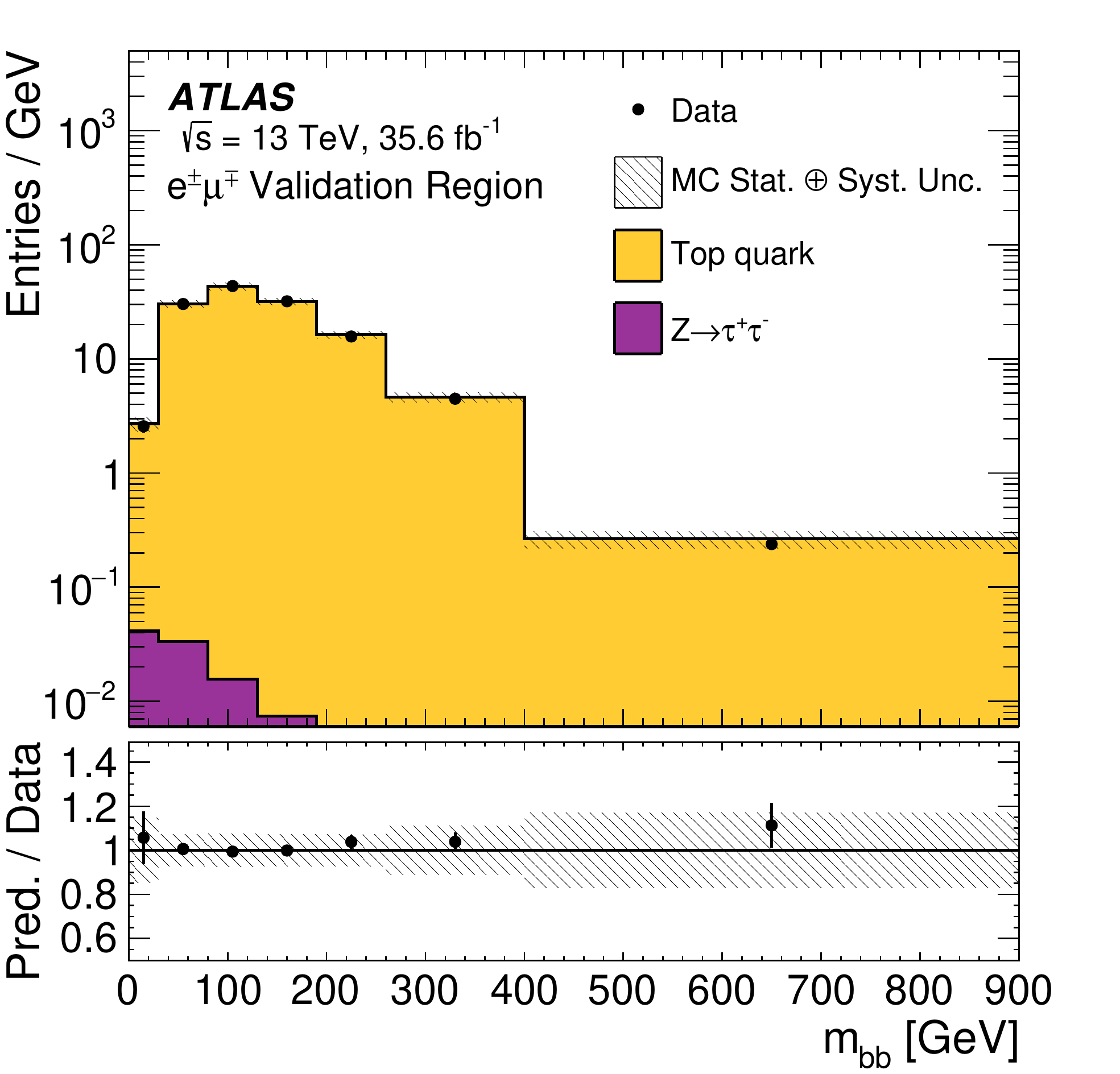}
\caption{
Transverse momentum  (left) and invariant mass (right) of the di-\bjet\ system built with the two  highest-\pt\ \bjets\ for events with at least two \bjets\
in the $t\bar{t}$ validation region.
Systematic uncertainties of the predicted distributions are combined with the statistical ones in the hatched band, and the statistical uncertainty of the data is shown as error bars.
The systematic uncertainties for the predictions account only for the yield and the shape of $t\bar{t}$ events.
}
\label{fig:topVR}
\end{center}
\end{figure}

The modelling of $t\bar{t}$ production in the simulation is validated using a $t\bar{t}$-enriched  region, which is selected by requiring
that  events have two leptons of different flavour ($e \mu$);   all other selections are the same as in the signal region.
As an example, Figure~\ref{fig:topVR} shows the \ptbb\  and the $m_{bb}$ distributions for events with at least two \bjets.
The total background from top quarks is the sum of $t\bar{t}$ and single-top events, where the  latter are about 3\% of the $t\bar{t}$ component in the validation region, and other backgrounds are negligible.
Data and simulation agree well within the uncertainties which account for both the yield and shape uncertainties of simulated
$t\bar{t}$  events and the statistical uncertainties of predictions and data.
 
Background contributions from multijet events in the electron and muon channels are estimated using a data-driven technique.
Multijet-enriched control regions without $b$-tag and $m_{\ell\ell}$ requirements are used to derive the expected shape of this background.
In the electron channel, the multijet-enriched control region is defined by applying the full signal event selection except for the electron identification and
the $d_0/\sigma_{d_0}$ cuts, and inverting the isolation selection for both electron candidates.
In the muon channel, the multijet-enriched control region is defined by applying the full signal event selection
but requiring both muon candidates to have the same charge.
In both channels, contributions from non-multijet sources in the control regions are estimated from simulation and subtracted from the data, with the remaining distributions used as shape templates.
A fit of the $m_{\ell\ell}$ distribution to data is then performed  within  the  window of  \mbox{60 GeV $<m_{\ell\ell}<$ 160 GeV} in the
one-jet and two-jets pre-tag regions separately and leaving the normalisation of the signal and of the multijet background templates
free to float in the fit, while the normalisation of the other processes is fixed in the fit.
The multijet background estimate in the pre-tag region is then extrapolated to the two signal regions using normalisation factors equal to the fraction of events in the multijet control region that satisfy the 1-tag and 2-tag requirements.
Contributions from non-multijet processes are  subtracted before estimating this fraction.
Systematic uncertainties are assessed by varying the $m_{\ell\ell}$  range and the binning of the fit, excluding the \Zbos\ peak from the fit, performing the fit in the tagged  regions in place of the pre-tag ones, and by allowing the other processes to be varied independently in the fit.
The estimated size of the multijet background  is consistent with zero within the statistical uncertainty even after considering all sources of systematic uncertainty.
It is therefore  neglected in the analysis.

\subsection{Extraction of the cross-section for \Zbos\ production in association with light-jets  and $c$-jets}
\label{sec:FlavFit}
The flavour fit used for the extraction  of the yields of $Z$ + light-jets and $Z$ + $c$-jets backgrounds for the 1-tag and 2-tag selections is  a maximum-likelihood fit to data based on  flavour-sensitive distributions.
The fit is done simultaneously in the electron and muon channels with templates derived from  simulation.
 
\begin{figure}[h!]
\begin{center}
\includegraphics[width=.49\linewidth]{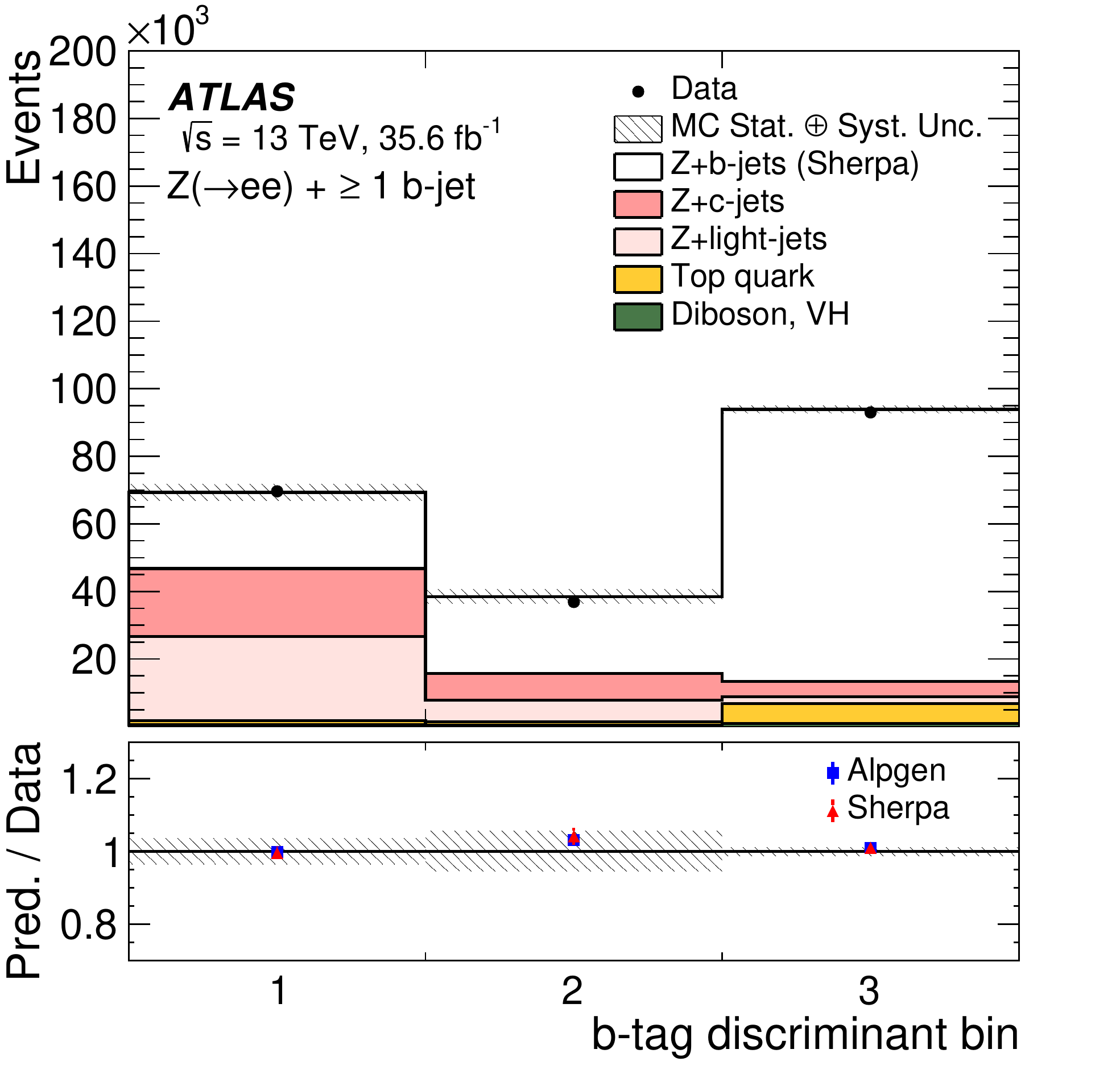}
\includegraphics[width=.49\linewidth]{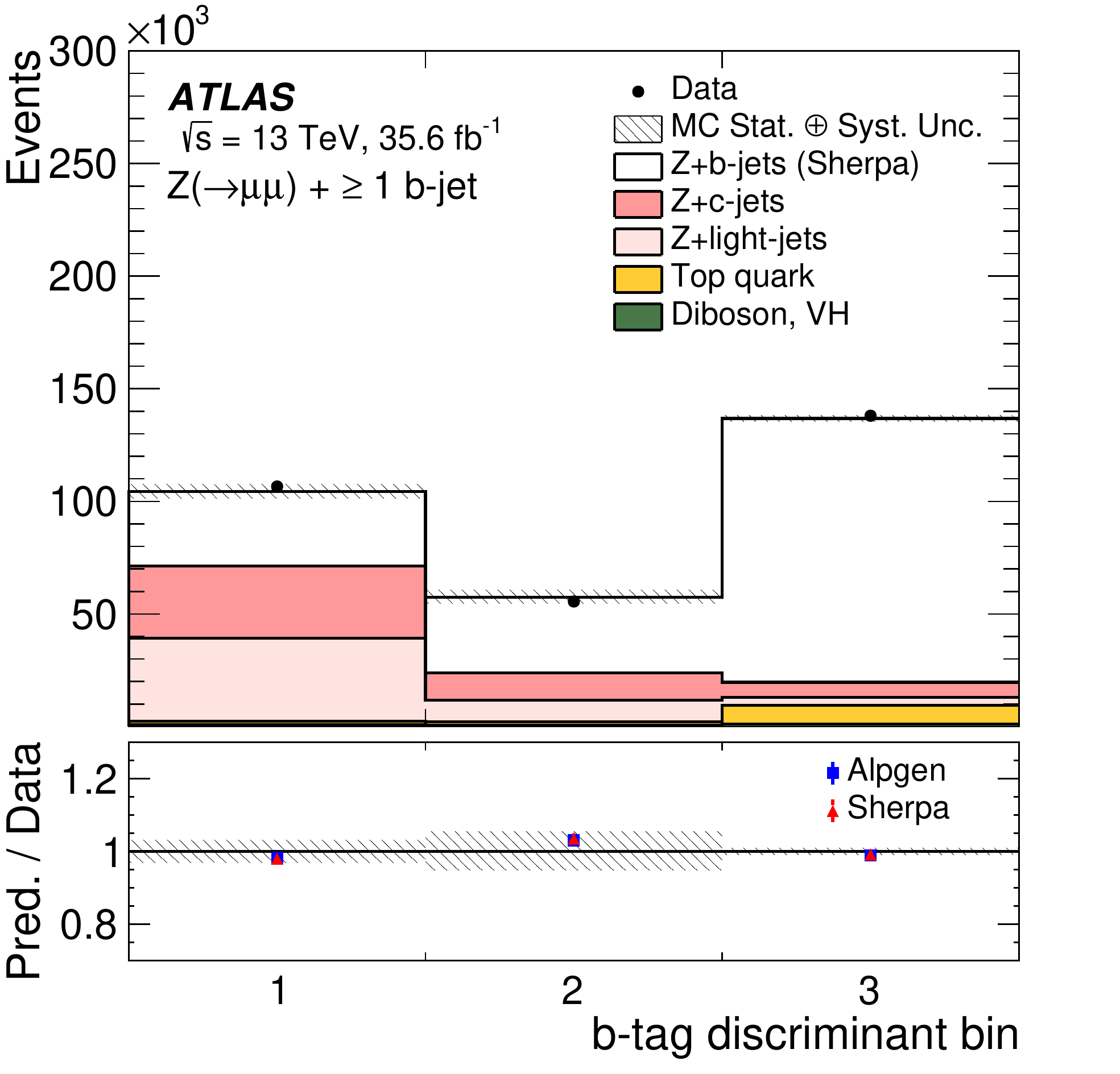}\\
\includegraphics[width=.49\linewidth]{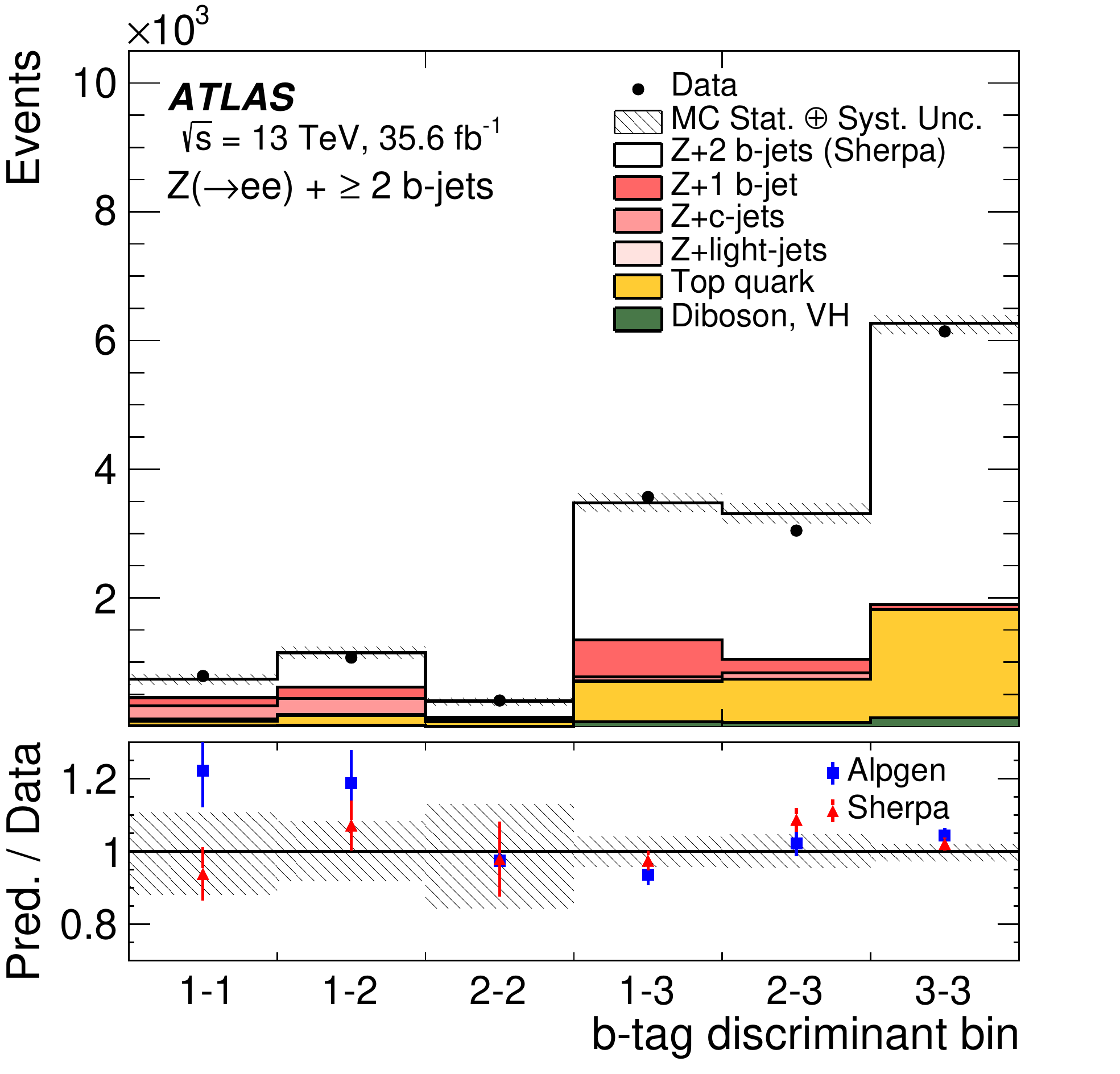}
\includegraphics[width=.49\linewidth]{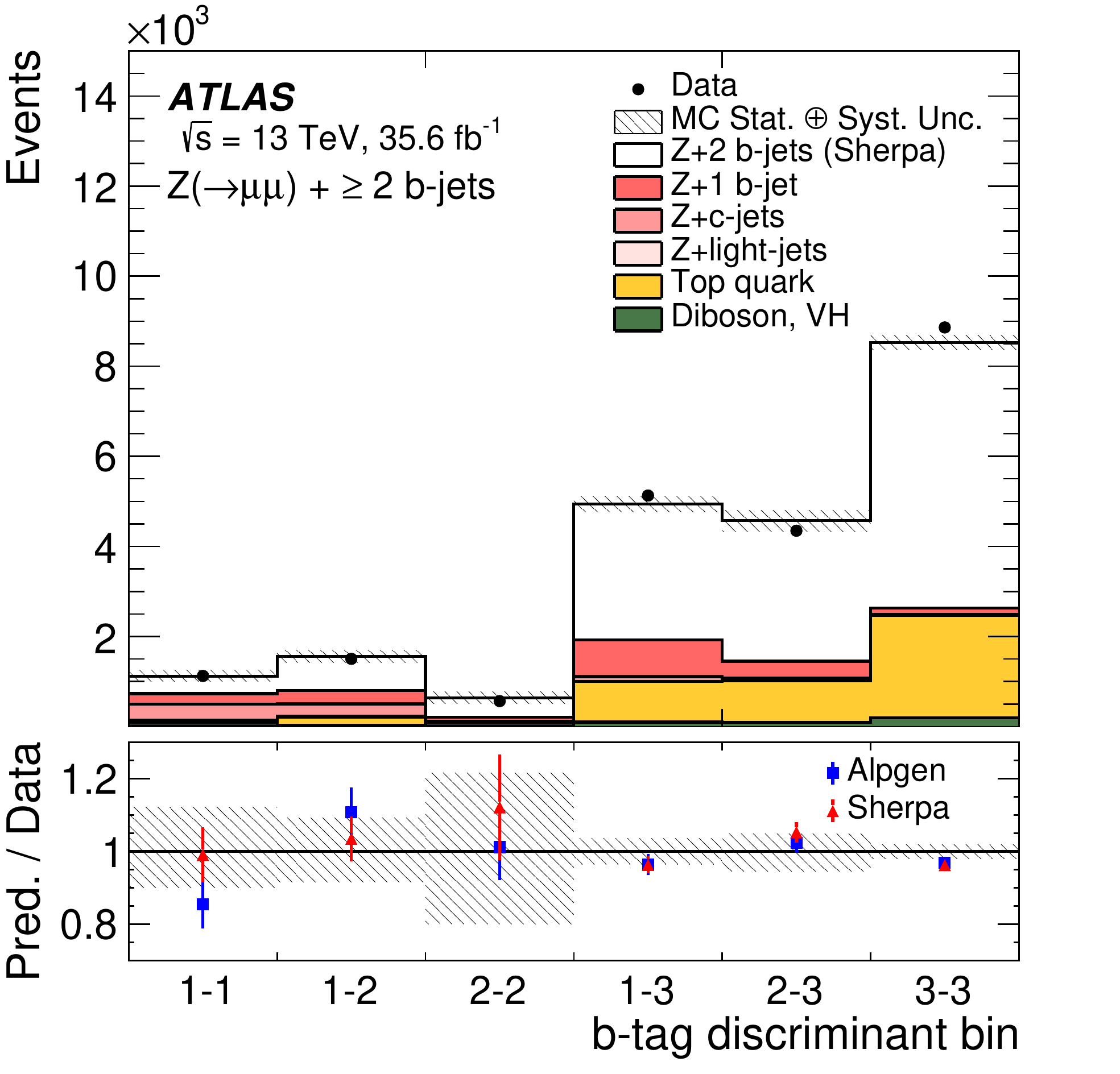}
\caption{
Post-fit $b$-tagging discriminant distributions for the electron (left) and muon (right) channels in the 1-tag (top) and 2-tag (bottom) signal regions.
The lower panels display the ratios of the predictions to data  using the  signal and $Z$+jet  background simulation either from \sherpa (red) or  \alpgen (blue).
Systematic and statistical uncertainties for the predicted distributions are combined in the hatched band, and the statistical uncertainty, estimated with pseudo-experiments, is shown on the data points.
The systematic uncertainties account for both the detector-level uncertainties and the theory uncertainty of the non-$Z$ backgrounds.
}\label{fig:12tagFit}
\end{center}
\end{figure}

In the 1-tag region, the $b$-tagging discriminant output of the leading \bjet\ is   used as the flavour-sensitive distribution.
This observable for events belonging to the signal region is distributed into three intervals that define the bins of the discriminant output distribution.
Each bin corresponds to a certain range of $b$-tagging efficiency. The bins are numbered from 1 to  3, corresponding respectively
to efficiencies of   60\%--70\% (bin 1), 50\%--60\% (bin 2) and $<$50\% (bin 3)  as estimated from simulated $t\bar{t}$ events.
The light-flavour jet ($c$-jet) misidentification rates for the three bins are  respectively
0.195\%     (5.4\%),     0.048\%   (1.96\%), and  $<$0.017\% ($<$0.94\%).
The signal template is built with simulated $Z+\geq 1 b$ events. The template shapes of the $Z+l$ and $Z+c$ samples are very similar (as shown in Figure~\ref{fig:12tagFit}), hence those samples are combined to form a  single template.
All non-\Zjet s backgrounds are combined into a single template, determined from the sum of their predicted contributions.
The normalisations of the signal and of the \Zjet s background
are free to float in the fit, while the normalisation of the sum of the non-\Zjet s backgrounds is fixed to their estimate.
 
In the 2-tag region the combination of the three bins of the $b$-tagging discriminant outputs of the leading and sub-leading \bjets\ produces a distribution with six bins that is used for the fit to data.
The signal template is built with simulated $Z+bb$  events. Templates built with  $Z+b$, $Z+c$ and $Z+l$ simulated events
are combined into a  single template.
Because of the  large rejection of light-flavour jets achieved in the 2-tag selection, the simulated $Z+l$ events in this region  are not subjected to the $b$-tagging requirement.
Instead they are weighted by a per-event probability that the jets pass the two-$b$-tags selection (procedure referred to as the \textit{truth}-tagging).
This probability is computed on the basis of the per-jet probabilities, which are assumed to be independent of each other~\cite{HIGG-2016-29}.
As for the fit in the 2-tag region, the normalisations of the signal and of the \Zjet s background
are also free to float, while the normalisation of the other backgrounds is fixed to their estimate.
 
Tables~\ref{tab:1tagFit} and \ref{tab:2tagFit} show the normalisation scale factors in the  1- and 2-tag regions
obtained from the fit, together with  the post-fit yields for the signal and  \Zjet\ background samples generated with \sherpa or \alpgen.
There is good agreement between the sum of the signal and background post-fit yields of   \sherpa and  \alpgen.
The differences between \sherpa and \alpgen in the modelling of the \Zjet\  backgrounds after the flavour fit are taken into account in the systematic uncertainties as described below.
The statistical uncertainty is estimated with  pseudo-experiments.
 
Figure~\ref{fig:12tagFit} shows the $b$-tagging discriminant bins after the fit in the 1-tag and 2-tag regions.
In the upper panel of each figure, data are compared with the fit results obtained using templates derived from \sherpa samples for signal and  \Zjet\ backgrounds. The lower panel shows the ratio of post-fit predictions to data
using the \sherpa or \alpgen samples for signal and  \Zjet\ backgrounds.
 
The  \Zjet s backgrounds predicted by \sherpa and corrected for the normalisation factor obtained from the fit are used as the nominal estimate in this analysis.
Systematic uncertainties due to the object selection efficiencies and calibrations, discussed in Section~\ref{sec:mcCorr},
affect the normalisation and the shape of  \Zjet s backgrounds. They  are assessed by repeating the fit with the templates varied according to each of the systematic uncertainties. The fit is also  repeated for  each of the uncertainties affecting the
$t \bar{t}$ and other backgrounds detailed above.
An  additional systematic uncertainty (referred to as the flavour fit uncertainty) in the normalisation of the  \Zjet s backgrounds is estimated by
repeating the fit after separating the $Z+c$ from the $Z+l$ template in the 1-tag region, and after separating the $Z+b$ from the $Z+c$ and $Z+l$ templates in the 2-tag region.
An uncertainty affecting the shape and rate of the \Zjet s background is derived by taking the difference between the post-fit \Zjet s background evaluations using \sherpa and \alpgen samples.
Another uncertainty accounts for potential jet--jet correlations that are not covered by the \textit{truth}-tagging procedure which mitigates the large statistical fluctuations in the 2-tag region for $Z+l$.
A 20$\%$ uncertainty is derived by taking the largest difference between the double-tagged event yields obtained with or without the weighting procedure being applied to
simulated samples of $Z+bb$, $Z+cc$, $W+bb$, and $W+cc$\footnote{Simulated \Zjet s events are  categorised as  $Z + cc$ ($W + cc$) if they belong to the $Z+c$ ($W+c$)
category and have at least two $c$-jets.}.
These samples suffer less from statistical limitations. The test is done with both the \sherpa and \alpgen samples.
 
\begin{table}[h!]
\centering
\begin{tabular}{l|c|c|c|c|c}
\hline
{\small Generator} &  {\small Signal}  & {\small $Z$+jets background }  & {\small Signal} & {\small $Z$+jets background } & {\small Signal + $Z$+jets } \\
&         {\small SF}      &     {\small  SF}     &     {\small post-fit yield}&{\small post-fit yield} &    {\small post-fit yield}\\
\hline
{\small \sherpa} & 1.109 $\pm$ 0.003 &  0.861 $\pm$ 0.004  &  309\,650 $\pm$ 810 & 166\,640 $\pm$ 650 & 476\,290 $\pm$ 750\\
\hline
{\small \alpgen} & 1.480 $\pm$ 0.004  & 1.015 $\pm$ 0.002 & 297\,670 $\pm$  740 & 178\,100 $\pm$  400 & 475\,810 $\pm$  480 \\
\hline
\end{tabular}
\caption{Scale factors obtained for the fitted signal and  \Zjet\ background for \sherpa and \alpgen fits, the total post-fit yields, and the statistical uncertainty, estimated with pseudo-experiments, from the fit for the 1-tag signal region.}
\label{tab:1tagFit}
\end{table}
 
\begin{table}[h!]
\centering
\begin{tabular}{l|c|c|c|c|c}
\hline
{\small Generator} &  {\small Signal}  & {\small $Z$+ jets background }  & {\small Signal} & {\small $Z$+ jets background } & {\small Signal + $Z$+jets } \\
&         {\small SF}      &     {\small  SF}     &     {\small post-fit yield}&{\small post-fit yield} &    {\small post-fit yield}\\
\hline
\sherpa &  1.18 $\pm$ 0.01 & 1.08 $\pm$ 0.04 & 23\,440 $\pm$ 250 & 4780 $\pm$ 180  &  28\,220 $\pm$ 200 \\
\hline
\alpgen &  1.18 $\pm$ 0.01  &     1.30 $\pm$ 0.05 & 23\,650 $\pm$ 240 &  4550 $\pm$  180 & 28\,200 $\pm$ 200 \\
\hline
\end{tabular}
 
\caption{Scale factors obtained for the fitted signal and  \Zjet\ background for \sherpa and \alpgen fits, the total post-fit yields, and the statistical uncertainty, estimated with pseudo-experiments, from the fit for the 2-tag signal region.}
\label{tab:2tagFit}
\end{table}

\begin{figure}[h!]
\begin{center}
\includegraphics[width=.49\linewidth]{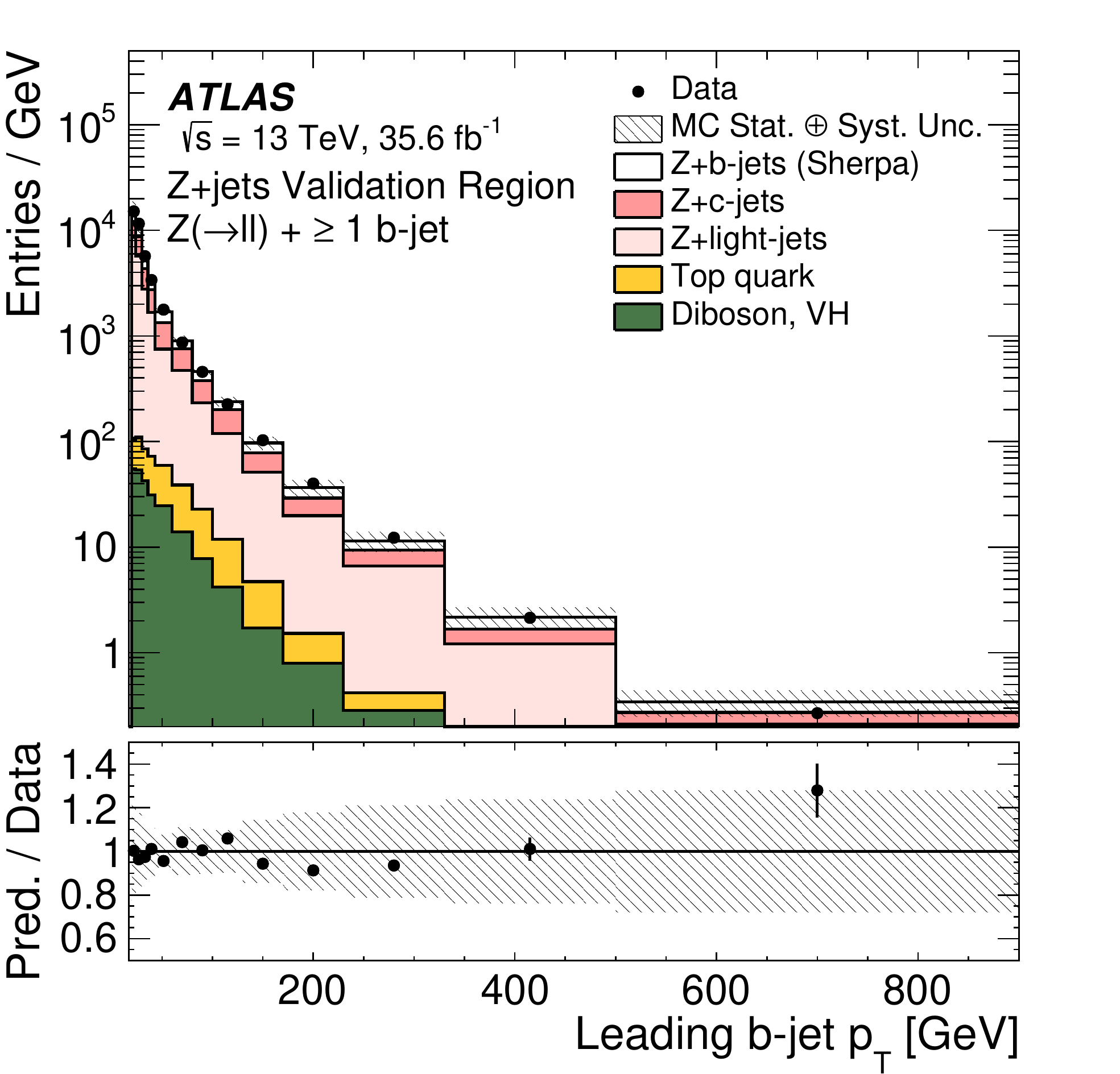}
\includegraphics[width=.49\linewidth]{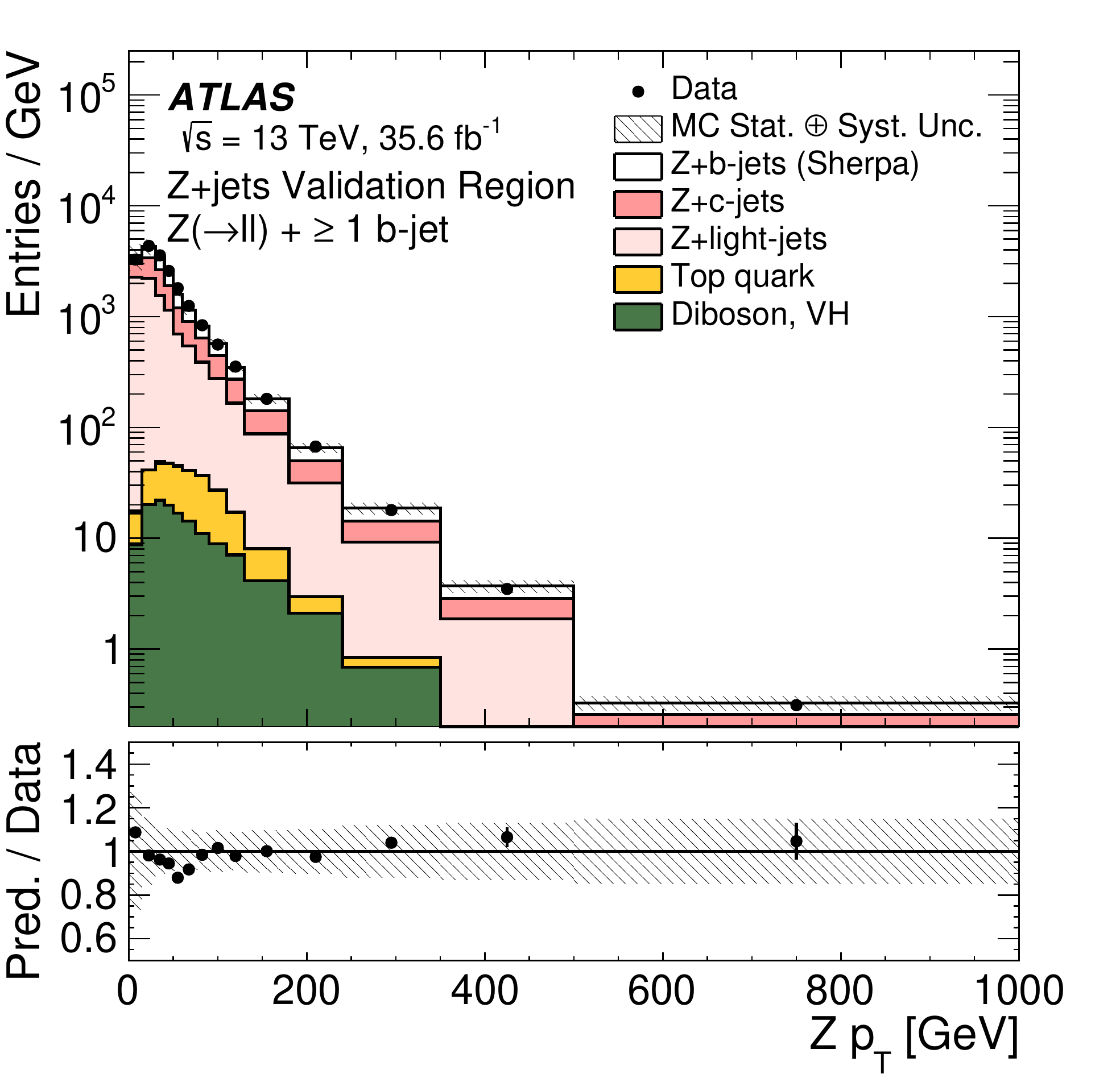}
\caption{
The \pt\ of the leading \bjet\ (left) and of the $Z$ boson  (right) for events with at least one \bjet\ in the $Z$+jets validation region defined in Table~\ref{tab:ZBJetSelection}.
Post-fit distributions for signal and $Z$ + jets backgrounds are shown.
Systematic and statistical uncertainties for the predicted distributions are combined in the hatched band, and the statistical uncertainty is shown on the data points.
The uncertainty in the predictions includes only the flavour-tagging efficiency uncertainty and flavour-fit uncertainty.}\label{fig:ftagVR}
\end{center}
\end{figure}

The post-fit estimate of  the \sherpa \Zjet s background  is validated in a region defined by applying the full signal event selection with the exception of $b$-tagging requirements.
Events with at least one \bjet, with the $b$-tagging discriminant output in the \bjet\ efficiency  range of 70\%--77\% and
light-flavour jet ($c$-jet) misidentification rates of  0.51\% (7.7\%),  are selected to provide a sample enriched in $c$-jets and light-flavour jets.
As an example, Figure~\ref{fig:ftagVR} shows the \pt\ of the leading \bjet\ and the \pt\ of the $Z$ boson in this region.
The $Z+l$ and  $Z+c$ backgrounds constitute 50\% and 28\% of the total prediction, respectively.
Agreement between data and estimated backgrounds is observed within
uncertainties. These include the uncertainties due to the flavour fit and $b$-tagging efficiency, and the statistical uncertainties of the predictions and data.
 
The normalisation factors of the signal samples, shown in Tables~\ref{tab:1tagFit} and \ref{tab:2tagFit}, are applied in Figures~\ref{fig:12tagFit} and~\ref{fig:ftagVR} in this section to demonstrate the robustness of this procedure,
while in the following sections, post-fit normalisation factors are applied only to $Z$+jets background.

\section{Kinematic distributions}
\label{sec:KinPlots}
 
After the signal selection criteria are applied, the measured and expected distributions are compared at the detector level.
The  $Z$+jets background is shown for the normalisation factors derived from the flavour fit. Pre-fit distributions are used for the signal samples.
 
\begin{figure}[htbp]
\begin{center}
\begin{tabular}{lr}
\includegraphics[width=0.49\textwidth]{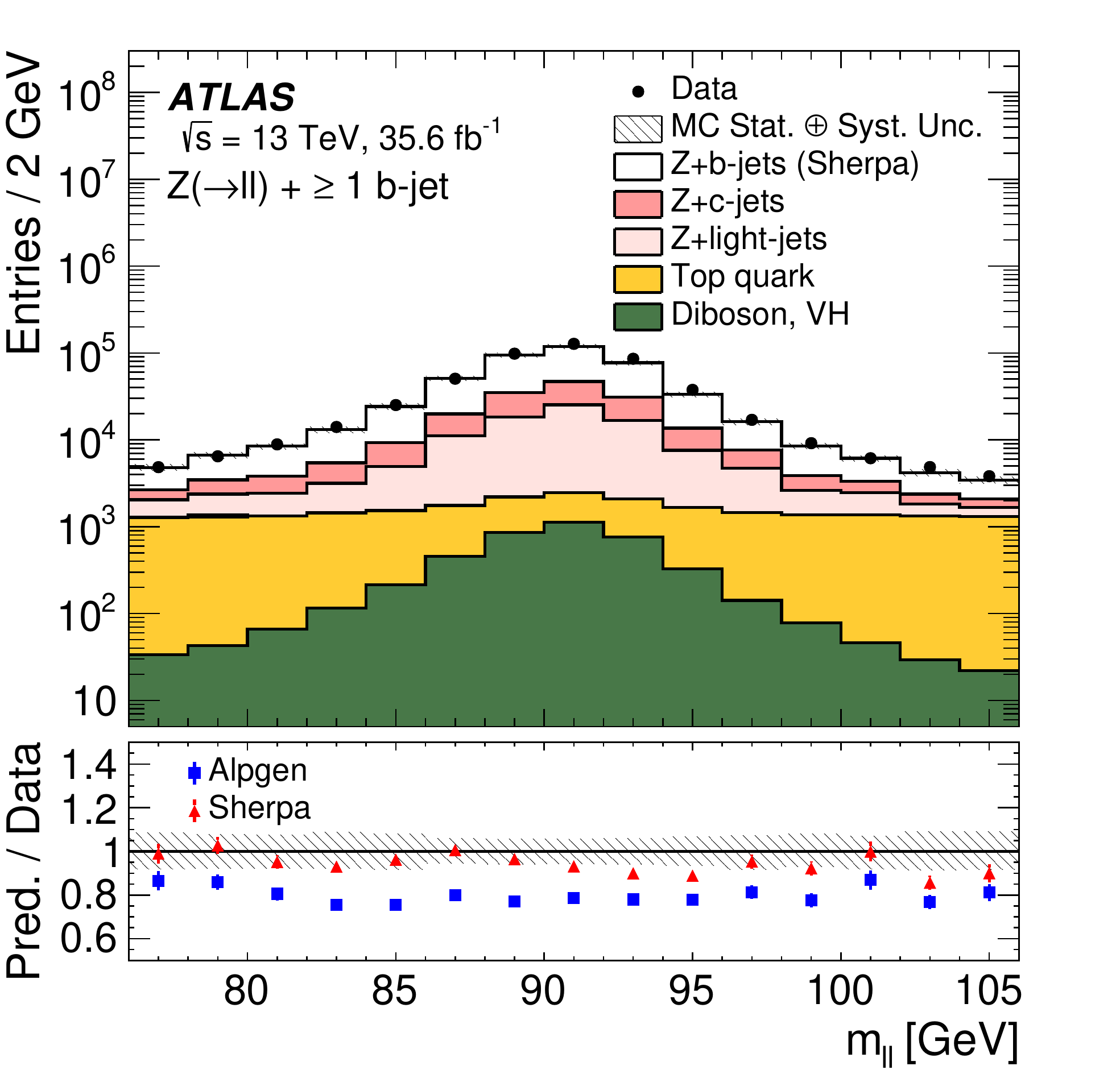}&
\includegraphics[width=0.49\textwidth]{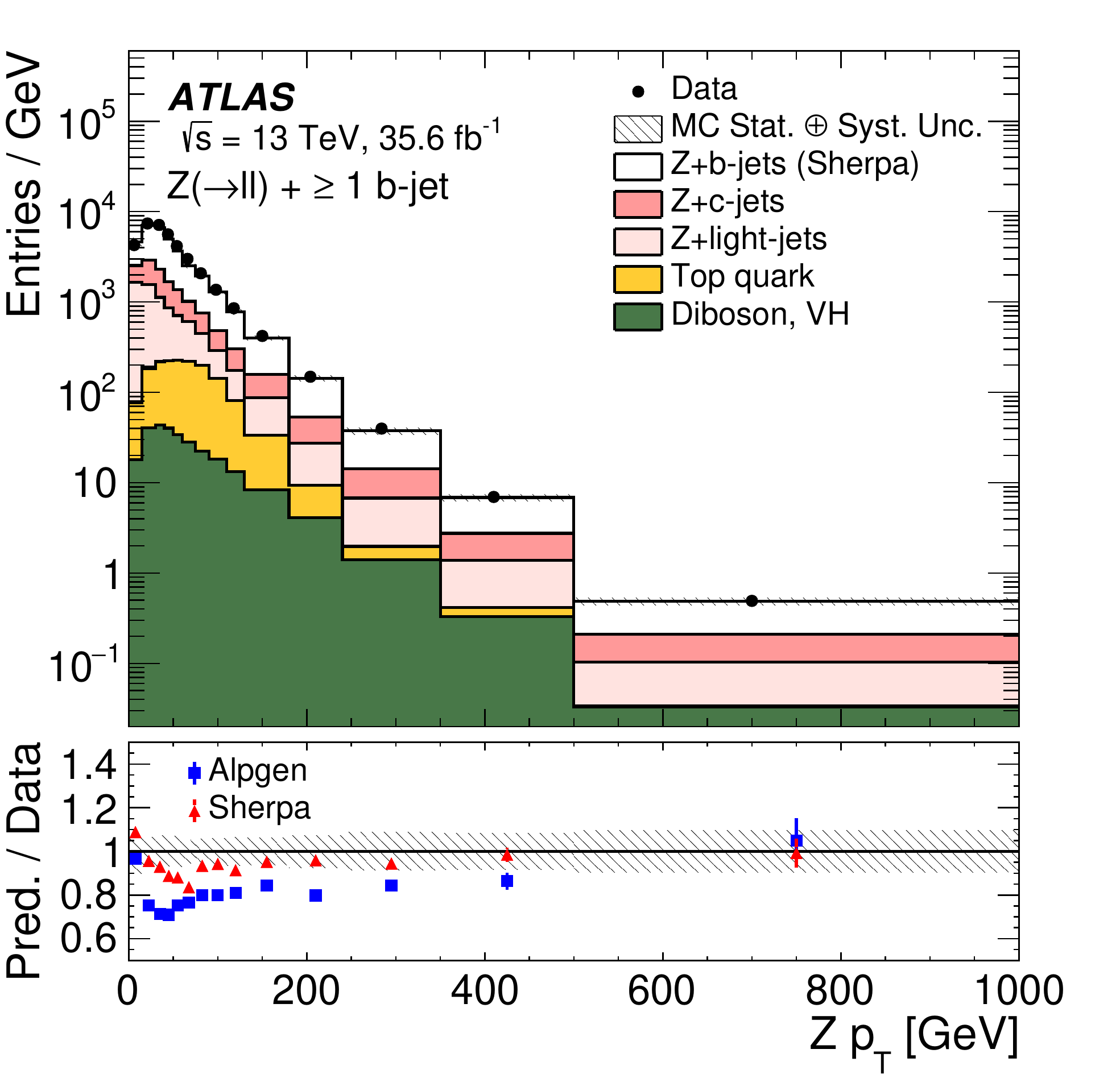}\\
\end{tabular}
\end{center}
\caption{
Distribution of events passing the signal selection as a function of $m_{\ell\ell}$ (left) and $p_{\textrm{T},Z}$ (right) for events with at least one \bjet. The lower panels display the ratio of the predictions for signal plus background to data using either \sherpa (red) or \alpgen+~\pythiasix (blue) as the signal simulation. The statistical uncertainty of the data is shown as black error bars and the total uncertainty of the prediction as a hatched band. The latter consists of the statistical uncertainty and all  systematic uncertainties from the predictions.
}
\label{fig:DLP_Z1b}
\end{figure}

\begin{figure}[htbp]
\begin{center}
\begin{tabular}{lr}
\includegraphics[width=0.49\textwidth]{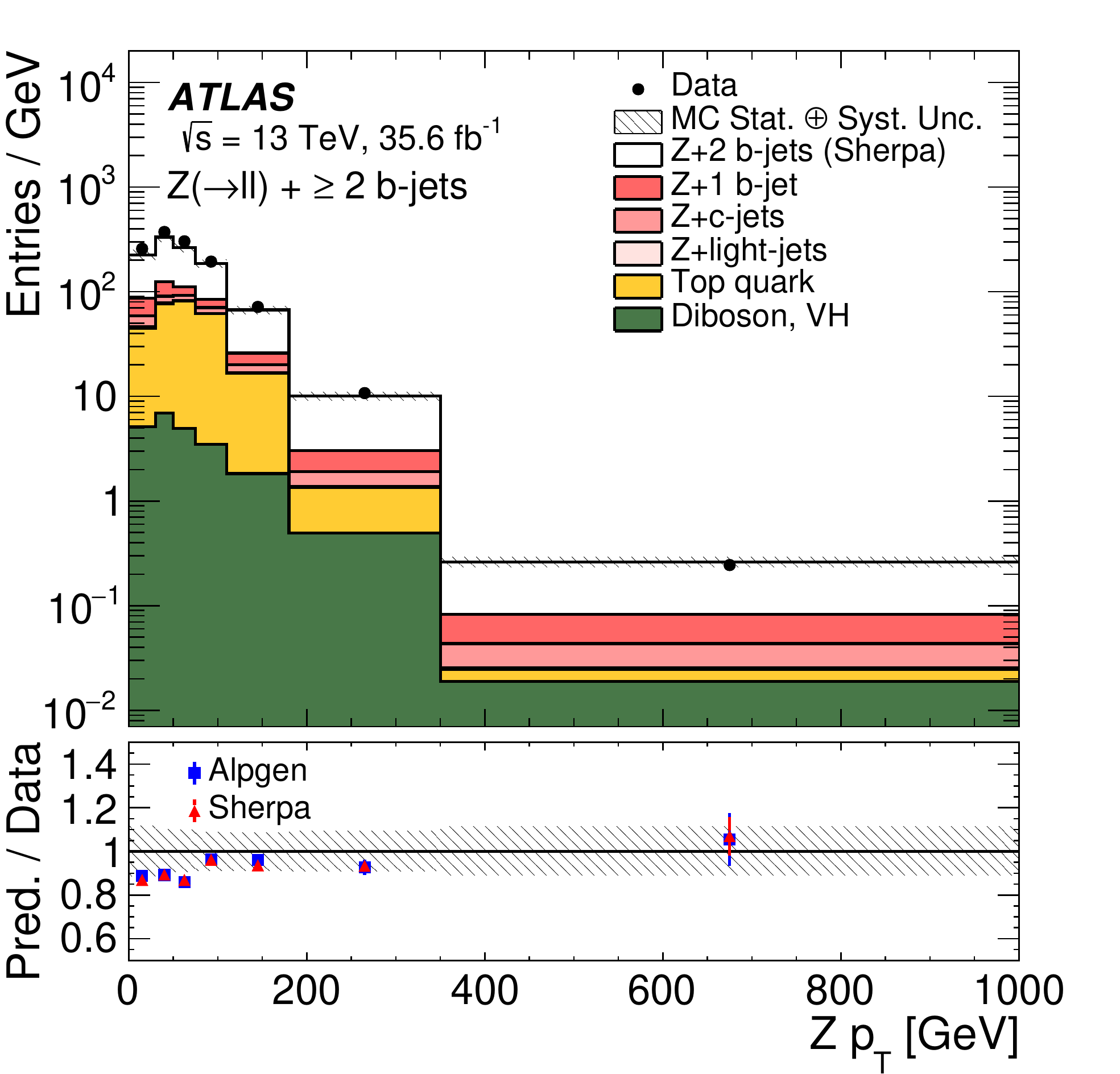}&
\includegraphics[width=0.49\textwidth]{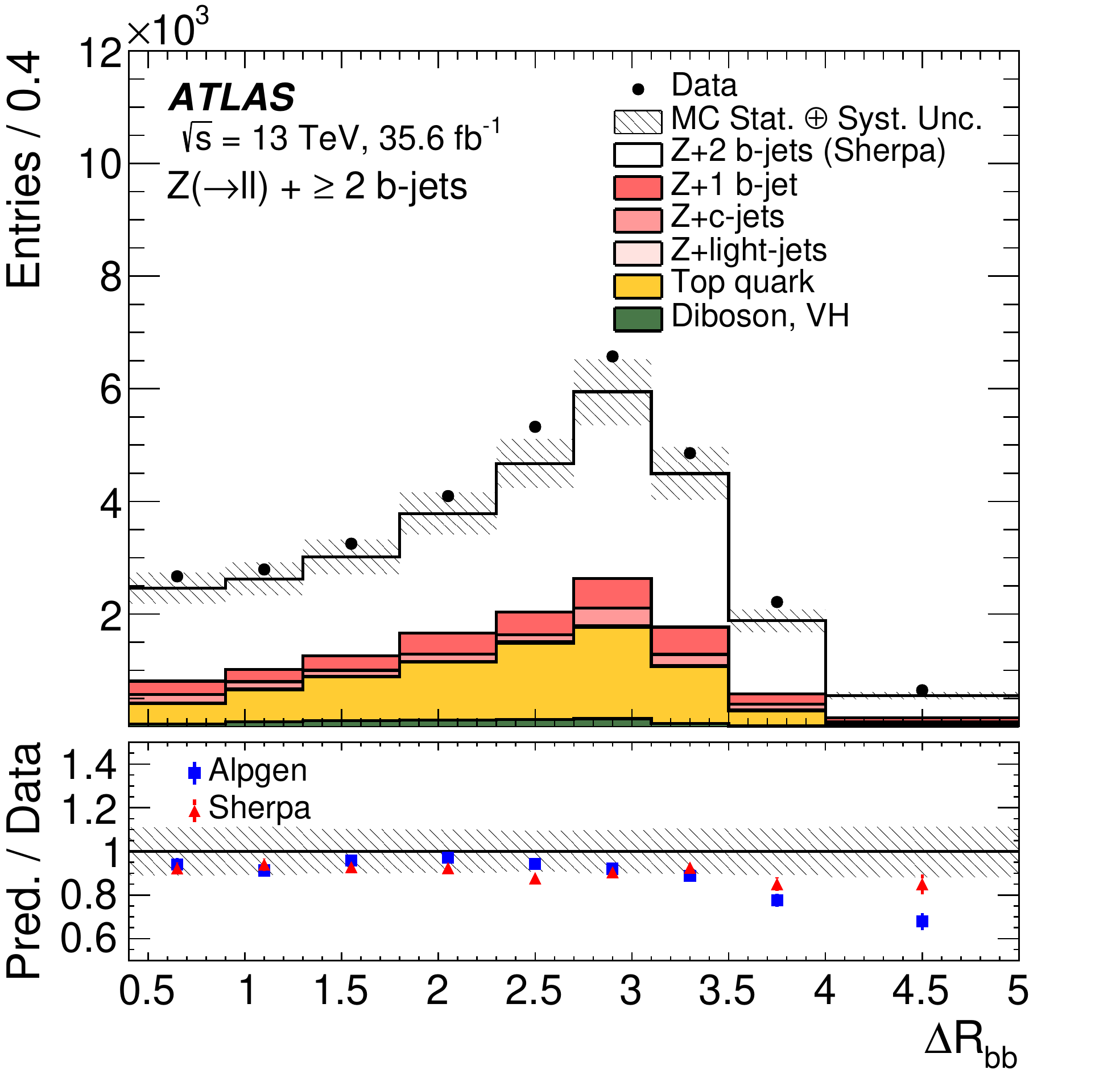}\\
\end{tabular}
\end{center}
\caption{
Distribution of events passing the signal selection as a function of $p_{\textrm{T},Z}$ (left) and  $\Delta R_{bb}$ (right)  for events with at least two \bjets. The lower panels display the ratio of the predictions for signal plus background to data using either \sherpa (red) or \alpgen+~\pythiasix (blue) as the signal simulation. The statistical uncertainty of the data is shown as black error bars and the total uncertainty of the prediction as the hatched band. The latter consists of the statistical uncertainty and all  systematic uncertainties from the predictions.
}
\label{fig:DLP_Z2b}
\end{figure}

Figure~\ref{fig:DLP_Z1b} shows, as an example, the distributions of the $m_{\ell\ell}$ and \pt\ of the $Z$ boson  for events in the 1-tag region.
Figure~\ref{fig:DLP_Z2b} shows  the \pt\ of the $Z$ boson and the $\Delta R_{bb}$ distributions for events in the 2-tag region.
The uncertainty bands include the statistical uncertainties of the simulated sample,
the event-selection uncertainties described in Section~\ref{sec:EvtSel} (omitting the common  luminosity uncertainty), and the background  uncertainties described in Section~\ref{sec:bkgEstimate}.
Both generators do not describe precisely  the data in the full range of the measurement,  although the \sherpa generator provides the best agreement with data.
 
The total numbers of selected events in data and in predictions are presented in Table~\ref{tab:DLP_ZpT},
together with the prediction of each process, expressed as a fraction of the total number of predicted events.
 
\begin{table}[h!]
\begin{minipage}{0.495\textwidth}
\centering
\begin{tabular}{l c}
\hline
\multicolumn{2}{c}{  1-tag region}  \\
\hline
\multicolumn{2}{c}{Signal}\\
\hline
$Z$ + $b$, $Z$ + $bb$& 59$\%$ \\
\hline
\multicolumn{2}{c}{Backgrounds}\\
\hline
$Z$ + $c$& 18$\%$\\
$Z$  + $l$ & 18$\%$  \\
Top & ~~4$\%$ \\
Diboson,  $VH$ & ~~1$\%$\\
Others & <1$\%$\\
\hline
Total predicted  & 470\,000 $\pm$ 650 \\
Data & 499\,645 \\
\hline
\end{tabular}

\end{minipage}
\begin{minipage}{0.495\textwidth}
\centering
\begin{tabular}{l c}
\hline
\multicolumn{2}{c}{  2-tag region}\\
\hline
\multicolumn{2}{c}{Signal}\\
\hline
$Z$ + $bb$ &  60$\%$\\
\hline
\multicolumn{2}{c}{Backgrounds}\\
\hline
$Z$ + $b$ & ~~9$\%$\\
$Z$ + $c$ &  ~~5$\%$\\
$Z$ + $l$ & <1$\%$\\
Top &  23$\%$\\
Diboson,  $VH$ &  ~~2$\%$\\
Others &  ~~1$\%$\\
\hline
Total predicted  & 33\,070 $\pm$ 180\\
Data &  36\,548\\
\hline
\end{tabular}

\end{minipage}
\caption{The expected size of the signal and backgrounds, expressed as a fraction of the total number of predicted events  for inclusive \bjet\ multiplicities for the signal selection.
The signal and $Z$+jets background predictions are from the \sherpa generator, with the $Z$+jets background estimate obtained after applying the normalisation scale factors obtained from the flavour fit.
The total numbers of predicted and observed events are also shown.  The uncertainty in the total predicted number of events is statistical only.
}
\label{tab:DLP_ZpT}
\end{table}
 

\section{Correction to particle level}
\label{sec:Unfolding}
The signal event yields are determined by subtracting the estimated background contributions from the data. The resulting distributions are corrected
for detector-level effects  to the fiducial phase space at particle level defined in Table~\ref{tab:fiducial_ps}.
The  procedure, based on simulated samples,  corrects for $Z$-boson, jet, and \bjet\ selection efficiencies, resolution effects, and small differences between the fiducial and detector-level phase spaces.
The pre-fit distributions of the \sherpa signal samples are used to perform the unfolding procedure.
The signal samples for the simulation of $Z$ events with at least one or at least two \bjets\ are defined in Section~\ref{sec:EvtSel}.
Particle-level objects are selected with requirements  close to the corresponding requirements for reconstructed signal candidate objects, in order to limit the dependence of the measurement on theoretical predictions. In this definition, the lepton kinematic variables
are computed using final-state leptons from the \Zbos\ decay. Photons radiated by the boson decay products within a cone of
size $\Delta R = 0.1$ around the direction of a final-state lepton are added to the lepton, and the sum is referred
to as the `dressed' lepton. Particle-level jets are identified by applying the anti-$k_t$  algorithm with $R = 0.4$
to all final-state particles with a lifetime longer than 30~ps, excluding the dressed \Zbos\ decay products.
A jet is  identified as $b$-tagged if it lies within $\Delta R = 0.3$ of one or more weakly decaying $b$-hadrons with $\pt > 5$~\GeV.
If  a $b$-hadron matches more than one jet, only the closest jet in $\Delta R$ is labelled as a \bjet.
 
\begin{table}[!h]
\begin{center}
\begin{tabular}{c|c}
\hline
Kinematic variable     & Acceptance cut  \\
\hline
Lepton \pt\          &   $\pt>27$~\GeV\\
Lepton  $\eta$       &    $|\eta|<2.5$\\
\hline
$m_{\ell\ell}$ & $m_{\ell\ell} = 91 \pm 15$~\GeV \\
\hline
$b$-jet \pt\          & $\pt > 20$~\GeV\ \\
$b$-jet rapidity & $|y|<2.5$\\
$b$-jet--lepton angular distance &  $\Delta R$(\bjet, $\ell$) > 0.4\\
\hline
\end{tabular}
\caption{Kinematic criteria defining the fiducial phase space of the measurement at particle level.}
\label{tab:fiducial_ps}
\end{center}
\end{table}

The correction of differential distributions is implemented using an iterative Bayesian method of unfolding~\cite{DAgostini} with two iterations.
Simulated events are used to generate a response matrix for each distribution  to account for bin-to-bin migration effects between the detector-level and particle-level distributions.
The matrix is filled with the events that pass both the detector-level and particle-level selections.
The particle-level prediction is used as the initial prior to determine the first estimate of the unfolded data distribution. For the second iteration,
the new estimate of unfolded data is obtained using the background-subtracted data and an unfolding matrix, which is derived on the basis of the Bayes' theorem from the response matrix and the current prior.
The background-subtracted  data are corrected for the expected fraction of events which pass the detector-level selection, but not the particle-level one (unmatched-events), before entering the iterative unfolding.
For each bin of each differential distribution, the unfolded event yields are divided by the integrated luminosity of the data sample and by the bin width, to obtain the  cross-section measurement.
The differential cross-section measurement of a given  observable in the $i$-th bin is given by:
\[
\sigma_i = \dfrac{1}{\epsilon _i L} \sum U_{ij} f_j   N_j^{\textrm{bsD}}\textrm{,}
\]
where $L$ is the integrated luminosity, $\epsilon _i$ is the reconstruction efficiency in $i$-th bin, $N_j^{\textrm{bsD}}$ is the number of background-subtracted data events in
the $j$-th bin, $f_j$ is the factor that corrects for  unmatched  events in the $j$-th bin, and $U_{ij}$ is the element $(i,j)$ of the unfolding matrix calculated after two iterations, using the updated prior from the first iteration and the response matrix.
 
The measurement of the inclusive cross-section for  \Zbos\ events with at least one or at least two \bjets\  is
obtained by applying a particle-level correction to the number of events in data with at least one or at least two \bjets, after background subtraction.
The  correction, which is applied as a divisor of the background-subtracted data, is derived from the ratio of the total number of reconstructed  events in the detector-level phase space to the number of  particle-level events in the fiducial phase space. It is  $0.399 \pm 0.001$ for \Zbos\ events with at least one \bjet\ and $0.258 \pm 0.002$ for \Zbos\ events with at least two \bjets, using \sherpa signal samples and quoting the statistical error.
 
Since the electron and muon decay channels are combined to increase the precision of the signal fits to data, the corrections and response matrices are made using electron and muon signal samples to obtain combined particle-level yields.
To validate this procedure, the analysis is performed for each of the two lepton channels separately.
The results obtained from the individual  channels are compatible within $1.4\sigma$ and  $1.6\sigma$ with the inclusive cross-section of  \Zbos\ events with at least one \bjet\ and at least two \bjets, respectively.
This comparison uses only the sum in quadrature of the statistical and uncorrelated systematic uncertainties.
The differential cross-section measurements in the two channels also agree over the full range of each distribution.


\section{Uncertainties in the cross-section measurements}
\label{sec:UnfoldingSys}
\begin{figure}[h!]
\begin{center}
\begin{tabular}{lr}
\includegraphics[width=.49\linewidth]{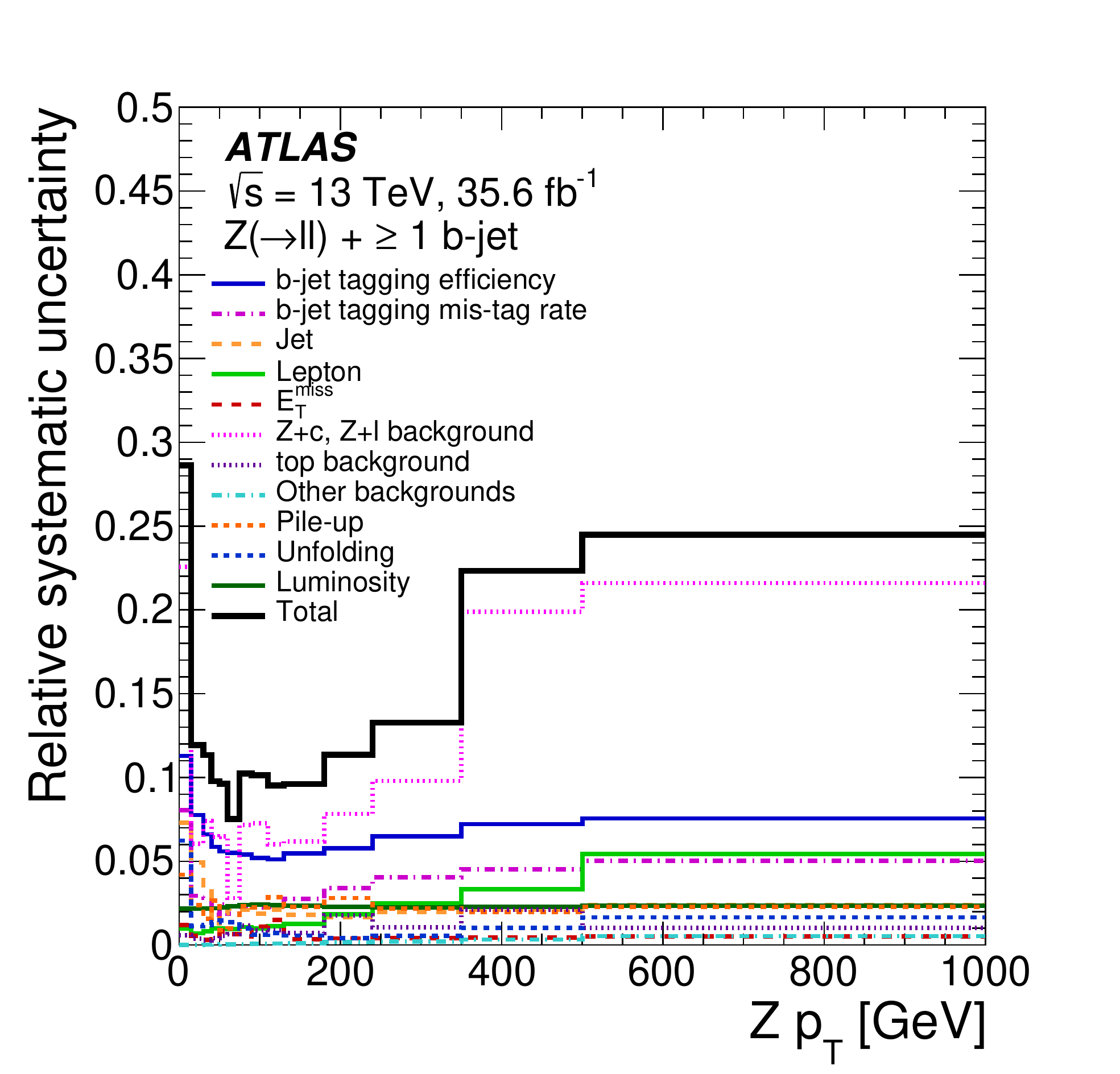}
\includegraphics[width=.49\linewidth]{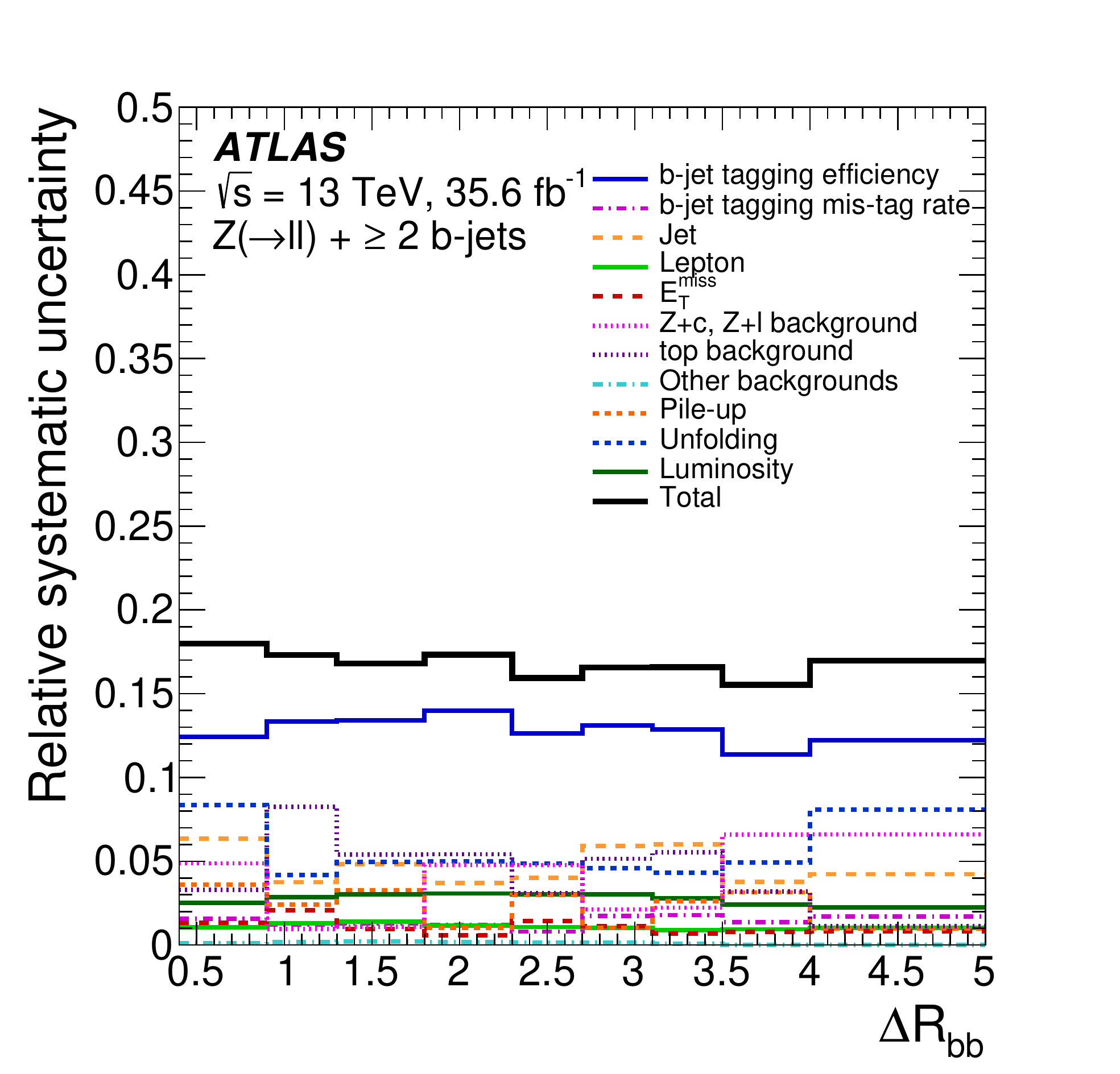}
\end{tabular}
\end{center}
\caption{
Relative systematic uncertainties in the fiducial cross-section as a function of the  \Zbos\ \pt\ in events with at least  one \bjet\ (left)
and as a function of the $\Delta R$ between the two leading \bjets\ in events with at least  two \bjets\ (right).
The total uncertainty is shown in black while the different components listed in Table~\ref{tab:unc_Zll} are shown in different colours.}
\label{fig:sys_break_example}
\end{figure}

Table~\ref{tab:unc_Zll} summarises the systematic uncertainties of the inclusive \Zbjet s cross-sections in the one- and two-$b$-tag regions.
Figure~\ref{fig:sys_break_example} shows as an example the breakdown of the systematic uncertainties in the cross-section as a function of \Zbos\ \pt\ for events with at least one $b$-jet and as a function of $\Delta R_{bb}$ for events with at least two $b$-jets.
 
\begin{table}[h!]
\centering
\begin{tabular}{l|c|c}\toprule
Source of uncertainty& $Z(\to \ell\ell)~+ \ge1$ \bjet & $Z(\to \ell\ell)~+ \ge2$ \bjet s\\
& [$\%$] &[$\%$] \\
\midrule
\bjet\ tagging  efficiency & 7.0 & 14~~~~~~\\
\bjet\ mistag rate & 2.4 & 1.1\\
Jet & 2.4 & 5.0\\
Lepton  & 0.8 & 1.2\\
$\MET$ &  0.6&1.3\\
$Z$ + $c$ and $Z$ + $l$ backgrounds & 4.5 & 1.1\\
Top  background &  0.5 & 3.8\\
Other backgrounds & <0.1~~ & 0.1\\
Pile-up &  1.7 & 2.6\\
Unfolding & 3.8& 4.1\\
Luminosity & 2.3 & 2.9\\
\midrule
Total [\%] & 10~~~~~~& 16~~~~~~\\
\bottomrule
\end{tabular}
 
\caption{Relative systematic uncertainties in the measured production cross-sections of $Z(\to \ell \ell)~+\geq$~1~\bjet\ and $Z(\to \ell \ell)~+\geq$~2~\bjet s events.
The ``Jet'' term includes the JES, JER and JVT uncertainties.
The ``Lepton'' term includes the lepton trigger, efficiency, scale and resolution uncertainties.
The ``$Z+c$ and $Z+l$ backgrounds'' term also includes the $Z+1b$ background in the  $Z$ $+ \ge$ 2 \bjets\ measurement.}
\label{tab:unc_Zll}
\end{table}
 
The systematic uncertainties in the cross-sections associated with the detector-level uncertainty sources described in Section~\ref{sec:mcCorr}
are derived for each observable by propagating systematic shifts from each  source through both the response matrices (unfolding factor) and the subtracted background contributions into the unfolded data for the differential (inclusive) cross-section measurements.
The dominant source of uncertainty  is the modelling of the $b$-tagging efficiency.
Its impact on the inclusive cross-section ranges from 7.0\%
for $Z$-boson events with at least one \bjet\  to 14\% for $Z$-boson events with at least two \bjets.
Its effect on differential cross-section measurements ranges from 5\%  to 10\% for $Z$-boson events with at least one \bjet\ and from 10\%  to 15\% for $Z$-boson events with at least two \bjets.
The impact of the mistag rate of $c$- and light-jets is smaller; it is 2.4\% for $Z$-boson events with at least one \bjet\  and 1\% for $Z$-boson events with at least two \bjets.
 
The uncertainty from each background source is determined by applying shifts to the subtracted background contributions and to the nominal response matrices or unfolding factors.
The sources of uncertainty considered  for  $Z+l$ and $Z+c$ (and $Z+1b$ in the $Z$ $+ \ge$ 2\bjets\ measurement), $t\bar{t}$ and single-top,  diboson and other minor backgrounds are  described in Section~\ref{sec:bkgEstimate}.
The dominant uncertainty in the background to events with at least one \bjet\ originates from $Z$+jets events. This uncertainty contributes 4.5\% to the uncertainty in the inclusive cross-section.
An uncertainty of 3.7\% derives from the difference between the modelling in \alpgen and \sherpa, while 2.6\% is due to the flavour fit uncertainty.
The impact of this uncertainty on the differential cross-sections ranges from a few per cent up to 25\% in the extreme corners of the phase space. For a \Zbos\ \pt\ value of about 500~\GeV, the   difference between the modelling in \alpgen and \sherpa contributes 18\% to this uncertainty, and the flavour fit uncertainty is 12\%.
 
In contrast, the uncertainty in the estimation of background from $t\bar{t}$ events is the dominant source of uncertainty in the background to $Z$-boson events with at least two \bjets.
It contributes 3.8\% to the inclusive cross-section and ranges from 1\% to 9\% in the differential cross-sections.
 
The uncertainty due to modelling of the \Zbjet s\ signal samples in the events with at least one and at least two \bjets\ are also accounted for.
This is evaluated for each observable by reweighting the generator-level distribution in the \sherpa samples to provide a better description of the data at detector level.
The modified \sherpa samples are then used to emulate data and are unfolded with  the nominal simulated sample.
An additional source accounts for the possible mismodelling of an observable  that is not one of the unfolded observables (i.e. a hidden variable).
This uncertainty is evaluated by reweighting, in the \sherpa samples, the generator-level distribution of the leading lepton's \pt, which  is one of the observables showing the largest mismodelling,
to provide a better description of the data at detector level. The modified \sherpa samples are used to unfold the data.
The effect of the hidden variable's mismodelling is negligible for all considered variables and all bins.
A third uncertainty source  accounts for the different hadronisation and parton-shower models used for the signal simulation.
This uncertainty is evaluated by unfolding the \alpgen signal samples, which emulate the background-subtracted data, with the \sherpa signal samples.
The generator-level distributions from the \alpgen samples are first reweighted to agree with
\sherpa in order to remove effects related to shape differences. The difference between the generator-level distribution and the unfolded \alpgen reweighted distribution is taken
as the uncertainty.
For the inclusive cross-section, the modelling uncertainty is estimated by replacing the unfolding factor computed with \sherpa with the one computed with \alpgen.
The dependence on the size of the simulated sample is derived using pseudo-experiments, and the spread of the results is taken as an uncertainty.
The  statistical term  is typically less than a few per cent. It reaches 5\% in the last bin of the  $\Delta R_{bb}$ distribution and
15\% only in the last bin of the $\Delta y_{bb}$ distribution.
 
The total unfolding uncertainty in the inclusive cross-sections is at the level of 4\% in each of the two signal regions.
In the differential distributions it is less than 5\% in the 1-tag region and at a level of 5\%--10\% in the 2-tag region,
except in some bins of the angular variables and in the tail of the \pt\ and $m_{bb}$ distributions, where it reaches 20\%.

\section{Results}
\label{sec:result}
 
 
The inclusive and differential cross-section measurements for $Z$ $+ \ge$ 1 \bjet\ and $Z$ $+ \ge$ 2 \bjets\ are shown in Figures~\ref{fig:pred_XS_incl}--\ref{fig:pred_diBJets_PtjjMjj}.
The statistical uncertainty of the data is  propagated through the unfolding by using 1000 pseudo-experiments, repeating the flavour fit for each of them. The statistical uncertainty in the inclusive cross-sections of
$Z$ $+ \ge$ 1 \bjet\ and $Z$ + $\ge$ 2 \bjets\ is 0.3$\%$ and 0.8$\%$ respectively.
As mentioned in Section~\ref{sec:UnfoldingSys}, the systematic uncertainties are propagated through the unfolding via the response matrices or the unfolding factors and via the variation of the subtracted background.
The measurements are compared with the predictions from \sherpa \textsc{5FNS (NLO)}, \alpgen + \pythiasixshort  \textsc{4 FNS (LO)}, \sherpaF \textsc{4FNS+5FNS (NLO)},
\sherpaZbbNLO , \textsc{MGaMC} + \pythiaeightshort \textsc{5FNS (LO)},
\textsc{MGaMC} + \pythiaeightshort \textsc{Zbb 4FNS (NLO)} and \textsc{MGaMC} + \pythiaeightshort \textsc{5FNS (NLO)}.
Theoretical uncertainties of \sherpa \textsc{5FNS (NLO)}, computed as described in Section~\ref{sec:DataMC}, are shown in the comparison with data.
In this section, all predictions are normalised to their own cross-section to allow an unbiased comparison among different generators.\footnote{The NNLO cross-section $K$-factor
applied to the inclusive \alpgen and \sherpa samples in previous sections is removed.}

\subsection{Inclusive cross-sections}
 
\begin{figure}[h!]
\begin{center}
\begin{tabular}{lr}
\includegraphics[width=.49\linewidth]{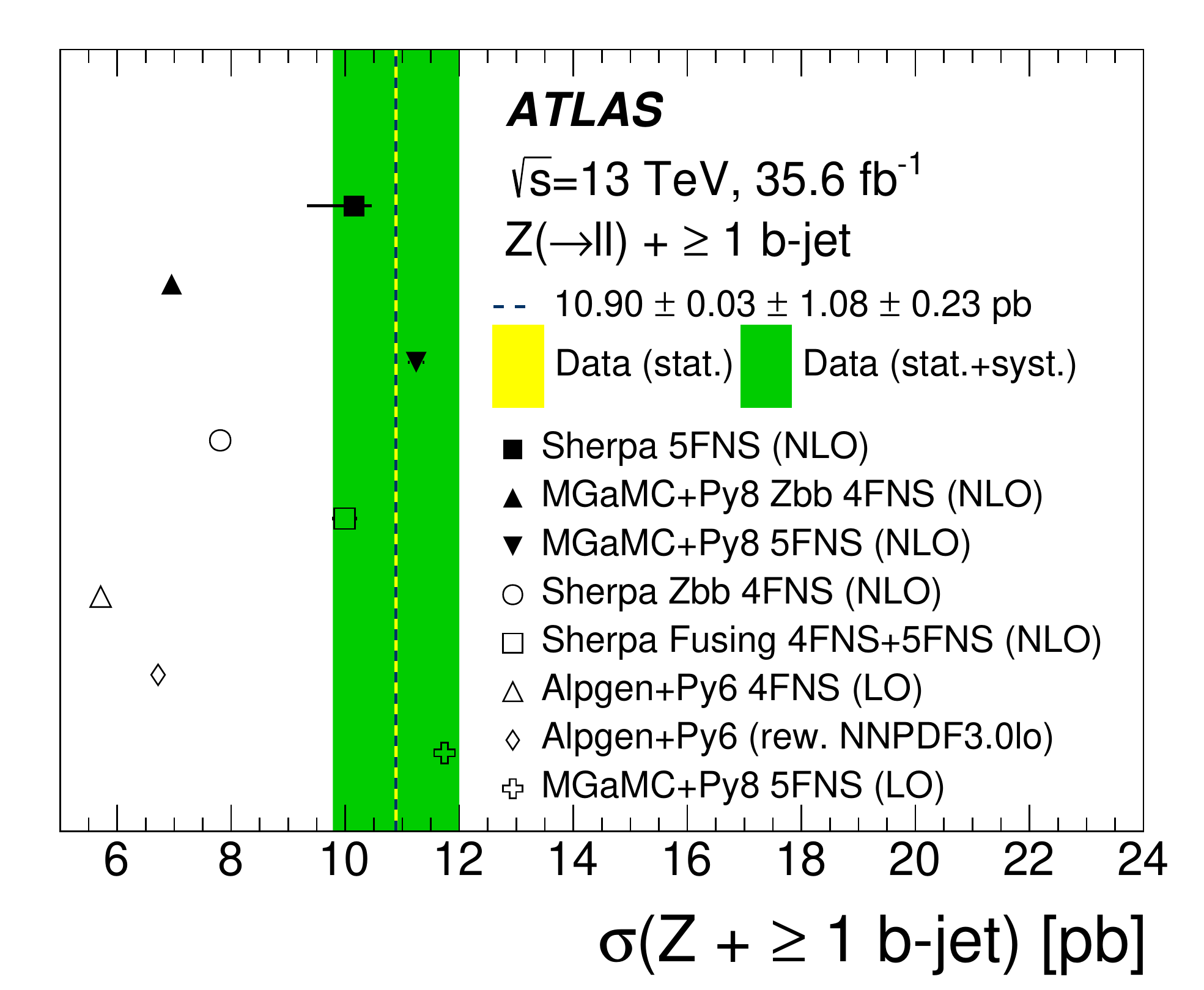}
\includegraphics[width=.49\linewidth]{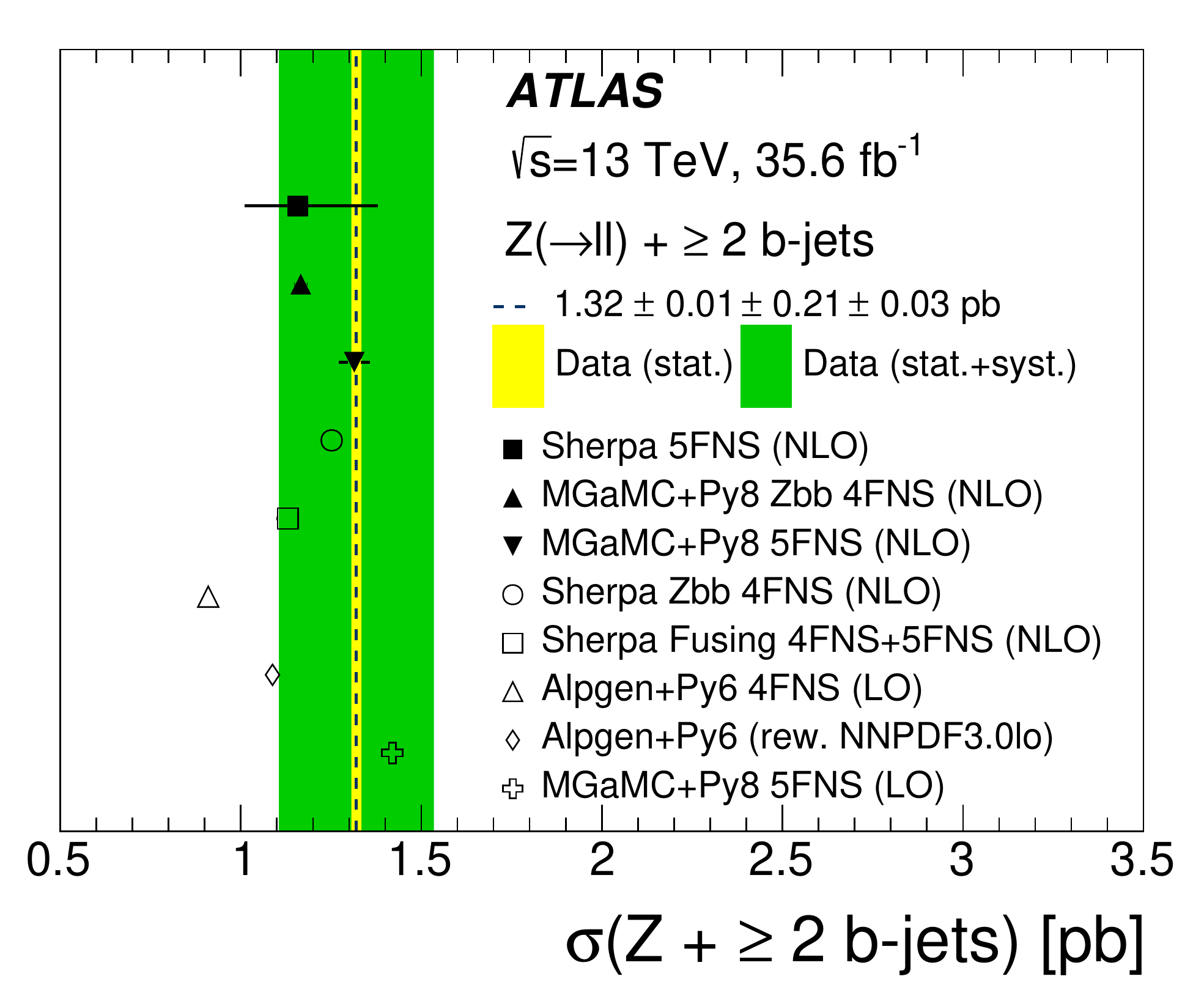}
\end{tabular}
\end{center}
\caption{
Measured cross-sections for $Z$ + $\ge$ 1 \bjet\ (left) and $Z$ + $\ge$ 2 \bjet s (right). The data are compared with the predictions from
\sherpa \textsc{5FNS (NLO)},  \alpgen + \pythiasixshort  \textsc{4 FNS (LO)},  \sherpaF \textsc{4FNS+5FNS (NLO)},
\sherpaZbbNLO , \textsc{MGaMC} + \pythiaeightshort \textsc{5FNS (LO)},  \textsc{MGaMC} + \pythiaeightshort \textsc{Zbb 4FNS (NLO)}  and
\textsc{MGaMC} + \pythiaeightshort \textsc{5FNS (NLO)}.
The yellow band corresponds to the statistical uncertainty of the data, and the green band to statistical and systematic uncertainties of the data, added in quadrature.
The error bars on the \sherpa \textsc{5FNS (NLO)} predictions correspond to the statistical and theoretical uncertainties added in quadrature. Only statistical uncertainties are shown for the other predictions.
}
\label{fig:pred_XS_incl}
\end{figure}

The measured inclusive cross-sections for $Z$ $+ \ge$ 1 \bjet\ and $Z$ + $\ge$ 2 \bjet s, shown in Figure~\ref{fig:pred_XS_incl}, are
$10.90\pm 0.03 (\textrm{stat.})\pm 1.08 (\textrm{syst.})\pm 0.25 (\textrm{lumi.})$~pb and $1.32\pm0.01 (\textrm{stat.})\pm 0.21 (\textrm{syst.})\pm 0.04(\textrm{lumi.})$~pb, respectively.
The  4FNS MC predictions are systematically lower than data in the  inclusive one-\bjet\ case, both for MC generators with LO matrix elements, as implemented in \alpgen + \pythiasixshort  \textsc{4FNS (LO)}, and for $Zbb$ predictions at NLO, as implemented in \sherpaZbbNLO and \textsc{MGaMC} + \pythiaeightshort \textsc{Zbb 4FNS (NLO)}.
The 4FNS predictions agree well with data in the inclusive two-\bjet\ case.
Even though the LO  \alpgen + \pythiasixshort  \textsc{4FNS (LO)} underestimates the data, the predictions and data agree within two standard deviations ($2\sigma$) of the experimental uncertainty.
Use of the NNPDF3.0lo PDF set in \alpgen predictions gives better agreement with data because of a higher acceptance in the fiducial region.
The 5FNS simulations, in general, adequately predict the inclusive cross-sections for both $Z$ $+ \ge$ 1 \bjet\ and $Z$ + $\ge$ 2 \bjets.
Overall, this is consistent with the results presented in the  ATLAS measurement at $\sqrt s = 7$~\TeV~\cite{STDM-2012-15}.

\subsection{Differential cross-sections for $Z$ $+ \ge$ 1 \bjet\ }
 
\begin{figure}[h]
\begin{center}
\begin{tabular}{lr}
\includegraphics[width=.49\linewidth]{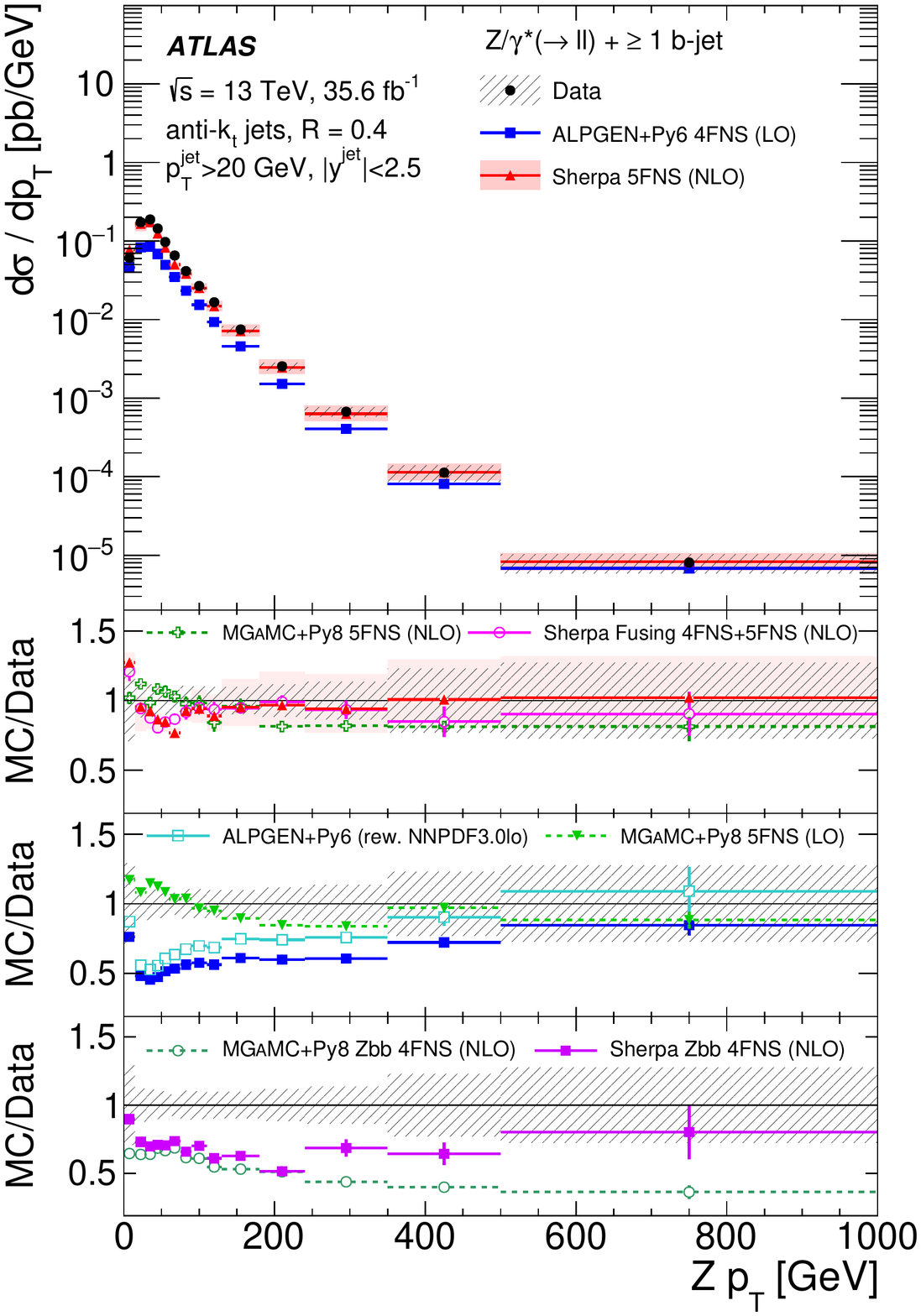}
\includegraphics[width=.49\linewidth]{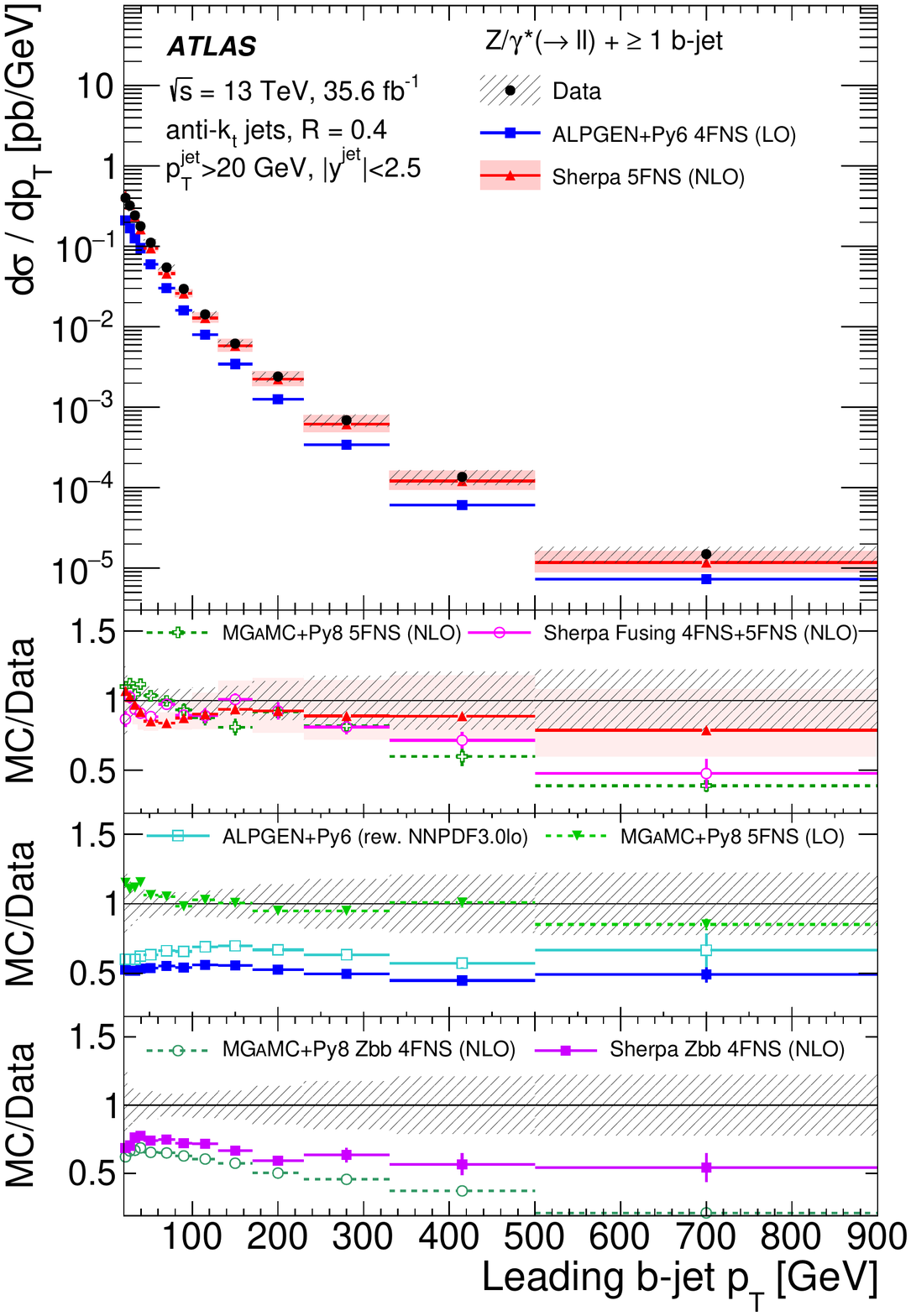}
\end{tabular}
\end{center}
\caption{
Measured cross-section as a function of \Zbos\ \pt\ (left) and leading \bjet\ \pt\ (right) in events with at least one \bjet.
The data are compared with the predictions from
\sherpa \textsc{5FNS (NLO)},  \alpgen + \pythiasixshort  \textsc{4 FNS (LO)},
\sherpaF \textsc{4FNS+5FNS (NLO)},
\sherpaZbbNLO , \textsc{MGaMC} + \pythiaeightshort \textsc{5FNS (LO)},
\textsc{MGaMC} + \pythiaeightshort \textsc{Zbb 4FNS (NLO)}  and
\textsc{MGaMC} + \pythiaeightshort \textsc{5FNS (NLO)}.
The error bars correspond to the statistical uncertainty, and the hatched bands to the data statistical and systematic uncertainties added in quadrature.
The red band corresponds to the statistical and theoretical uncertainties of \sherpa \textsc{5FNS (NLO)} added in quadrature.
Only statistical uncertainties are shown for the other predictions.
}
\label{fig:pred_1BTag_Z_Pt_and_leadBJet_Pt}
\end{figure}

\begin{figure}[h]
\begin{center}
\begin{tabular}{lr}
\includegraphics[width=.49\linewidth]{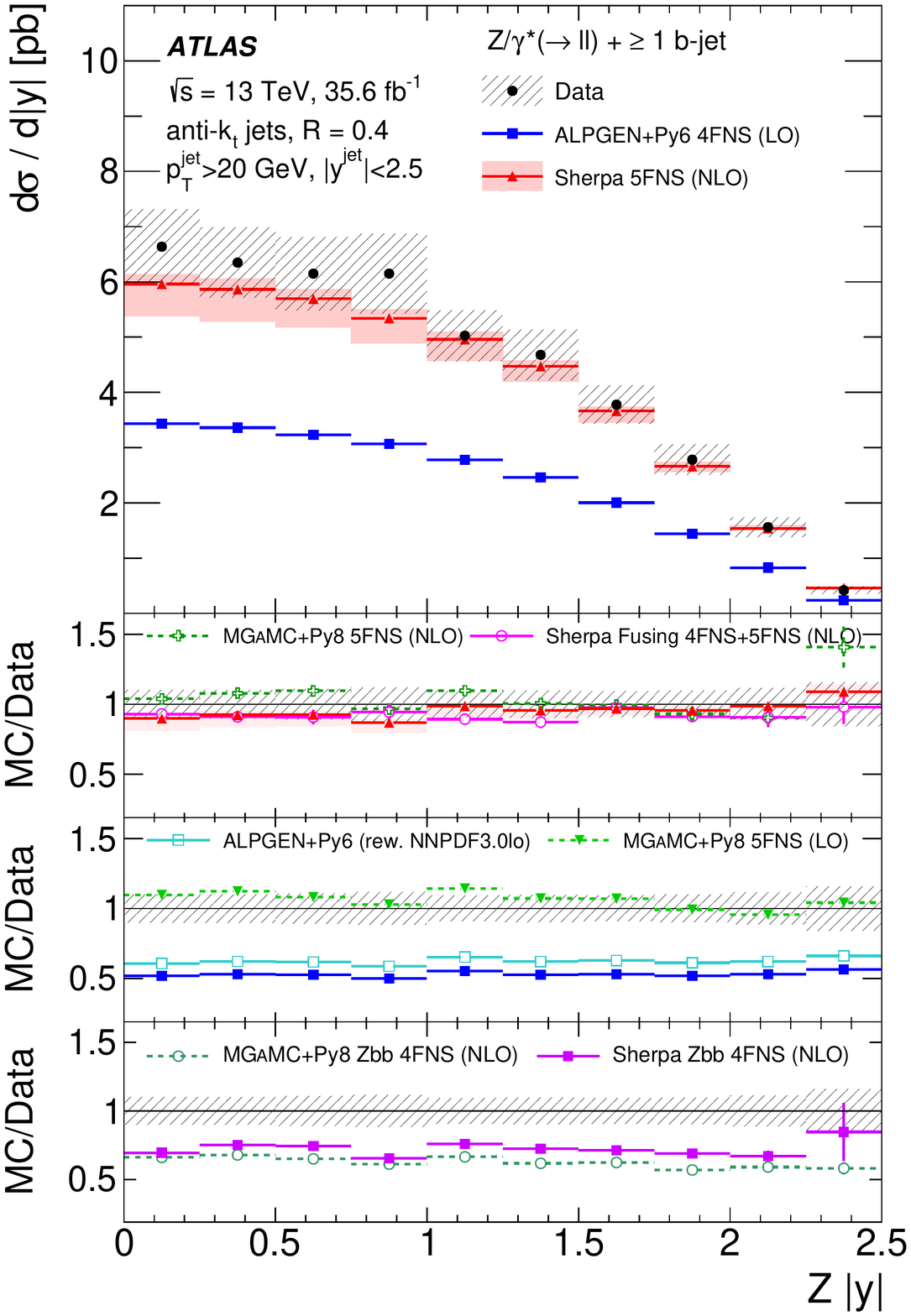}
\includegraphics[width=.49\linewidth]{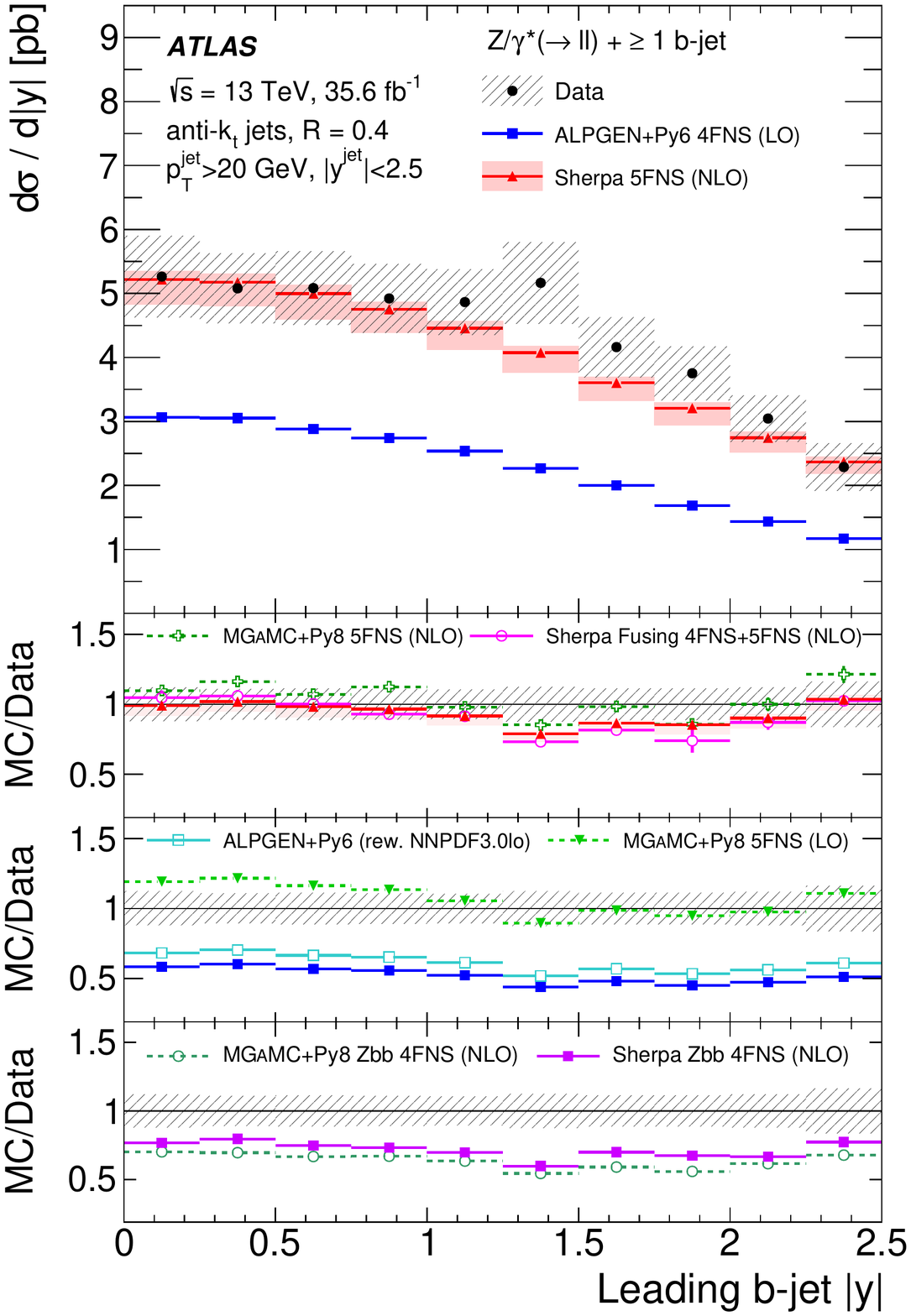}
\end{tabular}
\end{center}
\caption{
Measured cross-section as a function of \Zbos\ $|y|$ (left) and leading \bjet\ $|y|$ (right) in events with at least one \bjet.
The data are compared with the predictions from
\sherpa \textsc{5FNS (NLO)},  \alpgen + \pythiasixshort  \textsc{4 FNS (LO)},
\sherpaF \textsc{4FNS+5FNS (NLO)},
\sherpaZbbNLO , \textsc{MGaMC} + \pythiaeightshort \textsc{5FNS (LO)},
\textsc{MGaMC} + \pythiaeightshort \textsc{Zbb 4FNS (NLO)}  and
\textsc{MGaMC} + \pythiaeightshort \textsc{5FNS (NLO)}.
The error bars correspond to the statistical uncertainty, and the hatched bands to the data statistical and systematic uncertainties added in quadrature.
The red band corresponds to the statistical and theoretical uncertainties of \sherpa \textsc{5FNS (NLO)} added in quadrature.
Only statistical uncertainties are shown for the other predictions.
}
\label{fig:pred_1BTag_Z_Y_and_leadBJet_Y}
\end{figure}

\begin{figure}[h]
\begin{center}
\begin{tabular}{lr}
\includegraphics[width=.49\linewidth]{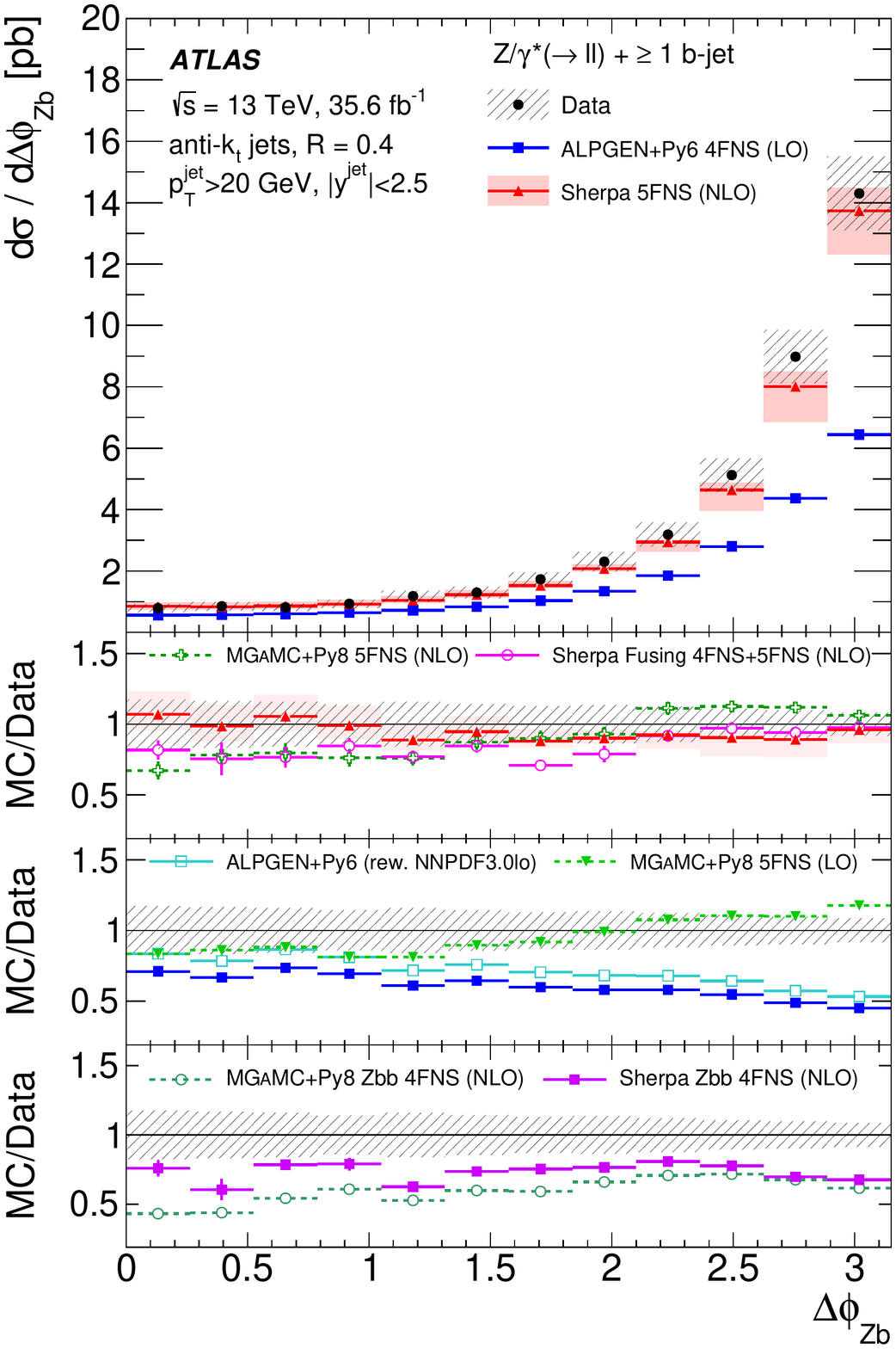}
\includegraphics[width=.49\linewidth]{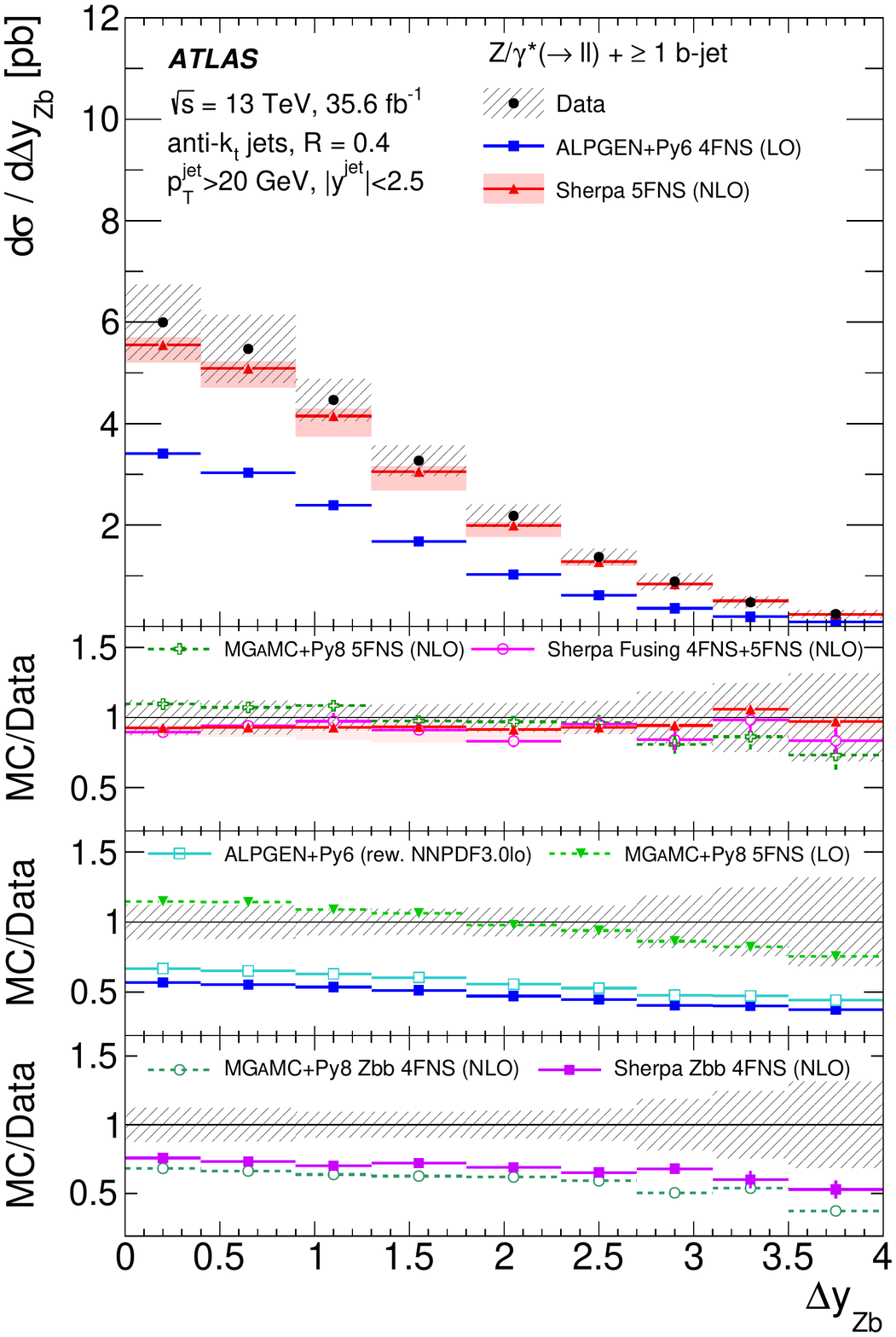}
\end{tabular}
\end{center}
\caption{
Measured cross-section as a function of $\Delta \phi$ (left) and $\Delta y$  between the \Zbos\ candidate and the leading \bjet\ (right) in events with at least one \bjet.
The data are compared with the predictions from
\sherpa \textsc{5FNS (NLO)},  \alpgen + \pythiasixshort  \textsc{4 FNS (LO)},
\sherpaF \textsc{4FNS+5FNS (NLO)},
\sherpaZbbNLO , \textsc{MGaMC} + \pythiaeightshort \textsc{5FNS (LO)},
\textsc{MGaMC} + \pythiaeightshort \textsc{Zbb 4FNS (NLO)}  and
\textsc{MGaMC} + \pythiaeightshort \textsc{5FNS (NLO)}.
The error bars correspond to the statistical uncertainty, and the hatched bands to the data statistical and systematic uncertainties added in quadrature.
The red band corresponds to the statistical and theoretical uncertainties of \sherpa \textsc{5FNS (NLO)} added in quadrature.
Only statistical uncertainties are shown for the other predictions.
}
\label{fig:pred_DPhi_Zb_and_DY_Zb}
\end{figure}

\begin{figure}[h!]
\begin{center}
\begin{tabular}{lr}
\includegraphics[width=.60\linewidth]{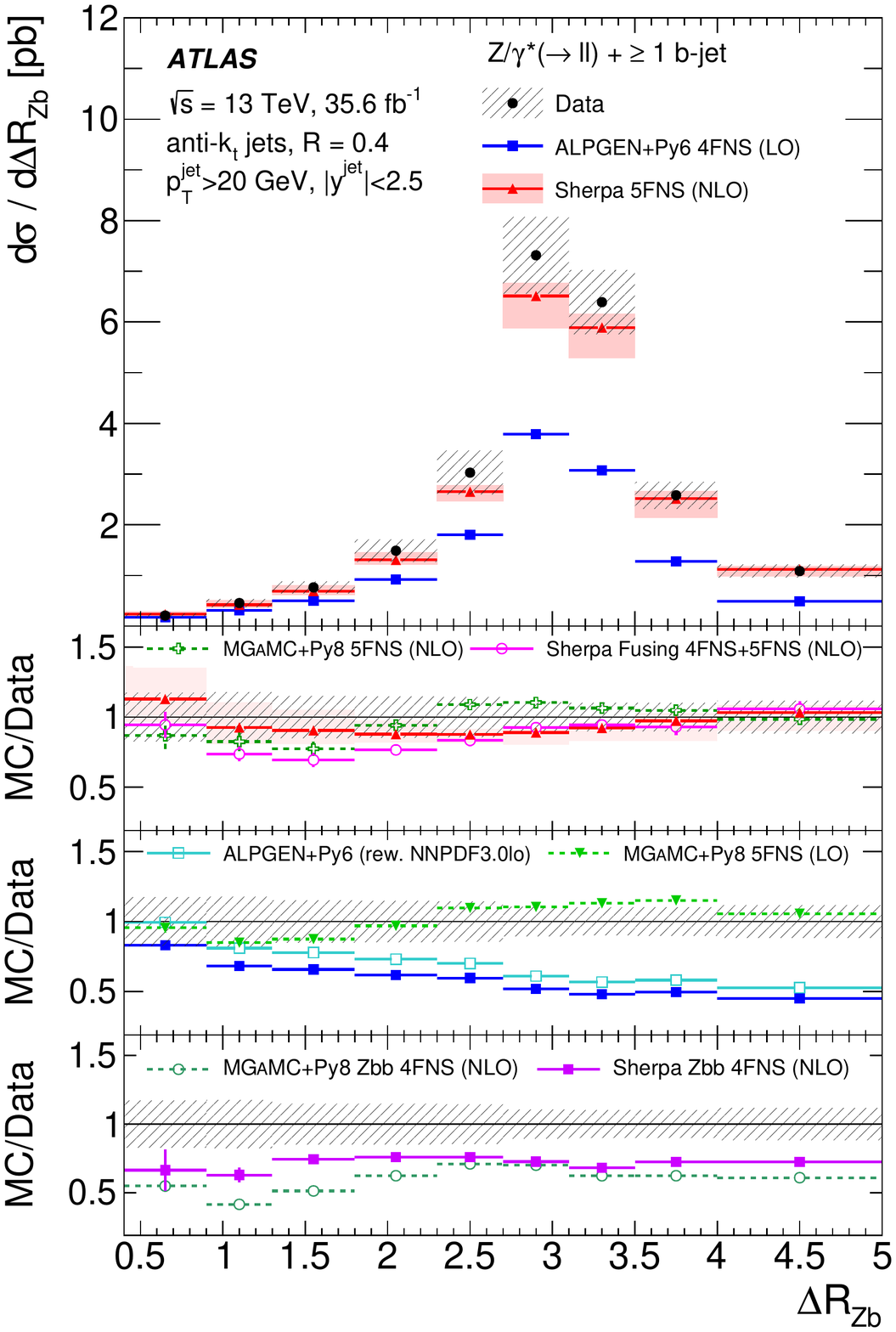}
\end{tabular}
\end{center}
\caption{
Measured cross-section as a function of the \DR\ between the \Zbos\ candidate and the leading \bjet\ in events with at least one \bjet.
The data are compared with the predictions from  \sherpa \textsc{5FNS (NLO)},  \alpgen + \pythiasixshort  \textsc{4 FNS (LO)},
\sherpaF \textsc{4FNS+5FNS (NLO)},
\sherpaZbbNLO , \textsc{MGaMC} + \pythiaeightshort \textsc{5FNS (LO)},
\textsc{MGaMC} + \pythiaeightshort \textsc{Zbb 4FNS (NLO)}  and
\textsc{MGaMC} + \pythiaeightshort \textsc{5FNS (NLO)}.
The error bars correspond to the statistical uncertainty, and the hatched bands to the statistical and systematic uncertainties of the data, added in quadrature.
The red band corresponds to the statistical and theoretical uncertainties of \sherpa \textsc{5FNS (NLO)} added in quadrature.
Only statistical uncertainties are shown for the other predictions.
}
\label{fig:pred_DR_Zb}
\end{figure}

The differential cross-section measurements for the $Z$ $+ \ge$ 1 \bjet\ process are shown in Figures~\ref{fig:pred_1BTag_Z_Pt_and_leadBJet_Pt}--\ref{fig:pred_DR_Zb}.
Each distribution is presented and discussed in detail in this section.
 
The distributions of the transverse momentum  of the $Z$ boson and of the jets probe pQCD over a wide range of scales  and
provide important input to the background prediction for other SM processes, including Higgs boson production and searches beyond the SM.
The differential cross-section as a function of the \Zbos\ \pt\  for events with at least one \bjet\ is shown in Figure~\ref{fig:pred_1BTag_Z_Pt_and_leadBJet_Pt} (left).
In the low \pt\ region, up to 100~\GeV,  where soft radiative effects play a role, all the predicted
shapes except that of \textsc{MGaMC} + \pythiaeightshort \textsc{Zbb 4FNS (NLO)} exhibit trends different from those in the data.
Overall, the predictions from \sherpa \textsc{5FNS (NLO)} and \sherpaF \textsc{4FNS+5FNS (NLO)}  show the best agreement with data.
Predictions from \textsc{MGaMC} + \pythiaeightshort \textsc{5FNS (LO)} and \textsc{MGaMC} + \pythiaeightshort \textsc{5FNS (NLO)} are within the experimental uncertainty band
for most of the bins. The harder \Zbos\ \pt\ in  \alpgen predictions than in data has already been reported by ATLAS for data collected at $\sqrt s = 7$~\TeV~\cite{STDM-2012-15}.
Figure~\ref{fig:pred_1BTag_Z_Pt_and_leadBJet_Pt} (right) shows the leading \bjet\ \pt.
\textsc{MGaMC} + \pythiaeightshort \textsc{5FNS (LO)}   provides a satisfactory description within the uncertainty of the data, while \textsc{MGaMC} + \pythiaeightshort \textsc{5FNS (NLO)}  underestimates the data in the high \pt\ region. This region is populated by additional hard radiation, which in  \textsc{MGaMC} + \pythiaeightshort \textsc{5FNS (NLO)} is simulated only
via parton shower.
\sherpa \textsc{5FNS (NLO)}  exhibits the best agreement with data.
The contrasting behaviour of \sherpaF \textsc{4FNS+5FNS (NLO)}, which underestimates the data at high \pt, may be interesting to investigate further in the future.
The NLO 4FNS predictions of $Zbb$, as implemented in \sherpa and \textsc{MGaMC}, show a softer leading \bjet\ \pt, while the
inclusive LO 4FNS prediction, as implemented in \alpgen, describes the shape of the data quite well despite the large underestimation of the normalisation already discussed for Figure~\ref{fig:pred_XS_incl}.
 
The distributions of the \Zbos\ rapidity, the leading \bjet\ rapidity, and their separation, ${\Delta y_{Zb}}$, are directly sensitive to the $b$-quark PDFs and
to higher-order diagram contributions, and they may show differences for different flavour schemes.
The differential cross-sections as a function of the \Zbos\ rapidity and of the leading \bjet\ rapidity for events with at least one \bjet\ are shown in Figure~\ref{fig:pred_1BTag_Z_Y_and_leadBJet_Y}.
All MC predictions provide a satisfactory description of the shape of the data.
Some modulation relative to data is observed in the leading \bjet\ $|y|$ distribution, in some cases beyond the experimental uncertainty.
Figure~\ref{fig:pred_DPhi_Zb_and_DY_Zb} (right) shows the differential cross-section as a function of ${\Delta y_{Zb}}$.
\sherpa \textsc{5FNS (NLO)} and \sherpaF \textsc{4FNS+5FNS (NLO)} describe the data quite well, while all other predictions exhibit a slightly smaller rapidity separation than data, even if within the uncertainty of the data.
Use of a different PDF set as in \alpgen predictions leads to a change in the distribution, but the differences are small compared with the experimental uncertainties.
 
The distribution of ${\Delta\phi_{Zb}}$ is sensitive to the presence of additional radiation in the event.
In fixed order calculations of the $Z+1b$ process, the LO matrix element provides contributions only for ${\Delta\phi_{Zb}}= \pi$, while the NLO matrix element is the first order which populates the region of ${\Delta\phi_{Zb}}< \pi$. In MC simulations the region below $\pi$ is populated via parton shower and via merging of parton shower with multi-parton matrix elements.
Therefore the region of
small azimuthal separation between the $Z$ boson and the leading $b$-jet is the most sensitive to additional QCD radiation and soft corrections.
It is also sensitive to the presence of boosted particles decaying into a $Z$ boson and $b$-quarks.
The differential cross-section as a function of  ${\Delta\phi_{Zb}}$ for events with at least one \bjet\ is shown in Figure~\ref{fig:pred_DPhi_Zb_and_DY_Zb} (left).
The \sherpa \textsc{5FNS (NLO)} generator provides the best agreement with data.  \sherpaF \textsc{4FNS+5FNS (NLO)} is still consistent with data within the experimental uncertainty in most of the bins, but a small difference between the two simulations is observed for small values.
This result is highly correlated with the difference observed in the leading \bjet\ \pt\ distribution.
It confirms that the current performance of \sherpaF \textsc{4FNS+5FNS (NLO)} in the regime of high-\pt\ jets with a $Z$ boson emitted collinearly is slightly worse than the \sherpa \textsc{5FNS (NLO)} configuration.
All \textsc{MGaMC} simulations predict too many large azimuthal separations, with a consequent deficit at small angles. Also, in this case
the modelling in \textsc{MGaMC} + \pythiaeightshort \textsc{5FNS (NLO)} is slightly worse than in \textsc{MGaMC} + \pythiaeightshort \textsc{5FNS (LO)}.  The differential cross-section as a function of ${\Delta R_{Zb}}$, as shown in Figure~\ref{fig:pred_DR_Zb},  contains the convolution of effects discussed for the ${\Delta y_{Zb}}$ and ${\Delta\phi_{Zb}}$ distributions.

\subsection{Differential cross-sections for $Z$ $+ \ge$ 2 \bjets\ }
 
\begin{figure}[h]
\begin{center}
\begin{tabular}{lr}
\includegraphics[width=.49\linewidth]{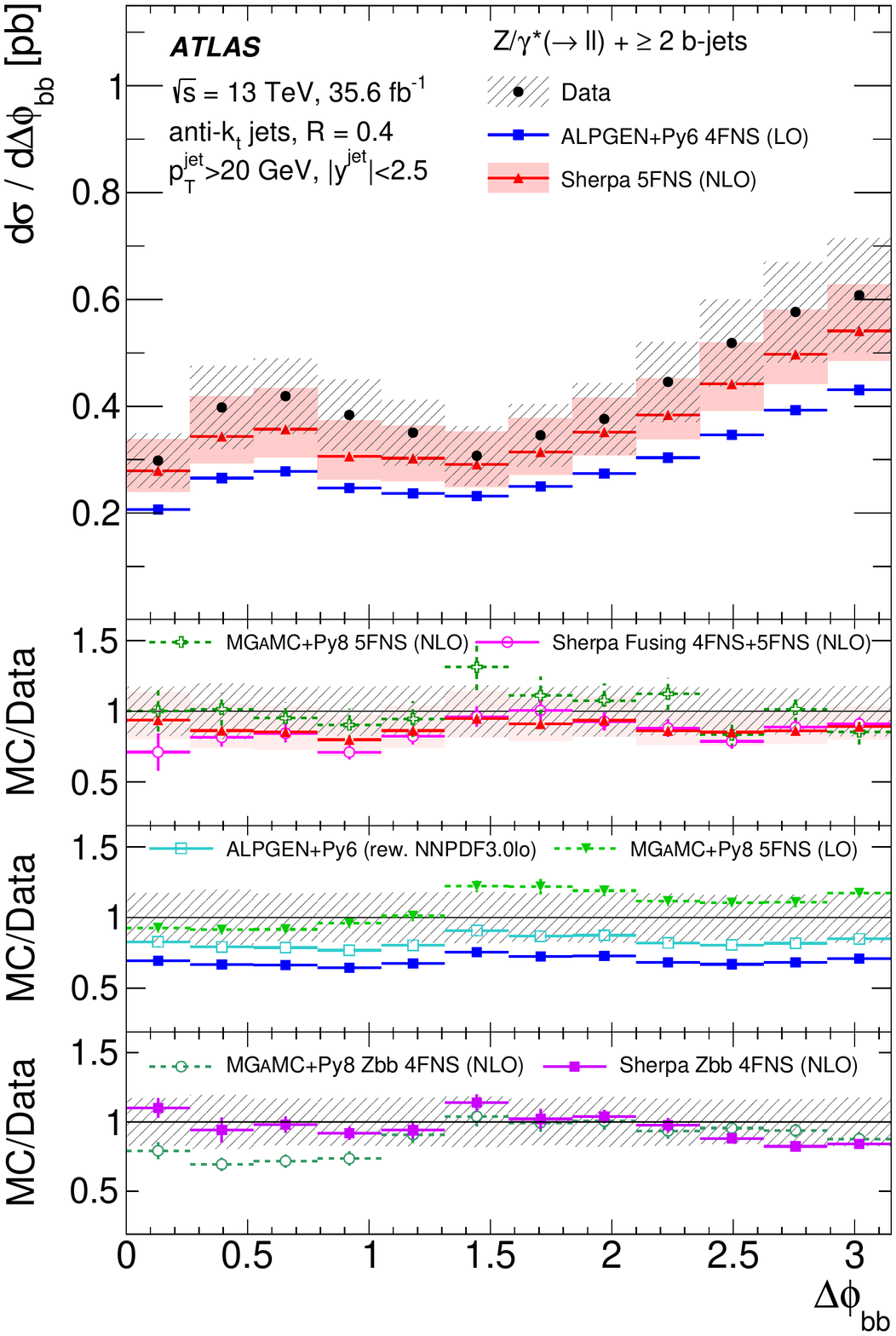}
\includegraphics[width=.49\linewidth]{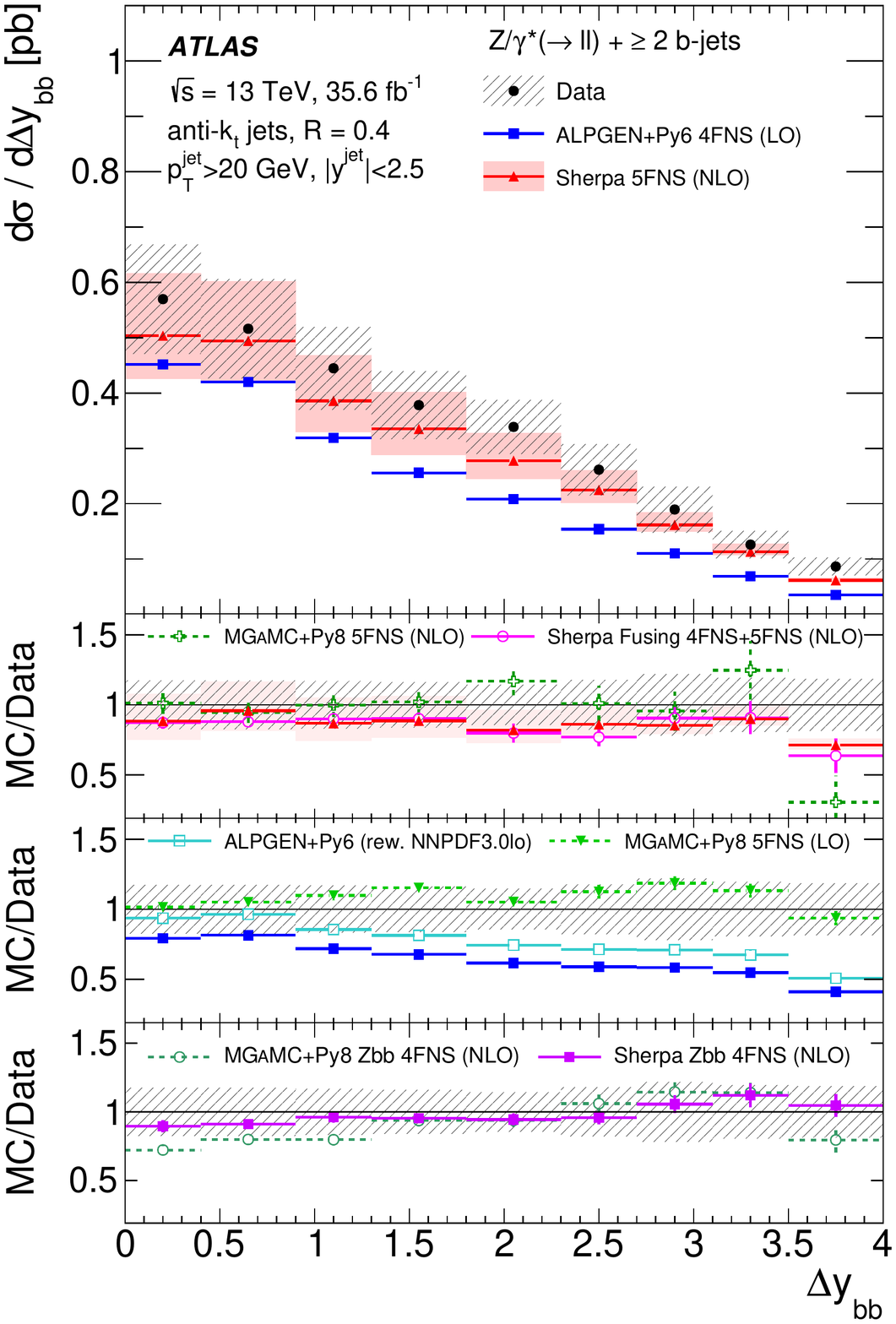}
\end{tabular}
\end{center}
\caption{
Measured cross-section as a function of $\Delta \phi$ (left) and $\Delta y$ between the two leading \bjet s (right) in events with at least two \bjet s.
The data are compared with the predictions from
\sherpa \textsc{5FNS (NLO)},  \alpgen + \pythiasixshort  \textsc{4 FNS (LO)},
\sherpaF \textsc{4FNS+5FNS (NLO)},
\sherpaZbbNLO , \textsc{MGaMC} + \pythiaeightshort \textsc{5FNS (LO)},
\textsc{MGaMC} + \pythiaeightshort \textsc{Zbb 4FNS (NLO)}  and
\textsc{MGaMC} + \pythiaeightshort \textsc{5FNS (NLO)}.
The error bars correspond to the statistical uncertainty, and the hatched bands to the data statistical and systematic uncertainties added in quadrature.
The red band corresponds to the statistical and theoretical uncertainties of \sherpa \textsc{5FNS (NLO)} added in quadrature.
Only statistical uncertainties are shown for the other predictions.
}
\label{fig:pred_DPhi_bb_and_DY_bb}
\end{figure}

\begin{figure}[h]
\begin{center}
\begin{tabular}{lr}
\includegraphics[width=.49\linewidth]{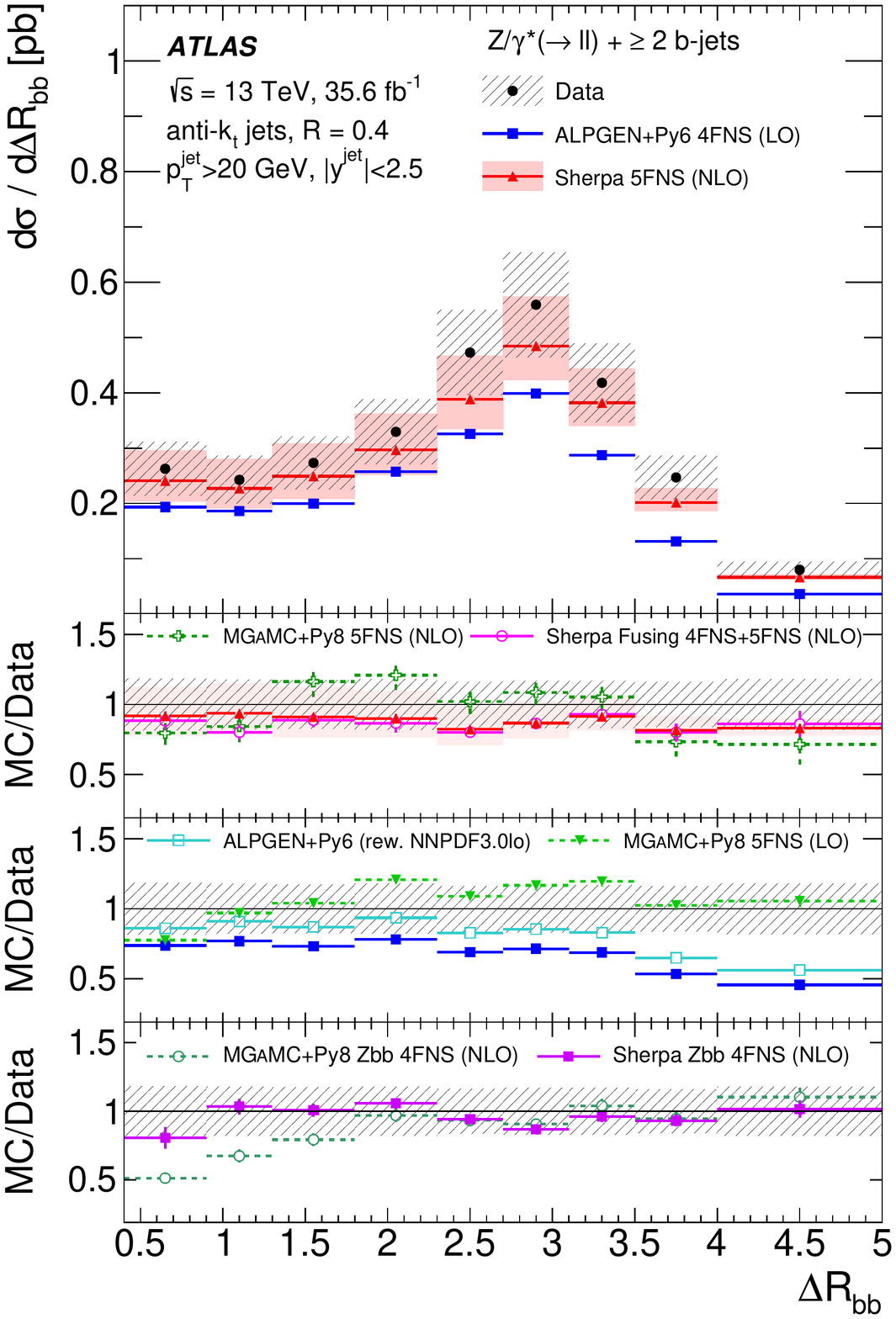}
\includegraphics[width=.49\linewidth]{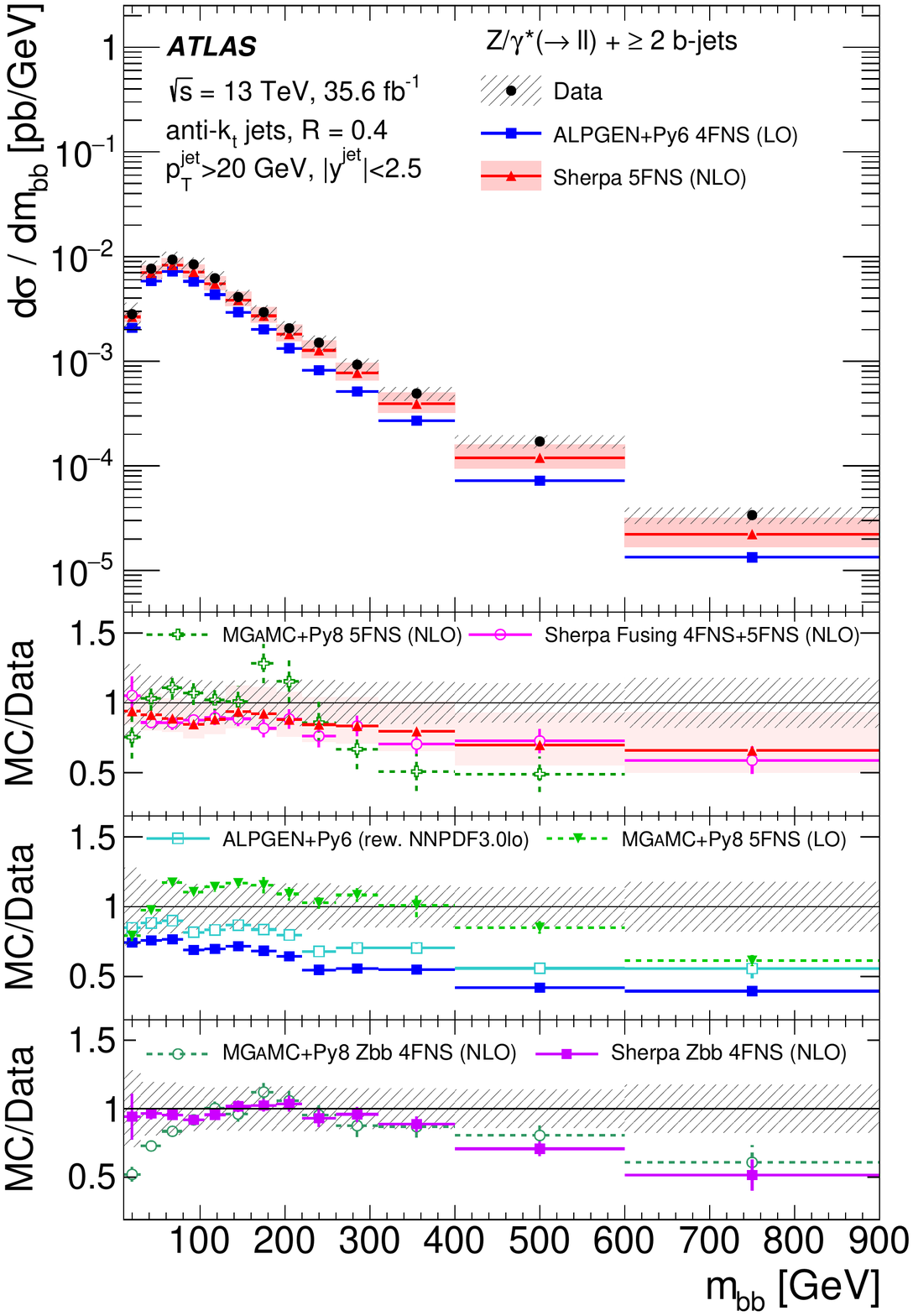}
\end{tabular}
\end{center}
\caption{
Measured cross-section as a function of $\Delta R$ between the two leading \bjet s (left) and invariant mass of the two leading \bjet s (right) in events with at least two \bjet s.
The data are compared with the predictions from
\sherpa \textsc{5FNS (NLO)},  \alpgen + \pythiasixshort  \textsc{4 FNS (LO)},
\sherpaF \textsc{4FNS+5FNS (NLO)},
\sherpaZbbNLO , \textsc{MGaMC} + \pythiaeightshort \textsc{5FNS (LO)},
\textsc{MGaMC} + \pythiaeightshort \textsc{Zbb 4FNS (NLO)}  and
\textsc{MGaMC} + \pythiaeightshort \textsc{5FNS (NLO)}.
The error bars correspond to the statistical uncertainty, and the hatched bands to the data statistical and systematic uncertainties added in quadrature.
The red band corresponds to the statistical and theoretical uncertainties of \sherpa \textsc{5FNS (NLO)} added in quadrature.
Only statistical uncertainties are shown for the other predictions.
}
\label{fig:pred_DR_bb_and_M_bb}
\end{figure}

\begin{figure}[h]
\begin{center}
\begin{tabular}{lr}
\includegraphics[width=.49\linewidth]{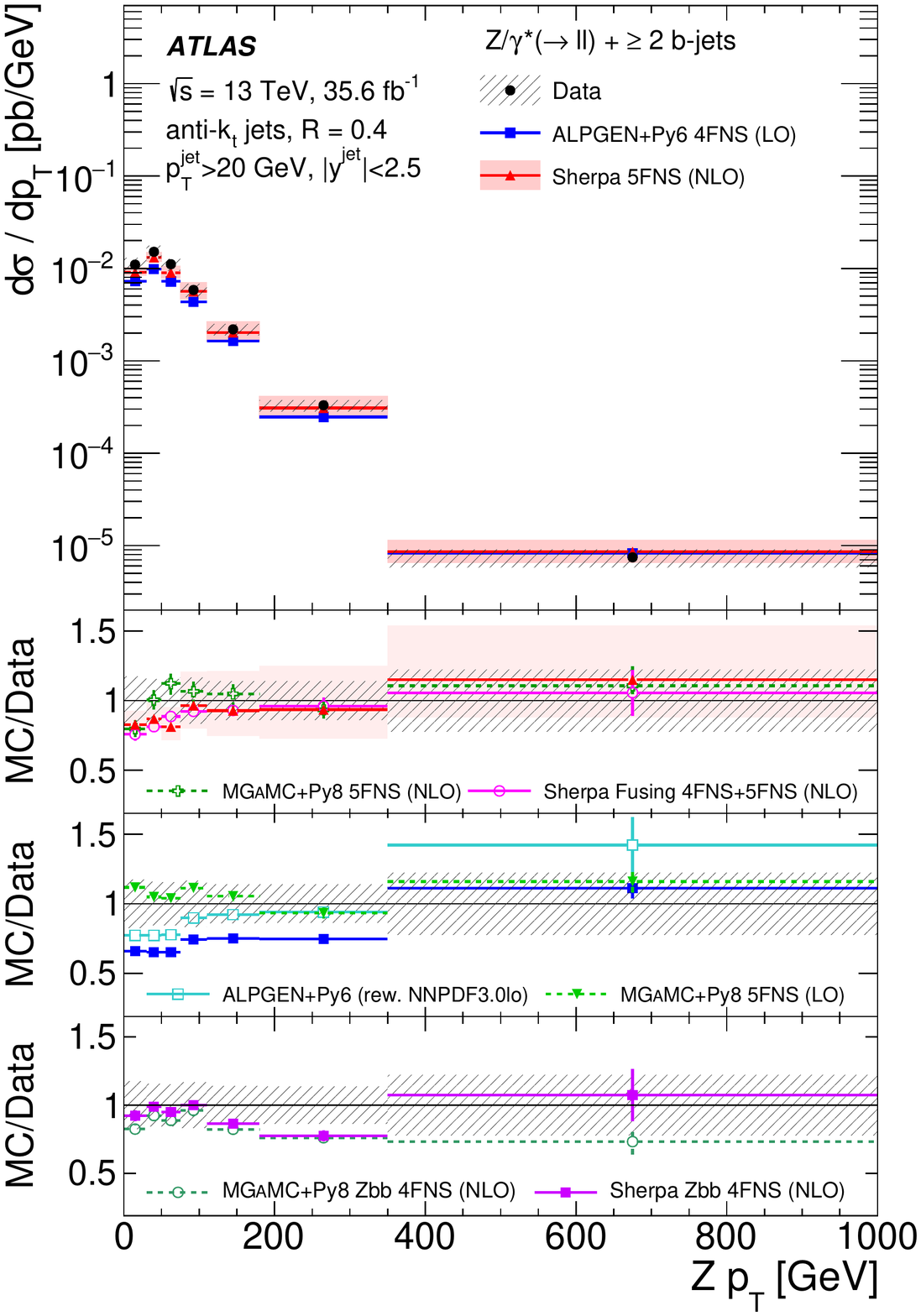}
\includegraphics[width=.49\linewidth]{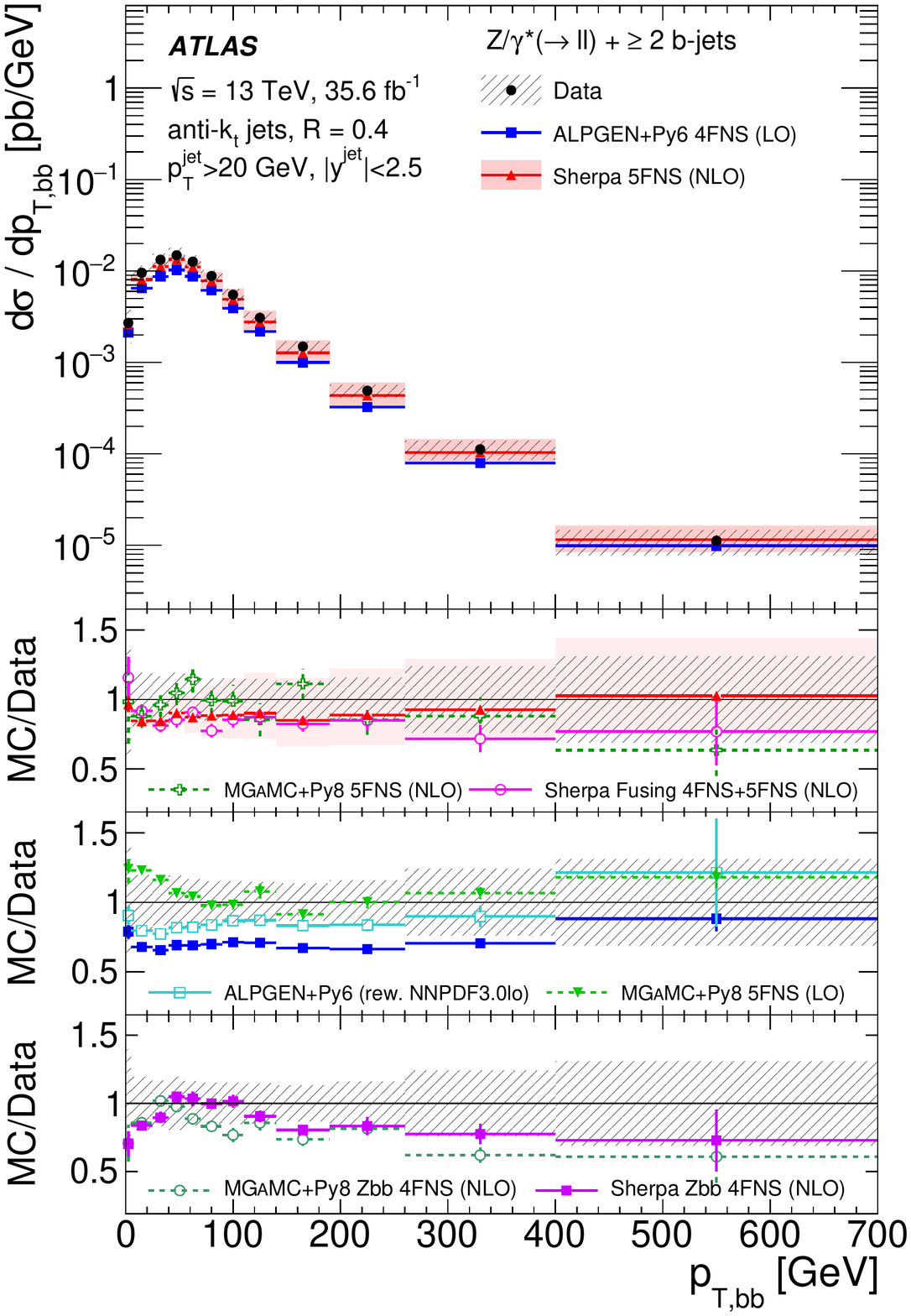}
\end{tabular}
\end{center}
\caption{
Measured cross-section as a function of \pt\ of the $Z$ boson (left) and  of the di-\bjet\ system (\ptbb ) (right) in events with at least two \bjet s.
The data are compared with the predictions from
\sherpa \textsc{5FNS (NLO)},  \alpgen + \pythiasixshort  \textsc{4 FNS (LO)},
\sherpaF \textsc{4FNS+5FNS (NLO)},
\sherpaZbbNLO , \textsc{MGaMC} + \pythiaeightshort \textsc{5FNS (LO)},
\textsc{MGaMC} + \pythiaeightshort \textsc{Zbb 4FNS (NLO)}  and
\textsc{MGaMC} + \pythiaeightshort \textsc{5FNS (NLO)}.
The error bars correspond to the statistical uncertainty, and the hatched bands to the data statistical and systematic uncertainties added in quadrature.
The red band corresponds to the statistical and theoretical uncertainties of \sherpa \textsc{5FNS (NLO)} added in quadrature.
Only statistical uncertainties are shown for the other predictions.
}
\label{fig:pred_2BTag_Z_Pt_diBJets_Pt}
\end{figure}

\begin{figure}[h!]
\begin{center}
\begin{tabular}{lr}
\includegraphics[width=.60\linewidth]{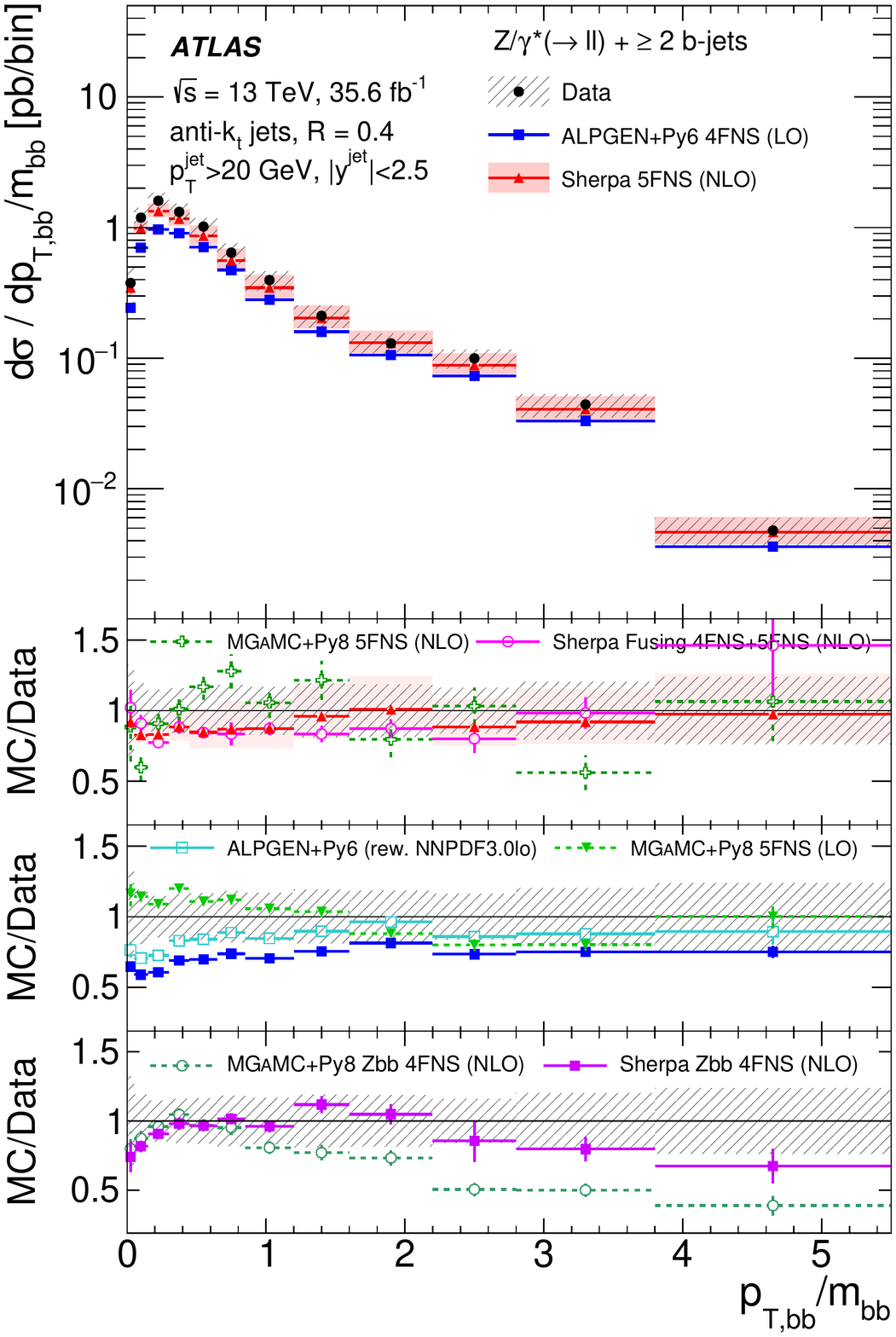}
\end{tabular}
\end{center}
\caption{
Measured cross-section as a function of the \pt\ of the di-\bjet\ system  divided by its invariant mass \mbox{(\ptmbb)} in events with at least two \bjet s.
The data are compared with the predictions from  \sherpa \textsc{5FNS (NLO)},  \alpgen + \pythiasixshort  \textsc{4 FNS (LO)},
\sherpaF \textsc{4FNS+5FNS (NLO)},
\sherpaZbbNLO , \textsc{MGaMC} + \pythiaeightshort \textsc{5FNS (LO)},
\textsc{MGaMC} + \pythiaeightshort \textsc{Zbb 4FNS (NLO)}  and
\textsc{MGaMC} + \pythiaeightshort \textsc{5FNS (NLO)}.
The error bars correspond to the statistical uncertainty, and the hatched bands to the statistical and systematic uncertainties of the data, added in quadrature.
The red band corresponds to the statistical and theoretical uncertainties of \sherpa \textsc{5FNS (NLO)} added in quadrature.
Only statistical uncertainties are shown for the other predictions.
}
\label{fig:pred_diBJets_PtjjMjj}
\end{figure}

Events with a $Z$ boson produced in association with two \bjets\ constitute an important background to other SM and beyond-SM processes.
Furthermore, they probe the mechanism of a gluon splitting into heavy quarks.
The differential cross-section measurements for $Z$ $+ \ge$ 2 \bjet\  are shown in Figures~\ref{fig:pred_DPhi_bb_and_DY_bb}--\ref{fig:pred_diBJets_PtjjMjj}.
Each distribution is presented and discussed in detail in this section.
 
The distributions of angular separation between the two leading \bjets\ allow characterisation of the hard radiation at large angles  and the soft radiation for collinear emissions.
The differential cross-sections as a function of  ${\Delta\phi_{bb}}$ and of ${\Delta y_{bb}}$ are shown in Figure~\ref{fig:pred_DPhi_bb_and_DY_bb}.
Most of the predictions provide satisfactory descriptions of the data within the large experimental uncertainties.
Disagreement between data and \textsc{MGaMC} + \pythiaeightshort \textsc{Zbb 4FNS (NLO)} is observed at low values of ${\Delta \phi_{bb}}$.
Mismodelling of ${\Delta y_{bb}}$  is observed for \alpgen.
This observable has some sensitivity to PDFs, but that is below the experimental uncertainties.
The ${\Delta R_{bb}}$ observable is sensitive to the various production mechanisms of the $Zbb$ final state. The region at low  ${\Delta R_{bb}}$ is dominated by the production of two \bjets\ from gluon splitting.
Probing this region requires two \bjets\ in the final state, so it is not sensitive to very small angles of the splitting.
The interplay of the modelling of ${\Delta\phi_{bb}}$ and ${\Delta y_{bb}}$ in \alpgen + \pythiasixshort  \textsc{4 FNS (LO)}
influences the prediction of the ${\Delta R_{bb}}$ distribution shown in Figure~\ref{fig:pred_DR_bb_and_M_bb} (left).
All \sherpa predictions describe the shape of this observable quite well, featuring a substantial improvement at low ${\Delta R_{bb}}$
relative to the LO version reported by ATLAS using data at $\sqrt s= 7$~\TeV.
Overall, this is consistent with the results presented in the  ATLAS measurement of gluon-splitting properties  at $\sqrt s = 13$~\TeV~\cite{STDM-2012-15}.
\textsc{MGaMC} + \pythiaeightshort \textsc{Zbb 4FNS (NLO)} presents a large mismodelling at low ${\Delta R_{bb}}$, which is the part of the phase space dominated by gluon splitting.
 
The invariant mass of the two leading \bjets\ is  an important observable in the measurement of associated $ZH$ production with Higgs boson decays into $b\bar{b}$, and in searches for physics beyond the SM in the same final state.
The differential cross-section as a function of ${m_{bb}}$ for events with at least two \bjets\ is shown in Figure~\ref{fig:pred_DR_bb_and_M_bb} (right).
All \sherpa predictions  provide a quite good model of the shape of this observable's distribution up to about  300~\GeV, while the other predictions show various discrepancies in this region.
This is particularly evident for \textsc{MGaMC} + \pythiaeightshort \textsc{Zbb 4FNS (NLO)}, and it is consistent with the mismodelling observed at low ${\Delta R_{bb}}$, the region dominated by gluon splitting.
In the high mass range all predictions underestimate the data, resulting in a sizeable mismodelling.
Hence the use of these predictions for the background estimate in searches for physics beyond the SM in this final state could be problematic.

The differential cross-sections as a function of the \Zbos\ \pt\ and of the \pt\ of the di-\bjet\  system (\ptbb) for events with at least two \bjets\
are shown in Figure~\ref{fig:pred_2BTag_Z_Pt_diBJets_Pt}.
Most of the predictions agree with data within the large experimental uncertainties, which are about 25$\%$ in most of the bins, and large statistical uncertainties of the predictions, which for some MC samples reach 25$\%$ in the highest bins.
\alpgen shows a harder \Zbos\ \pt\ spectrum than data, as was observed in the distribution of events with at least one \bjet.
The $Zbb$ simulation at NLO with 4FNS, as implemented in \textsc{MGaMC} + \pythiaeightshort \textsc{Zbb 4FNS (NLO)}
and \sherpaZbbNLO, shows better agreement with data with respect to the \pt\ distributions for events with at least one \bjet, but significant disagreement is still observed.
 
Finally, the ratio of the \pt\ of the  di-\bjet\ system to its invariant mass (\ptmbb) is sensitive to gluon splitting: a small value indicates a hard splitting and a large value is a consequence of soft splitting.
The differential cross-section as a function of \ptmbb\ is shown in Figure~\ref{fig:pred_diBJets_PtjjMjj}.
\sherpa \textsc{5FNS (NLO)} and \sherpaF \textsc{4FNS+5FNS (NLO)}   show quite good agreement with data,
while  \textsc{MGaMC} + \pythiaeightshort \textsc{Zbb 4FNS (NLO)} agrees less well.
 
\FloatBarrier

\section{Conclusion}
\label{sec:conclusion}
 
This paper presents  a measurement of the cross-sections for \Zbos\ production in association with one or more \bjets\
in $pp$ collisions at $\sqrt{s} = 13$~\TeV. The analysed data correspond to an integrated luminosity of 35.6~fb$^{-1}$ recorded by the ATLAS detector at the LHC.
 
The cross-sections are measured using the electron and muon decay modes of the $Z$ boson in a fiducial phase space.
In addition to the inclusive cross-sections, differential cross-sections  of several kinematic observables are measured, extending the range of jet transverse momenta to higher values than reported in previous ATLAS publications, which used data at lower centre-of-mass energies.
 
The measurements are compared with predictions from a variety of Monte Carlo generators.
In general, 5-flavour number scheme (5FNS) calculations at NLO accuracy predict the inclusive cross-sections well, while inclusive 4-flavour number scheme (4FNS) LO calculations largely underestimate the data.
Predictions of $Zbb$ at NLO accuracy agree with data only in the two-\bjets\ case, and underestimate the data in the case of events with at least one \bjet.
Overall, \sherpa \textsc{5FNS (NLO)}, a 5FNS generator with matrix elements at NLO for up to two partons and matrix elements at LO for up to four partons, describes the various differential distributions within the experimental uncertainties.
A significant discrepancy, common to all generators, is found for large values of $m_{bb}$.
The \sherpaF \textsc{4FNS+5FNS (NLO)} simulation, which combines 4FNS with 5FNS  at NLO accuracy using a novel technique, agrees with \sherpa \textsc{5FNS (NLO)}, showing that in general at the scales tested by this measurement the effects of this merging are minor.
A disagreement of about 20 30\%  is observed for large values of the leading \bjet\ transverse momentum, and for small angular separations between the $Z$ boson and the leading \bjet.

The 5FNS simulation with matrix elements for up to four partons at LO, as implemented in \textsc{MGaMC} + \pythiaeightshort (LO), describes the data
within the experimental uncertainties in most cases.
In some cases this simulation is even better than predictions from \textsc{MGaMC} + \pythiaeightshort \textsc{5FNS (NLO)}, which  has matrix elements with only one parton at NLO. This indicates the importance of simulations with several partons in the matrix element for a fair description of the data.
The pure $Zbb$ simulation at NLO in the 4FNS, as generated by \sherpa  and \textsc{MGaMC}, shows significant deviations from the data even in the two-\bjets\ configuration, and this is more pronounced in \textsc{MGaMC}.
 
This measurement provides essential input for the improvement of theoretical predictions and Monte Carlo generators of $Z$-boson production in association with \bjets, allowing a better quantitative understanding of perturbative QCD.
 
\section*{Acknowledgements}
 

We thank CERN for the very successful operation of the LHC, as well as the
support staff from our institutions without whom ATLAS could not be
operated efficiently.
 
We acknowledge the support of ANPCyT, Argentina; YerPhI, Armenia; ARC, Australia; BMWFW and FWF, Austria; ANAS, Azerbaijan; SSTC, Belarus; CNPq and FAPESP, Brazil; NSERC, NRC and CFI, Canada; CERN; CONICYT, Chile; CAS, MOST and NSFC, China; COLCIENCIAS, Colombia; MSMT CR, MPO CR and VSC CR, Czech Republic; DNRF and DNSRC, Denmark; IN2P3-CNRS and CEA-DRF/IRFU, France; SRNSFG, Georgia; BMBF, HGF and MPG, Germany; GSRT, Greece; RGC and Hong Kong SAR, China; ISF and Benoziyo Center, Israel; INFN, Italy; MEXT and JSPS, Japan; CNRST, Morocco; NWO, Netherlands; RCN, Norway; MNiSW and NCN, Poland; FCT, Portugal; MNE/IFA, Romania; MES of Russia and NRC KI, Russia Federation; JINR; MESTD, Serbia; MSSR, Slovakia; ARRS and MIZ\v{S}, Slovenia; DST/NRF, South Africa; MINECO, Spain; SRC and Wallenberg Foundation, Sweden; SERI, SNSF and Cantons of Bern and Geneva, Switzerland; MOST, Taiwan; TAEK, Turkey; STFC, United Kingdom; DOE and NSF, United States of America. In addition, individual groups and members have received support from BCKDF, CANARIE, Compute Canada and CRC, Canada; ERC, ERDF, Horizon 2020, Marie Sk{\l}odowska-Curie Actions and COST, European Union; Investissements d'Avenir Labex, Investissements d'Avenir Idex and ANR, France; DFG and AvH Foundation, Germany; Herakleitos, Thales and Aristeia programmes co-financed by EU-ESF and the Greek NSRF, Greece; BSF-NSF and GIF, Israel; CERCA Programme Generalitat de Catalunya and PROMETEO Programme Generalitat Valenciana, Spain; G\"{o}ran Gustafssons Stiftelse, Sweden; The Royal Society and Leverhulme Trust, United Kingdom.
 
The crucial computing support from all WLCG partners is acknowledged gratefully, in particular from CERN, the ATLAS Tier-1 facilities at TRIUMF (Canada), NDGF (Denmark, Norway, Sweden), CC-IN2P3 (France), KIT/GridKA (Germany), INFN-CNAF (Italy), NL-T1 (Netherlands), PIC (Spain), ASGC (Taiwan), RAL (UK) and BNL (USA), the Tier-2 facilities worldwide and large non-WLCG resource providers. Major contributors of computing resources are listed in Ref.~\cite{ATL-GEN-PUB-2016-002}.
 
 
\clearpage

\printbibliography

\clearpage
 
\begin{flushleft}
\hypersetup{urlcolor=black}
{\Large The ATLAS Collaboration}

\bigskip

\AtlasOrcid[0000-0002-6665-4934]{G.~Aad}$^\textrm{\scriptsize 102}$,    
\AtlasOrcid[0000-0002-5888-2734]{B.~Abbott}$^\textrm{\scriptsize 128}$,    
\AtlasOrcid{D.C.~Abbott}$^\textrm{\scriptsize 103}$,    
\AtlasOrcid[0000-0002-2788-3822]{A.~Abed~Abud}$^\textrm{\scriptsize 36}$,    
\AtlasOrcid[0000-0002-1002-1652]{K.~Abeling}$^\textrm{\scriptsize 53}$,    
\AtlasOrcid[0000-0002-2987-4006]{D.K.~Abhayasinghe}$^\textrm{\scriptsize 94}$,    
\AtlasOrcid[0000-0002-8496-9294]{S.H.~Abidi}$^\textrm{\scriptsize 166}$,    
\AtlasOrcid[0000-0002-8279-9324]{O.S.~AbouZeid}$^\textrm{\scriptsize 40}$,    
\AtlasOrcid{N.L.~Abraham}$^\textrm{\scriptsize 155}$,    
\AtlasOrcid[0000-0001-5329-6640]{H.~Abramowicz}$^\textrm{\scriptsize 160}$
% ...
\bigskip
\\

% ...
$^{36}$CERN, Geneva; Switzerland.\\
% ...
$^{40}$Niels Bohr Institute, University of Copenhagen, Copenhagen; Denmark.\\
% ...
$^{53}$II. Physikalisches Institut, Georg-August-Universit\"{a}t G\"ottingen, G\"ottingen; Germany.\\
% ...
$^{94}$Department of Physics, Royal Holloway University of London, Egham; United Kingdom.\\
% ...
$^{102}$CPPM, Aix-Marseille Universit\'e, CNRS/IN2P3, Marseille; France.\\
$^{103}$Department of Physics, University of Massachusetts, Amherst MA; United States of America.\\
% ...
$^{128}$Homer L. Dodge Department of Physics and Astronomy, University of Oklahoma, Norman OK; United States of America.\\
% ...
$^{155}$Department of Physics and Astronomy, University of Sussex, Brighton; United Kingdom.\\
% ...
$^{160}$Raymond and Beverly Sackler School of Physics and Astronomy, Tel Aviv University, Tel Aviv; Israel.\\
% ...
$^{166}$Department of Physics, University of Toronto, Toronto ON; Canada.\\
% ...

\end{flushleft}

 
\end{document}